\documentclass[11pt,a4paper]{article}  
\usepackage{jheppub}
\usepackage{url,ulem}
\usepackage{afterpage}
\usepackage{latexsym}
\usepackage{amsmath}
\usepackage{lscape}
\usepackage{epstopdf}
\usepackage{subfigure}

\newcommand{\lsim}
{\;\raisebox{-.3em}{$\stackrel{\displaystyle <}{\sim}$}\;}

\def\text{\textstyle}

\def\bc{\begin{center}}
\def\ec{\end{center}}
\def\bi{\begin{itemize}}
\def\ei{\end{itemize}}

\title{Constrained Supersymmetry after two years of LHC data: a global
  view with Fittino} 
\author[a]{Philip Bechtle,} 
\author[b]{Torsten Bringmann,} 
\author[a]{Klaus  Desch,} 
\author[a,c]{Herbi Dreiner,} 
\author[d]{Matthias Hamer,}
\author[d]{Carsten Hensel,}
\author[e]{Michael Kr\"amer,} 
\author[f]{Nelly Nguyen,}
\author[g]{Werner Porod,} 
\author[h]{Xavier Prudent,}
\author[i]{Bj\"orn Sarrazin,}
\author[a]{Mathias Uhlenbrock,}
\author[a]{and Peter Wienemann}
 \affiliation[a]{Physikalisches
  Institut, University of Bonn, Bonn, Germany}
\affiliation[b]{{II.} Institute for Theoretical Physics, University of Hamburg, 
Hamburg, Germany}
\affiliation[c]{Bethe
  Center for Theoretical Physics, University of Bonn, Bonn, Germany}
\affiliation[d]{II.\ Physikalisches Institut, University of
  G\"ottingen, G\"ottingen, Germany} 
\affiliation[e]{Institute for
  Theoretical Particle Physics and Cosmology, RWTH Aachen University,
  Aachen, Germany} 
\affiliation[f]{Institute for Experimental Physics,
  University of Hamburg, Hamburg, Germany} 
\affiliation[g]{Institut
  f\"ur Theoretische Physik und Astrophysik, University of W\"urzburg,
  W\"urzburg, Germany}
\affiliation[h]{Institut f\"ur Kern- und
  Teilchenphysik, TU Dresden, Dresden, Germany}
\affiliation[i]{Deutsches Elektronen-Synchrotron DESY, Hamburg,
  Germany}

\abstract{We perform global fits to the parameters of the Constrained
  Minimal Supersymmetric Standard Model (CMSSM) and to a variant with
  non-universal Higgs masses (NUHM1). In addition to constraints from
  low-energy precision observables and the cosmological dark matter
  density, we take into account the LHC exclusions from searches in
  jets plus missing transverse energy signatures with about
  5\,fb$^{-1}$ of integrated luminosity.  We also include the most
  recent upper bound on the branching ratio $B_s\to\mu\mu$ from
  LHCb. Furthermore, constraints from and implications for direct and
  indirect dark matter searches are discussed. The best fit of the
  CMSSM prefers a light Higgs boson just above the experimentally
  excluded mass. We find that the description of the low-energy
  observables, $(g-2)_{\mu}$ in particular, and the non-observation of
  SUSY at the LHC become more and more incompatible within the
  CMSSM. A potential SM-like Higgs boson with mass around 126~GeV can
  barely be accommodated. Values for ${\cal B}(B_s\to\mu\mu)$ just
  around the Standard Model prediction are naturally expected in the
  best fit region. The most-preferred region is not yet affected by
  limits on direct WIMP searches, but the next generation of
  experiments will probe this region. Finally, we discuss implications
  from fine-tuning for the best fit regions.}


\begin{document}

\maketitle

\section{Introduction}
\label{sec:intro}

Supersymmetry (SUSY)~\cite{Wess:1974tw} is one of the theoretically
best-motivated extensions of the Standard Model of particle physics
(SM). Furthermore, there are intriguing complementary experimental
findings: the measured coupling constants of the electro-magnetic,
weak, and strong forces unify when extrapolated to high energies using
the renormalization group equations (RGEs) of SUSY, but not with the
RGEs of the SM \cite{Amaldi:1991cn}. The observed dark matter (DM)
density in the Universe \cite{Komatsu:2010fb} can easily be explained
by a stable neutralino lightest supersymmetric particle (LSP), acting
as a weakly interacting massive particle (WIMP)
\cite{Goldberg:1983nd}. The measured anomalous magnetic moment of the
muon \cite{Bennett:2006fi} appears to deviate slightly from its SM
prediction \cite{MuonAnom:Davier} but can be naturally accommodated
within SUSY with relatively light SUSY particles.  Moreover, several
precision observables like the mass of the $W$ boson and the effective
weak mixing angle $\sin^2\theta_{\rm eff}$, and some rare B-meson
decay branching ratios are also sensitive to the parameters of SUSY
models. On the other hand, SUSY particles have not been directly
observed at colliders \cite{pre-LHC-SUSY-limits, Aad:2011hh,
  daCosta:2011qk, Aad:2011yf, Aad:2011ks, Aad:2011kta, Aad:2011xk,
  Aad:2011xm, Aad:2011hz, Aad:2011kz, Aad:2011zb, Aad:2011qr,
  Aad:2011ib, ATLAS:2011ad, Aad:2011qa, Aad:2011yh, Aad:2011cwa,
  Aad:2011zj, Aad:2011cw, Aad:2012zn, :2012jk, Khachatryan:2011tk,
  Chatrchyan:2011wc, Chatrchyan:2011bz, Chatrchyan:2011wba,
  Chatrchyan:2011ah, Chatrchyan:2011ff, Chatrchyan:2011bj,
  Chatrchyan:2011ek, Chatrchyan:2011qs, Chatrchyan:2011zy,
  Collaboration:2011ida} and there is no convincing evidence for the
existence of WIMPs from direct or indirect
searches~\cite{Aprile:2011hi,Ahmed:2011gh,talk-collar,Cirelli:2012tf}.

The simplest supersymmetric extension of the SM is the Minimal
Supersymmetric Standard Model
(MSSM)~\cite{Nilles:1983ge,Haber:1984rc,Drees:2004jm}. Here the proton
is stable due to conserved R-parity, which also guarantees the
stability of the LSP. In its most general form (but with massless
neutrinos) the MSSM has 124 free parameters, including the 19 of the
SM.\footnote{Employing the same minimal particle content but
  guaranteeing proton stability instead by baryon triality, there are
  more than 200 free parameters \cite{Dreiner:1997uz,Barbier:2004ez}.}
However, current data is insufficient to place meaningful constraints
on this general model. For phenomenological studies it is thus
essential to consider more restrictive models. The most widely
considered is the constrained MSSM (CMSSM)
\cite{Nilles:1983ge,Drees:2004jm} which has only 5 new free parameters
beyond the SM\footnote{Models with 19-21 new parameters were for
  example considered in
  \cite{Djouadi:2002ze,Berger:2008cq,Bechtle:2005vt}.}
\begin{equation}
M_0,\,M_{1/2},\,A_0,\,\tan\beta,\,\mathrm{sgn}(\mu)\,.
\end{equation}
Here $M_0$ and $M_{1/2}$ denote the universal soft supersymmetry
breaking scalar and gaugino masses at the unification scale,
respectively. $A_0$ is the universal soft supersymmetry breaking
trilinear scalar coupling, $\tan\beta$ is the ratio of the vacuum
expectation values of the two $C\!P$-even neutral Higgs fields and $\mu$
is the Higgs mixing parameter in the superpotential.

Ignoring the direct searches at the ATLAS and CMS experiments at the
Large Hadron Collider (LHC) for now, all data is compatible with the 5
parameters of this highly simplified model (see \textit{e.g.}\
\cite{deAustri:2006pe,Allanach:2007qk,Buchmueller:2008qe,Buchmueller:2009fn,Buchmueller:2010ai,Bechtle:2009ty}). The
strictest constraints come from interpreting the WMAP measurement in
terms of the neutralino relic density~\cite{Komatsu:2010fb} and from the
anomalous magnetic moment of the muon. The best fits prefer $M_0 \lsim
100$\,GeV and $M_{1/2}\lsim 400$\,GeV
\cite{deAustri:2006pe,Allanach:2007qk,Buchmueller:2008qe,Buchmueller:2009fn,Buchmueller:2010ai,Bechtle:2009ty},
see also \cite{AbdusSalam:2011fc}.

In a minimal non-universal Higgs mass model (NUHM1) \cite{Baer:2004fu}
the free parameters are
\begin{equation}
M_0,\,M_H,\,M_{1/2},\,A_0,\,\tan\beta,\,\mathrm{sgn}(\mu)\,.
\end{equation}
In addition to the CMSSM parameters there is a universal scalar Higgs mass parameter at the
unification scale, $M_{H_u}=M_{H_d}=M_H$. It can differ from the other
supersymmetric scalar masses at the unification scale, $M_0$. In NUHM1
fits prior to the LHC data, light supersymmetric masses well below a
TeV are also preferred
\cite{Buchmueller:2008qe,Roszkowski:2009sm,Buchmueller:2009fn,Buchmueller:2010ai}.

In 2011, the LHC experiments ATLAS and CMS have each taken approximately 5\,fb$^{-1}$
of proton-proton collision data at $\sqrt{s} = 7$~TeV. The search for
SUSY final states within these data has not yet revealed any signal.
Instead, rather stringent bounds on the parameters of the CMSSM are
published in preliminary analyses~\cite{ATLAS-CMS-Moriond}. The
strictest bounds come from inclusive searches for events with missing
transverse energy and jets in the final state. 

In this paper, we investigate in detail the consequences of these
exclusions for the global interpretation of all existing data using
our framework \texttt{Fittino}~\cite{Bechtle:2004pc}.
\texttt{Fittino} constructs a global $\chi^2$ variable as a function
of the model parameters and performs a Markov-Chain-Monte-Carlo (MCMC)
scan of the parameter space. For the work presented here, 138~million
points in the parameter space have been found with a $\chi^2$ below~30.

In analyzing these parameter scans we address the following questions:
\begin{itemize}
\item What is the still allowed region in the CMSSM and NUHM1
  parameter space after two years of LHC data?  
\item To what extent are the interpretation of the non-LHC
  measurements within the CMSSM/NUHM1 and the non-observation of SUSY
  at the LHC in mutual tension?
\item What would be the impact of a light SUSY Higgs boson with a mass
  of around $126$\,GeV, compatible with the latest LHC
  results~\cite{ATLAS:2012ae,Collaboration:2012tx}, on the CMSSM/NUHM1 fits,
  and what would be the implication for its branching fractions and
  ratios of branching fractions?
\item What are the implications of the CMSSM/NUHM1 best fits for direct and indirect searches for
  WIMP dark matter?
\item To what extent is the remaining preferred parameter region in
  the CMSSM/NUHM1 fine-tuned?
\end{itemize}

Technically, the methods employed to obtain the results presented in
this paper differ in various aspects from our previous studies
\cite{Bechtle:2009ty,Bechtle:2011dm,Fittino:2011} and from those used
in other global fits with LHC exclusions~\cite{Allanach:2011ut,
  Buchmueller:2011aa, Allanach:2011wi, Bertone:2011nj,
  Buchmueller:2011ki, Buchmueller:2011sw, Fowlie:2011mb,
  Buchmueller:2011ab, Strege:2011pk, Roszkowski:2012uf,
  Beskidt:2012bh}. In contrast to our previous paper, the theoretical
calculations no longer rely on the combination of codes in the
\texttt{Mastercode} package \cite{mastercode} but make use of various
individual codes for the different theoretical calculations (see
Sec.~\ref{sec:ParameterDetermination}). Particular emphasis is placed
on an improved modeling of the LHC exclusions also in parameter
regions away from the published 95\% confidence level (CL) contours.
This is achieved through a fast simulation of the signal for each
tested point in parameter space instead of a simple parametrization
of the CL.  The experimental and theoretical input from direct and
indirect WIMP searches is obtained from
\texttt{AstroFit}~\cite{AstroFit}. In addition to the standard
quantitative evaluation of fine-tuning~\cite{Barbieri:1987fn} we
develop a new, phenomenological method based on parameter
correlations.

The paper is organized as follows. In
Sec.~\ref{sec:ParameterDetermination}, the \texttt{Fittino} framework
and the statistical methods to interpret the global $\chi^2$ variable
are introduced. A comparison of frequentist and Bayesian
interpretation of an example fit is also given. In
Sec.~\ref{sec:input}, we present the experimental input employed in
the fits. In particular, we consider indirect information from
laboratory-based precision experiments, information from dark matter
searches, and constraints from the exclusion of SUSY particles and
Higgs bosons at previous colliders and at the LHC.  We explain in some
detail the method to estimate the LHC confidence levels at arbitrary
points in parameter space through fast simulation. The results of
various fits are presented in Sec.~\ref{sec:results}. We first discuss
global fits with updated observables excluding direct searches at the
LHC, and then global fits including the current LHC exclusions. Then
the impact of a potential Higgs signal is evaluated both within the
CMSSM and the NUHM1 models.  This is followed by a discussion on the
implications of the obtained fit results for direct and indirect dark
matter searches. Subsequently, the fine-tuning of the best-fit regions
is discussed. In addition, we explore parameter correlations as a
measure of the required tuning of parameters with respect to each
other. Furthermore, we present a new way to estimate theory
systematics from a variation of the energy scale $Q$ at which the weak
scale mass parameters are evaluated after the RGE running. Our
conclusions are given in Sec.~\ref{sec:conclusion}.



\section{Obtaining constraints on model parameters}
\label{sec:ParameterDetermination}

In this section, we give an overview of the codes and methods used for the
global fits of the CMSSM and NUHM1 models. First, an overview over the
codes used for scanning the parameter space and predicting the
observables is given, followed by an explanation of the statistical methods
used to derive the results.

\subsection{The Fittino framework}\label{sec:fittino}

The \texttt{Fittino}~\cite{Bechtle:2004pc} framework is used to
perform a global Markov Chain Monte Carlo (MCMC) scan (for earlier
implementations and references see~\cite{Bechtle:2009ty}) of the
supersymmetric parameter space in all relevant parameter
dimensions. For the calculation of the SUSY particle spectrum, the
program \texttt{SPheno}~\cite{SPheno} version~3.1.4
(\texttt{SoftSUSY}~\cite{Allanach:2001kg} version~3.1.7 as a
cross-check) is interfaced. The resulting spectrum is used in
\texttt{micrOMEGAs}~\cite{Belanger:2008sj} version~2.2 for the
prediction of the dark matter relic density, in
\texttt{FeynHiggs}~\cite{Hahn:2010te} version~2.8.2 for the prediction
of the Higgs masses, the $W$ boson mass $m_{W}$, the effective weak
mixing angle $\sin^2\theta_{\mathrm{eff}}$, and the anomalous magnetic
moment of the muon $a_{\mu}$, in
\texttt{SuperISO}~\cite{Mahmoudi:2008bx} version~3.1 for the flavor
physics observables, and in \texttt{AstroFit}~\cite{AstroFit} for the
evaluation of the direct and indirect detection of dark matter
observables. 
A cross-check of the relic density computation between
\texttt{micrOMEGAs} versions~2.2 and 2.4 \cite{Belanger:2010gh} and
\texttt{DarkSUSY}~\cite{darksusy} version~5.0.5 (via the {\tt
  AstroFit} interface) is performed, which yields compatible results
in terms of allowed parameter space. For the prediction of the
branching fractions of the lightest MSSM Higgs boson,
\texttt{HDECAY}~\cite{Djouadi:1997yw} version~4.41 is used. The
available limits on SM and non-SM Higgs bosons, including most
available limits up to and including the ones presented by the LHC and
Tevatron collaborations at the Spring Conferences 2011, are evaluated
using \texttt{HiggsBounds}~\cite{Bechtle:2011sb} version~3.2. The
translation of the excluded CL into a $\chi^2$ contribution is
explained in Section~\ref{sec:HiggsBounds}.

The global $\chi^2$ is calculated in three steps: For all measurements
$O_{\rm meas}^{i}$ given in Tab.~\ref{tab:leobserables},
\begin{equation}
\chi^2_{\rm meas}=\sum_{i=1}^{N_{\rm meas}}\left(\frac{O_{\rm meas}^i
-O_{\rm pred}^i (\vec{P})}{\sigma^{i}}\right)^2
\end{equation}
is calculated for each parameter point $\vec{P}$ where the sum runs
over all $N_{\mathrm{meas}}$ measurements. In case of upper bounds
(e.g.{}, the bound on ${\cal B}(B_s\to\mu\mu)$)
\begin{equation}
\chi^2_{\rm meas+bound}=\chi^2_{\rm meas}+\sum_{i=1}^{N_{\rm bound}}
\left\{
\begin{array}{cl}
\left(\frac{O_{\rm limit}^i-O_{\rm pred}^i(\vec{P})}{\sigma^{i}}\right)^2
& \mathrm{for}\,O_{\rm pred}^i(\vec{P})>O_{\rm limit}^i\\[3mm] 0 &
\mathrm{otherwise}\end{array}
\right.
\end{equation}
is calculated, where $\sigma^i$ is the assumed theoretical uncertainty
of the prediction. Finally, the $\chi^2$ contributions from
\texttt{HiggsBounds}, from the LHC SUSY search constraint, and from the
direct and indirect detection of dark matter constraints are
calculated, as outlined in Section~\ref{sec:input}, and added to the
global $\chi^2$.

The MCMC algorithm employed uses a continuous optimisation of the
width of the Gaussian proposal density functions (pdf) based on the
variance of the accepted parameter points in all dimensions. Different
settings for the ratio between the variance of recent accepted
parameter points and the pdf widths are employed in parallel to
combine fine scans of the minima with the ability to cover the
complete parameter range. These methods and settings were developed
specifically for the application in the fits presented here. They were
found superior to our previous MCMC scan implementations
\cite{Bechtle:2009ty,Bechtle:2011dm,Fittino:2011}, where the pdf was
optimised in a separate study before performing the global MCMC
fit. At least 3~million points are obtained within $\Delta\chi^2<5.99$
from the minimum for each individual fit.

\subsection{Statistical interpretation}\label{sec:statistics}
In the frequentist interpretation of the MCMC fit, first the point
with the smallest $\chi^2$ is identified. 
The 1-dimensional $1\sigma$ (2-dimensional $2\sigma$) boundaries
are defined by $\Delta\chi^2<1$ ($\Delta\chi^2<5.99$) above the
minimal $\chi^2$ and hidden dimensions are treated by the profiling
technique, \textit{i.e.}\ the hidden dimensions are scanned until the point
with the lowest $\chi^2$ for the given visible dimensions is
found. The best fit point is the point with the smallest $\chi^2$.
Due to the excessive computing time necessary to find a reliable exact
minimum, it is computationally prohibitive to perform toy
fits~\cite{Bechtle:2009ty}, from which the exact CL coverage of the
$\Delta\chi^2<1$ and $\Delta\chi^2<5.99$ contours could be derived
reliably. Therefore, we cannot claim an exact match of the $1\sigma$
uncertainty with 68\,\%\,CL. 

In the Bayesian interpretation, the full posterior pdf for all
parameters is extracted from the MCMC local density. The
$m$-dimensional marginalised posterior pdf for $m$ parameters is drawn
from the full $n$-dimensional posterior pdf by integrating out all
other parameters,
\begin{eqnarray}
p^{m}_{\textrm{marg}}(\mathbf{P_i}) = \int p^{n}_{\textrm{full}} (\mathbf{P_i}, \mathbf{P_{j}}) d^{n-m}P_j .
\end{eqnarray}
The corresponding $1\sigma$ ($2\sigma$) boundaries are
defined by the smallest interval covering $68\%$ ($95\%$). The
interval is built iteratively from the binned, marginalised posterior pdf by
ordering all bins according to their probability $p_i$, starting with
the highest probability. Then the $p_i$ are subsequently added, until
$68\,\%$ ($95\,\%$) is reached:
\begin{eqnarray}
  \sum_{i=1}^{\textrm{max}_{\textrm{nbin}}} p_{i} < 0.68\,(0.95)\,. 
\end{eqnarray}
The allowed region is the one which contains all bins in the
range $[1,\textrm{max}_{\textrm{nbin}}]$. The remaining bins are
outside of the allowed region. In this way, both the 2-dimensional 
$2\sigma$ areas as well as 
the 1-dimensional (local) modes and the corresponding $1\sigma$ 
intervals are constructed from the full posterior pdf.

\begin{table}
  \caption{Results of the example fit for the frequentist
    and the Bayesian interpretation.  For the frequentist interpretation
    the point with the smallest $\chi^2$ is given with the corresponding
    1-dimensional $1\sigma$ uncertainties. For the Bayesian
    interpretation, the global mode of the full posterior pdf in the case
    of flat priors is given, \textit{i.e.}\ the point with the highest local 4D point density. In addition the maxima of the marginalised
    1-dimensional posteriors are shown with the boundaries of the
    smallest interval covering $68\%$ around the maximum.}  \begin{center}
    {\renewcommand{\arraystretch}{1.5} \begin{tabular}{lr|c|c}
        \hline\hline parameter & best fit (freq.) & global mode (Bayesian) & mode
        (marg., Bayesian) \\ \hline 
        $\tan \beta$ & $10.3^{+9.5}_{-4.7}$ &
        $8.2$ & $3.5^{+21.3}_{-1.3}$ \\ 
        $M_{1/2}\, \mbox{[GeV]}$ &
        $288.1^{+99.0}_{-58.3}$& $270.3$ &
        $143.5^{+377.5}_{-30.5}$\\ 
        $M_0\,\mbox{[GeV]}$ & $58.3^{+87.0}_{-14.9}$ & $52.4$ &
        $58.5^{+5055.5}_{-42.5}$ \\ 
        $A_0\,\mbox{[GeV]}$ & $259.8^{+686.9}_{-570.1}$
        & $23.5$ & $403.0^{+2379.0}_{-2027.0}$ \\ \hline\hline
  \label{tab:BayesVsFrequentist} \end{tabular} } \end{center}
\end{table}

\begin{figure}
  \subfigure[]{
    \includegraphics[width=0.49\textwidth,clip=]{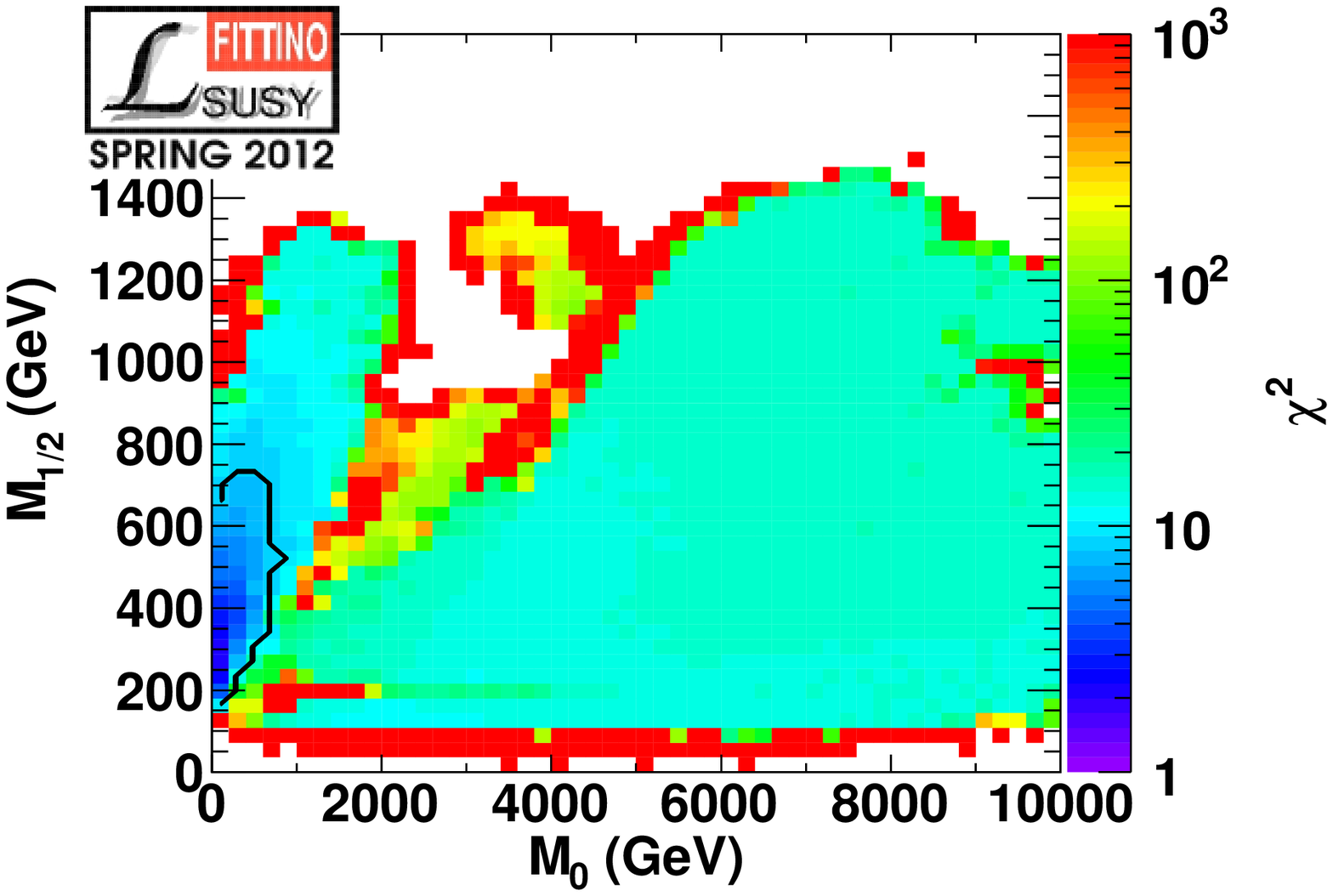}
    \label{fig:BayesVsFreq_Freq}
  }
  \subfigure[]{
    \includegraphics[width=0.49\textwidth,clip=]{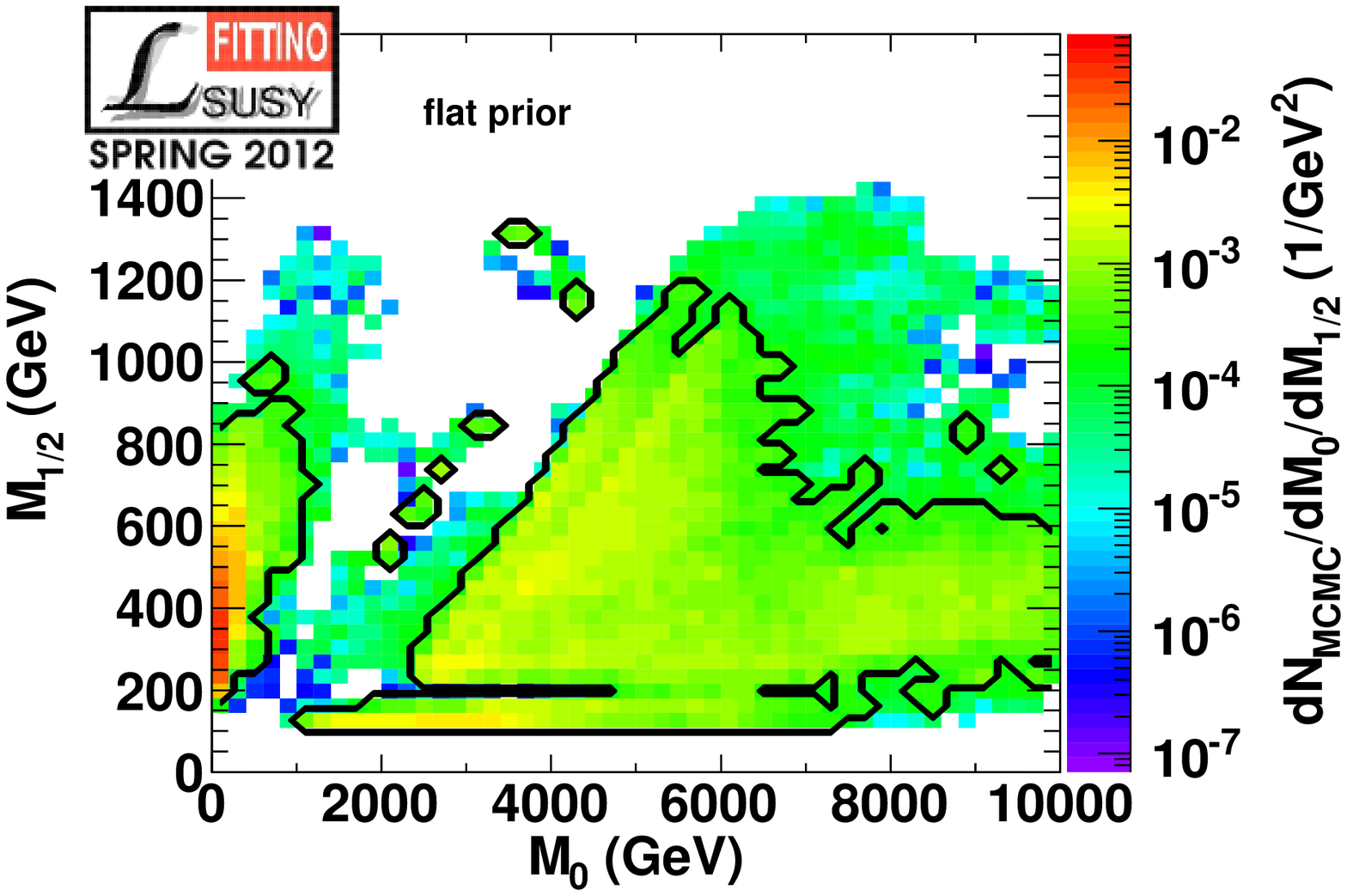}
    \label{fig:BayesVsFreq_BayesFlat}
  }
  \subfigure[]{
    \includegraphics[width=0.49\textwidth,clip=]{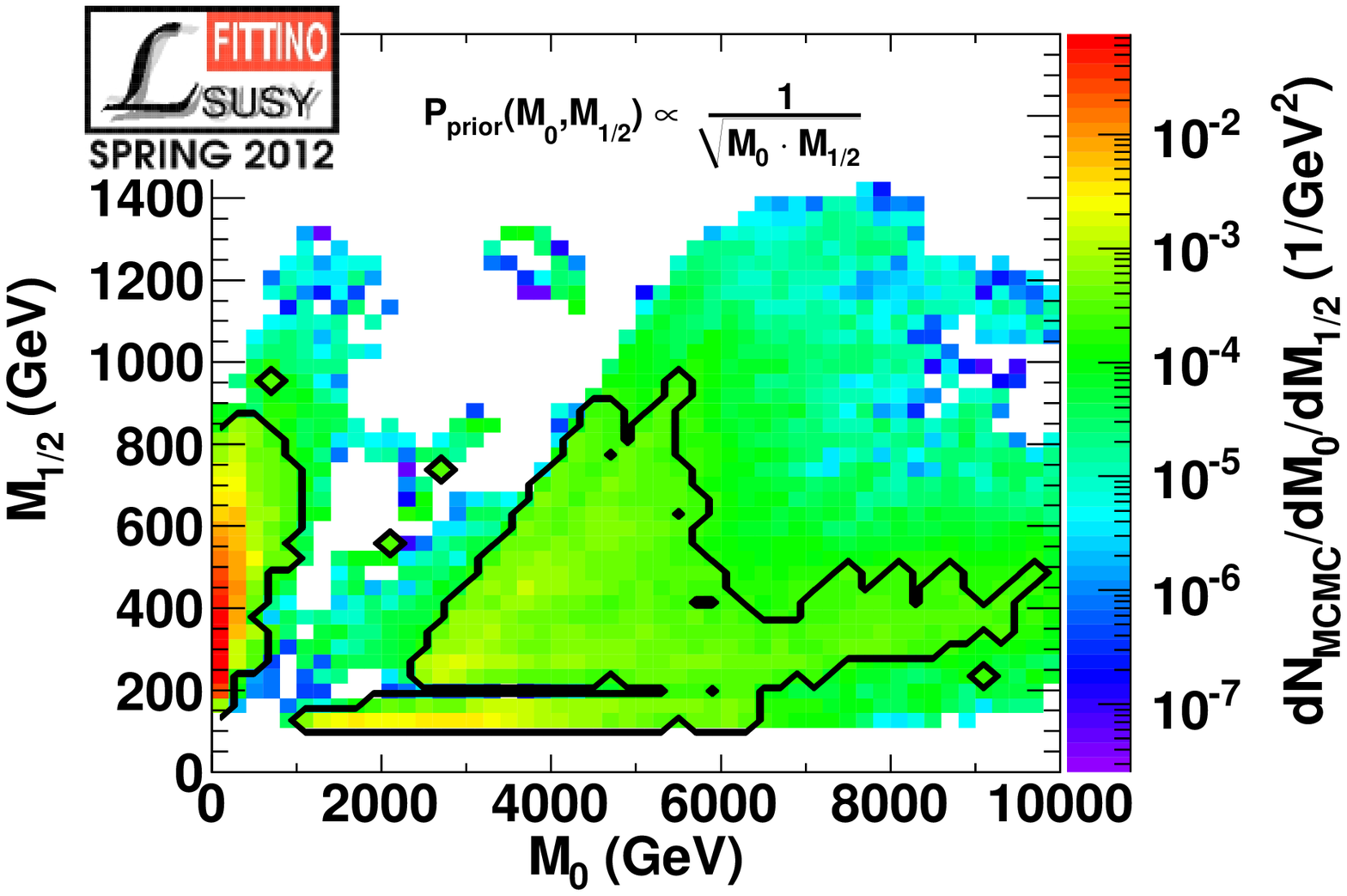}
    \label{fig:BayesVsFreq_BayesProd}
  }
  \subfigure[]{
    \includegraphics[width=0.49\textwidth,clip=]{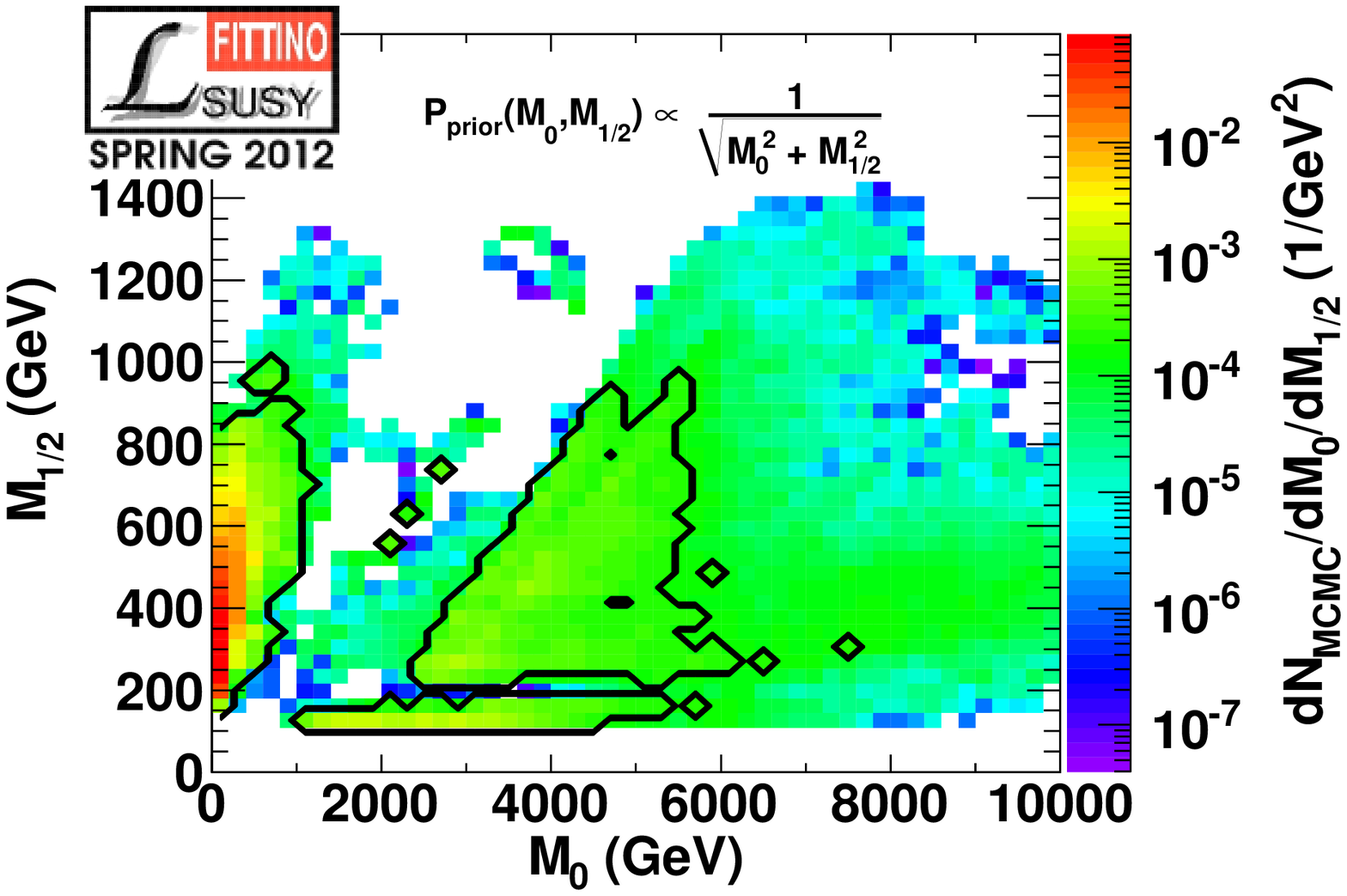}
    \label{fig:BayesVsFreq_BayesSum}
  }
  \caption{Comparison of different statistical interpretations of the
    same example fit based on~\cite{Bechtle:2011dm}.
    \subref{fig:BayesVsFreq_Freq} frequentist profile likelihood;
    \subref{fig:BayesVsFreq_BayesFlat} Bayesian marginalisation with
    flat prior;  \subref{fig:BayesVsFreq_BayesProd} and
    \subref{fig:BayesVsFreq_BayesSum}: two different non-flat priors,
    as described in the Text. The black lines indicate the 2$\sigma$ (frequentist) and 95\% CL
(Bayesian) contours, respectively.}
  \label{fig:BayesVsFreq}
\end{figure}

In Fig.~\ref{fig:BayesVsFreq}, both interpretations are shown for an example
fit in order to illustrate the differences. The fit is performed using
the same setup as in~\cite{Bechtle:2011dm}. Only the observables
listed in Tab.~\ref{tab:leobserables} are used, in contrast
to~\cite{Bechtle:2011dm}. This example fit is only used for testing
the statistical procedure, but not used for the results presented in
Section~\ref{sec:results}.  Figure~\ref{fig:BayesVsFreq_Freq} shows
the 2-dimensional $\chi^2$ profile in the ($M_{0}, M_{1/2}$) plane. The small black contour on the left indicates the
2-dimensional $2\sigma$ region in the frequentist interpretation.
Figure~\ref{fig:BayesVsFreq_BayesFlat} instead shows the MCMC scan
point-density in the same plane. The black contour indicates the
2-dimensional $95\%$-CL region for the same fit using a flat
prior. In the white area, no points have been reached by the MCMC
chain. Only
scans with constant proposal density functions are used, in contrast
to the continuous optimisation used for the other fits as described in
Section~\ref{sec:fittino} in order to keep the MCMC point density proportional
to the posterior pdf.
Note that for the frequentist interpretation also points which
were rejected by the MCMC algorithm have been taken into account.
This explains why there are small scanned
areas in the ($M_{0},M_{1/2}$) plane which appear in the plot for the
frequentist interpretation but not in the plot for the Bayesian
interpretation. The results of the frequentist and Bayesian
interpretation for the example fit are summarised in
Tab.~\ref{tab:BayesVsFrequentist}. Figures
~\ref{fig:BayesVsFreq_BayesProd} and ~\ref{fig:BayesVsFreq_BayesSum}
show the Bayesian interpretation in the case of two different
priors. Both priors have been chosen flat in $A_{0}$ and $\tan\beta$,
while the dependence on $M_{0}$ and $M_{1/2}$ is chosen to be
\begin{eqnarray}
  P\left( M_{0}, M_{1/2} \right )& \propto& \frac{1}{\sqrt{M_{0} \, M_{1/2}}} \hspace{0.4cm} \textrm{[Fig.~\ref{fig:BayesVsFreq_BayesProd}]}, \hspace{0.55cm} 
  P\left( M_{0}, M_{1/2} \right ) \propto \frac{1}{\sqrt{M_{0}^2 + M^2_{1/2}}} \hspace{0.4cm} \textrm{[Fig.~\ref{fig:BayesVsFreq_BayesSum}]}\,. \nonumber \\
&&
\end{eqnarray}
Given the significant prior
dependence of the results, we employ the frequentist interpretation for the results
given in this paper. In addition, the frequentist interpretation allows us to increase the
point density of the MCMC scans by also including rejected points and by
optimising the proposal density function continuously during the
fit. The appearance of the so-called focus point region (allowed
points at rather low $M_{1/2}$ and very high $M_0$) in the Bayesian
interpretation for the exactly same fit and the same MCMC chain as in
the frequentist interpretation also helps to understand the
significantly different allowed regions found in the literature in
previous publications, see \textit{e.g.}\ Fig.\,1 in \cite{AbdusSalam:2011fc}.

\section{Experimental constraints}
\label{sec:input}

In this section, we describe the present and potential future
experimental measurements and searches which we employ to constrain
the supersymmetric parameter space. These can be grouped into three
classes:
\begin{enumerate}
\item Indirect constraints involving supersymmetric loop corrections;
\item Constraints from astrophysical observations;
\item Direct sparticle and Higgs boson search limits from high-energy colliders.
\end{enumerate}
We discuss these groups of observables in turn below.

\subsection{Indirect constraints}

\label{subsec:leos}
Supersymmetric particles may contribute to various low-energy
observables through loop corrections. The various types of indirect
constraints that can be utilized to explore the SUSY parameter space
are
\begin{itemize}
\item rare decays of B-mesons;
\item the anomalous magnetic moment of the muon;
\item electroweak precision observables.
\end{itemize}
The observables and their measured values as used in this paper are listed
in Tab.~\ref{tab:leobserables}. The different observables are
briefly described in Sections \ref{sec:rareBdecays} to \ref{sec:EW}.
The Standard Model parameters that have been fixed are collected in
Tab.~\ref{tab:sminputs}. In one particular fit, the top quark mass is allowed to float.

\begin{table*}
  \caption{Low-energy observables employed. 
    In general, experimental and theoretical uncertainties have been added 
    in quadrature. The top quark mass $m_t$ is only used as an observable for the  
   fit where $m_t$ is also floating as SM input parameter.
  }\label{tab:leobserables}
  \begin{center}
    \begin{tabular}{l|c|c}
      \hline\hline
      $ {\cal B}(b\to s\gamma)$ & $(3.55 \pm  0.34) \times 10^{-4}$ & \cite{LEO:HFAG} \\
      $ {\cal B}(B_s\to\mu\mu) $ & $< 4.5\times10^{-9}$ &  \cite{LHCb-Moriond} \\   
      $ {\cal B}(B\to\tau\nu)$ & $(1.67 \pm 0.39) \times 10^{-4}$ & \cite{LEO:HFAG}  \\
      $ \Delta m_{B_s}$ & $17.78 \pm 5.2\,\mathrm{ps}^{-1}$                                   &  \cite{LEO:HFAG} \\
      $ a_{\mu}^{\mathrm{exp}}-a_{\mu}^{\mathrm{\mathrm{SM}}} $  & $(28.7 \pm 8.2)\times10^{-10}$ & \cite{Bennett:2006fi,MuonAnom:Davier} \\
      $m_W$ & $(80.385 \pm 0.015)$\,GeV  &  \cite{MW-Moriond} \\ 
      $ \sin^2\theta_{\mathrm{eff}}$ & $0.23113 \pm 0.00021$ & \cite{EW:Zres} \\
      $ \Omega_{\mathrm{CDM}} h^2  $ & $0.1123 \pm 0.0118$ &\cite{Komatsu:2010fb} \\
      $ m_t $         & $(173.2 \pm 1.34)$~GeV &  \cite{PDG2010}  \\
      \hline
    \end{tabular}
  \end{center}
\end{table*}

\begin{table*}
  \caption{Standard Model parameters that have been fixed, apart from $m_t$ in one particular fit.}\label{tab:sminputs}
  \begin{center}
    \begin{tabular}{l | c | c}
      \hline\hline
      $ 1/\alpha_{\rm em}  $ & $128.962$     &  \cite{MuonAnom:Davier}  \\
      $ G_{\rm F} $ & $1.16637\times 10^{-5}$  &  \cite{PDG2010}  \\
      $ \alpha_{\rm s}$ &   $0.1176$   &  \cite{PDG2010}  \\
      $ m_Z$ &  $91.1876$~GeV &  \cite{PDG2010}  \\
      $ m_b$ &  $4.19$~GeV  &  \cite{PDG2010}  \\
      $ m_t $         & $(173.2 \pm 1.34)$~GeV &  \cite{PDG2010}  \\
      $ m_{\tau}        $ &  $1.77682$~GeV &  \cite{PDG2010}  \\
      $ m_c  $           &   $1.27$ ~GeV &  \cite{PDG2010}  \\
      \hline
    \end{tabular}
  \end{center}
\end{table*}

\subsubsection{Rare decays of B mesons}\label{sec:rareBdecays}
Strong constraints on new physics models can be obtained from flavor
observables, including, in particular, B-meson decays.  The precise
measurement of branching fractions of rare decays which are helicity
suppressed or which are mediated only at the loop level by heavy
particles, places important restrictions on the supersymmetric
parameter space. Significant constraints come from $B_s$
oscillations, the branching fraction of $B\to\tau\nu$ and the
inclusive branching fraction of $b\to s\gamma$. Recently there has
been a substantial improvement in the limit on the branching ratio of
the decay $B_s\to\mu\mu$. The best limit is now ${\cal
  B}(B_s\to\mu\mu)<4.5 \times10^{-9}$~\cite{LHCb-Moriond}, which is
used as a default in our fits. In addition, we study the impact of a potential
observation of $B_s\to\mu\mu$ by LHCb \cite{LHCb:BsmumuProj} at the SM
rate of ${\cal B}_{\rm SM}(B_s\to\mu\mu)=(3.2\pm0.2)\times 10^{-9}
$~\cite{Buras:2010wr}, \textit{i.e.}\ assuming that no new physics is
observed in this rare B-decay. Note that compared to our previous
analyses~\cite{Bechtle:2009ty,Bechtle:2011dm,Fittino:2011}, various
observables from B- and K-meson decays have been discarded, as their
contributions to the global fit have been found to be negligible.

\subsubsection{The anomalous magnetic moment of the muon}\label{sec:muonAnom}

Even though the anomalous magnetic moment of the electron is measured
approximately two hundred times more precisely than that of the
muon, $(g-2)\equiv2a_{\mu}$, the sensitivity of $a_{\mu}$ to new
physics is enhanced by a factor of $(m_{\mu}/m_{e})^2\sim 43\,000$,
and thus represents a much more powerful constraint on the SUSY
parameter space. While the measurement of $a_{\mu}$~\cite{Bennett:2006fi} 
is undisputed, there is an ongoing debate about the accuracy
of its Standard Model prediction \cite{MuonAnom:Davier}. In
particular, the non-perturbative contribution to $a_{\mu}$ from the
hadronic vacuum polarization has to be extracted from experiment, via
the data from $e^+e^-$ annihilation to hadrons or from $\tau$-lepton
decays. For our base-line fit we have chosen the value of
$a_{\mu}^{\rm SM}$ based on $e^+e^-$ data, which has been argued
to be theoretically cleaner than that based on $\tau$ decays.  Note
that in the CMSSM there is a strong correlation between ${\cal
  B}(B_s\rightarrow\mu\mu)$ and $a_\mu$ \cite{Dedes:2001fv}.

\subsubsection{Electroweak observables}\label{sec:EW}
In contrast to our previous
analyses~\cite{Bechtle:2009ty,Bechtle:2011dm,Fittino:2011}, the
measurements of the Z-boson width and couplings, and the hadronic
cross-section on the Z pole have been excluded from the fit. Indeed,
despite their high precision and the absence of any ambiguity in the
interpretation of the measurement, the contributions of these
observables to the fit have been found to be negligible. Relevant
constraints only come from the effective weak mixing angle
$\sin^2\theta_{\mathrm{eff}}$ and the W-boson mass. The values of the
SM gauge couplings, and of the bottom, charm, and $\tau$-lepton masses
are fixed inputs to the fit, see Tab.~\ref{tab:sminputs}.

\subsection{Astrophysical constraints}
\label{sec:Astro}

Further constraints arise from astrophysical and cosmological
observations such as the cold dark matter relic density as well as
from direct and indirect dark matter searches. We use
\texttt{AstroFit}~\cite{AstroFit} for the evaluation of the direct and
indirect detection of dark matter observables.  The relic density is
provided by \texttt{micrOMEGAs}~\cite{Belanger:2008sj}.
\texttt{AstroFit} provides an interface to
\texttt{DarkSUSY}~\cite{darksusy} and adds an extensive database of
relevant astrophysical data to the {\tt Fittino} fit process. In
particular, the particle spectrum and couplings of a given SUSY model
are passed from \texttt{Fittino} using the SLHA standard
\cite{Skands:2003cj,Allanach:2008qq} to \texttt{AstroFit} which uses
various {\tt DarkSUSY} applications to calculate the predictions
$O_{\rm theo}$ for the requested observables $O$.  This result is then
compared to actual experimental data and contributes to the $\chi^2$
as $\Delta \chi^2 =\sum{\left [(O_{\rm exp} - O_{\rm
      theo})/\sigma_{\rm {exp}}\right]^2}$, where $O_ {\rm exp}$ is
the experimental measurement and $\sigma_{\rm {exp}}$ the $1\sigma$
(combined statistical and systematic) error on it. For exclusion
limits stated at a certain confidence level (\textit{e.g.}~from
gamma-ray observations of dwarf spheroidal galaxies) or experimental
results given in terms of confidence contours for claimed signals (as
in the case of direct detection), \texttt{AstroFit} returns
$\Delta\chi^2=0$ below the exclusion limit and inside the innermost
confidence contour, respectively; outside this region, it interpolates
between the $\Delta\chi^2$ values that correspond to the stated
confidence levels and extrapolates quadratically beyond the largest
explicitly given CLs (for limits, this is done by assuming that $O$
follows a Gaussian distribution around zero).  On return,
\texttt{AstroFit} passes the $\Delta\chi^2$ contribution from every
included observable back to \texttt{Fittino}. In this paper, we use
the dark matter relic density as well as the neutralino scattering and
annihilation rates that are relevant for direct and indirect matter
searches.

\subsubsection{The cold dark matter relic density}\label{sec:CDM}

Evidence for a considerable cold dark matter component in the
composition of the Universe derives from a great number of
observations that cover a large range of distance scales.  In
particular, the WMAP satellite has determined its cosmological
abundance from observations of the cosmic microwave background to an
impressive accuracy of $\Omega_{\rm DM}h^2 = 0.1123 \pm 0.0035 $
\cite{Komatsu:2010fb}. Weakly interacting massive particles (WIMPs),
if neutral and stable on cosmological time scales, are very good
candidates for the cosmological dark matter -- the prime example being
the lightest neutralino~\cite{Jungman:1995df}. Thermally produced in
the early Universe, the WIMP relic density is set by the
(co-)annihilation rate $\langle \sigma v\rangle_{\rm DM}$, see
\textit{e.g.}~Refs.~\cite{Gondolo:1990dk,Edsjo:1997bg}; requiring that
all dark matter is constituted by thermally produced neutralinos makes
the relic density one of the most constraining observables for the
CMSSM parameter space.  In our fits, we calculate the relic density
for comparison with both {\tt micrOMEGAs}~\cite{Belanger:2008sj} and
{\tt DarkSUSY}~\cite{darksusy} (via the {\tt AstroFit} interface).  We
add a 10\% theory uncertainty in the calculation of the relic density
due to missing higher order corrections in the anihilation cross
sections \cite{Herrmann:2007ku,Herrmann:2009wk,Boudjema:2011ig}.

\subsubsection{Direct dark matter detection}\label{sec:directDM}
Direct detection instruments look for elastic scattering events of
WIMPs with nuclei of the according target material in underground
laboratories. They detect signals via scintillation, phonons or ionization,
depending on the experiment (see \textit{e.g.}~\cite{Cerdeno:2010jj}
for a review of direct dark matter detection). The spin-independent
elastic scattering cross-section per nucleon, $\sigma_{\rm SI}$,
conventionally adopted for a model-independent comparison between
experimental results, is calculated with {\tt AstroFit}, taking properly 
into account the potentially
different scattering cross-sections on protons and neutrons.
Theoretical uncertainties of approximately 20\% from the so-called
sigma-term~\cite{Gasser:1990ce} affect the calculation through changes
to the form factor~\cite{Cerdeno:2010jj}. These uncertainties are
estimated by varying the form factor within \texttt{DarkSUSY} and
evaluating the resulting spread of the predicted value of $\sigma_{\rm
SI}$ for different points in the CMSSM parameter space along the
expected exclusion. A resulting systematic uncertainty on $\sigma_{\rm
SI}$ of up to 50\,\% was found and added in quadrature to the experimental
uncertainty obtained from the exclusion contour as described above.

Recently, controversial results have been reported from different
direct detection experiments. While some of the collaborations,
DAMA/LIBRA~\cite{dama}, CoGeNT~\cite{cogent,talk-collar} and
CRESST~\cite{Angloher:2011uu}, promulgate the detection of a signal,
others, primarily XENON100~\cite{Aprile:2011hi}, have shown upper
limits nominally excluding the regions of signal detection.  We
adopt the current XENON limit as well as various projections
including the XENON1T experiment as given in
Ref.~\cite{Baudis:2012bc}. For an upper limit such as from XENON, the
application of the aforementioned theoretical uncertainty on
$\sigma_{\rm SI}$ of 50\,\% translates into a maximally achievable
$\Delta\chi^2$ contribution of 4 for the XENON limit. For the positive
signals claimed by DAMA/LIBRA, CoGeNT and CRESST, no such limit
applies directly. For those, we observe prohibitively high
$\Delta\chi^2$ contributions and thus do not provide further fit results
in Section~\ref{sec:results}.

\subsubsection{Indirect dark matter detection}\label{sec:indirectDM}
Indirect detection instruments -- tracing products from dark matter
annihilation such as photons, antiprotons, positrons, and neutrinos in
the corresponding cosmic-ray fluxes -- have been included via the
gamma-ray channel (see \textit{e.g.} \cite{Cirelli:2010xx} for a
review of indirect detection methods).  Upper gamma-ray flux limits
from dwarf spheroidal galaxies, preferred observation targets due to
their high mass-to-light ratio and low background, from observations
with Fermi-LAT~\cite{fermi} and H.E.S.S.~\cite{hess} have been
included in the fit through {\tt AstroFit} \cite{AstroFit}.


We note that gamma-ray limits from dwarf spheroidals lead to the
currently strongest model-independent bounds on the annihilation rate
(at least for the range of dark matter masses that is relevant here).
In this work, we therefore do not include available photon flux
information from other sky regions, such as the galactic halo.  The
same holds true for other indirect detection channels, in particular
antiprotons and positrons.

\subsection{Direct search limits from high-energy colliders}\label{subsec:direct}
Searches for supersymmetric particles have been performed at various
high-energy colliders, with strong constraints on the MSSM particle
spectrum, including supersymmetric Higgs bosons, 
specifically from LEP, the Tevatron, and the LHC. In the fits we 
include the limit on the lightest chargino mass from LEP,  limits on
the supersymmetric neutral and charged Higgs boson masses from LEP and 
the Tevatron, and, in particular, limits from recent LHC searches for
supersymmetry in channels with jets and missing transverse energy. 

\subsubsection{The limit on the lightest chargino mass from LEP}\label{sec:LEPchargino}
The LEP collaborations have searched for charginos within the
MSSM. The analyses assume gaugino mass unification at the GUT scale,
$M_1=M_2=M_3 = M_{1/2}$, which leads to the relation
$M_1=\frac{5}{3}\tan^2\theta_W M_2$ at the electroweak scale. This
holds for both the CMSSM and the NUHM1 considered here. The LEP
collaborations searched for the dominant decay mode $\chi^{\pm}_1\to
W^{\pm*}\chi^0_1$~\cite{LEPSUSYWG}. We have used the most conservative
chargino mass limit of $m_{\chi^{\pm}_1}>102.5 $\,GeV, including a
conservative estimate of the theoretical uncertainty of $\Delta
m_{\chi^{\pm}_1}=1$\,GeV. Because of gaugino mass unification, the
limit on the chargino mass indirectly also excludes neutralinos with
masses $m_{\chi^{0}_1}<50$\,GeV in the constrained models considered
here \cite{PDG2010,Dreiner:2009ic}.

\subsubsection{Limits on Higgs boson masses}\label{sec:HiggsBounds}

The program \texttt{HiggsBounds}~\cite{Bechtle:2011sb} is used to
implement the limits on the supersymmetric neutral and charged Higgs
boson masses from LEP, the Tevatron and the LHC. \texttt{HiggsBounds}
first determines the strongest search channel for the various Higgs
signatures at every tested model point, based on the expected
limit. For the SUSY models investigated here, the LEP limits on
neutral Higgs bosons prevail in most of the relevant parameter
space. The $\mathrm{CL}_{s+b}$ and $\mathrm{CL}_{s}$ values
corresponding to these limits are tabulated in \texttt{HiggsBounds}
for all signal strengths. Thus, even though the likelihood of the data
for a given Higgs model is not directly available, the $\chi^2_h$
contribution from the Higgs searches can be calculated from the
observed $\mathrm{CL}_{s+b}$ value for the most sensitive search
channel and for any given Higgs mass and signal strength, with the
valid assumption of a Gaussian distributions of
$\mathrm{CL}_{s+b}$. The relation
\begin{equation}
\chi^2_{h}=2[\mathrm{erf}^{-1}(1-2\mathrm{CL}_{s+b})]^2
\end{equation}
is used to determine the full $\chi^2_{h}$ contribution disregarding
theoretical uncertainties of $m_h$. The contribution of the Higgs
limits to the total $\chi^2$ of the fit is then obtained from the
$\chi^2_{h}(m_h)$ distribution after folding with a Gaussian of width
$\Delta m_{h}=3$\,GeV to take into account the theory
error~\cite{Degrassi:2002fi, Allanach:2004rh, Heinemeyer:2004xw} of
the Higgs mass calculation. This procedure is implemented as an
addition to \texttt{HiggsBounds}~\cite{KarinaPrivateComm,Ellis:2007fu}.

In addition to the LEP results discussed above, the LHC collaborations
have now presented an updated result on the combination of searches
for the Standard Model Higgs boson~\cite{ATLAS:2012ae,
  Collaboration:2012tx} with a combined allowed range at the 95\,\%~CL
limit of $117.5<m_{h}<127.5$\,GeV, exceeding the LEP limit. No
detailed translation of the $\mathrm{CL}_{s+b}$ values from the
LHC-experiments into $\chi^2$ contributions is possible given the
published values.  Since the lightest Higgs boson in the CMSSM is
SM-like, we can nevertheless study the effect of the new limit on the global fit
in an approximate way by cutting on $m_{h}$, taking into account also
the theoretical uncertainty $\Delta m_{h}=3$\,GeV. Furthermore, in a
separate study we discuss the impact of a hypothetical SM-like Higgs
boson signal at around
$m_{h}=126$\,GeV~\cite{ATLAS:2012ae,Collaboration:2012tx} on global
fits in the MSSM and the NUHM1.

\subsubsection{SUSY searches at the LHC}\label{LHCSUSYanalysis}\label{subsec:lhc}

At the LHC, the most stringent limits on supersymmetric models
with R-parity conservation are obtained from searches in channels with
jets and missing transverse energy, $E_T^{\rm
  miss}$~\cite{ATLAS-CMS-Moriond}, where squarks $\tilde{q}$ and
gluinos $\tilde{g}$ are produced in pairs and decay through
$\tilde{q}\rightarrow q\tilde{\chi}_1^0$ and $\tilde{g}\rightarrow
qq\tilde{\chi}_1^0$ to purely hadronic final states with missing
energy from the weakly interacting and stable lightest neutralino
$\tilde{\chi}_1^0$.

In order to properly include the SUSY exclusions from the LHC searches
in the global SUSY fit, it is not sufficient to only consider the
95$\%\,$CL bounds published by the experimental collaborations for
specific models and particular choices of parameters. Instead, it is
mandatory to determine the corresponding $\chi^2$ contribution of the
LHC observables, $\chi^2_{\rm LHC}$, for every point in parameter
space. To that end we emulate the experimental SUSY search analyses by
using state-of-the-art simulation tools and a public detector
simulation to approximate the signal expectation for the specific model
under consideration and any given set of model parameters. Details of
the analysis are described below.

We determine the LHC SUSY exclusions from the search for squarks and
gluinos using final states with 2, 3 or 4 jets, zero leptons, and
missing transverse momentum, following a recent ATLAS
analysis~\cite{LHC:SUSYsearch}.
The signal cross-section is simulated on a grid in the $(M_0,M_{1/2})$
plane, fixing $\tan\beta=10$, $A_0=0$ and $\mu>0$. (We discuss
the dependence of the signal on $\tan\beta$ and $A_0$ below.) First,
for each point of the $(M_0,M_{1/2})$ grid, the SUSY mass spectrum is
calculated with {\tt SPheno 3.1.0}\,\cite{SPheno}. Events are then
generated using {\tt Herwig 2.4.2}~\cite{Bahr:2008pv} and passed on
to the fast detector simulation {\tt DELPHES.1.9}~\cite{Ovyn:2009tx}
where reconstruction and detector smearing are performed. The
efficiencies and fake rates of particle identification given by {\tt
DELPHES} are corrected using public measurements of detector
resolutions by ATLAS~\cite{RECSIM}. Finally, the events have to pass
the selection cuts specified in \cite{LHC:SUSYsearch}.

\begin{figure}
  \subfigure[]{
  \includegraphics[width=0.49\textwidth,clip=]{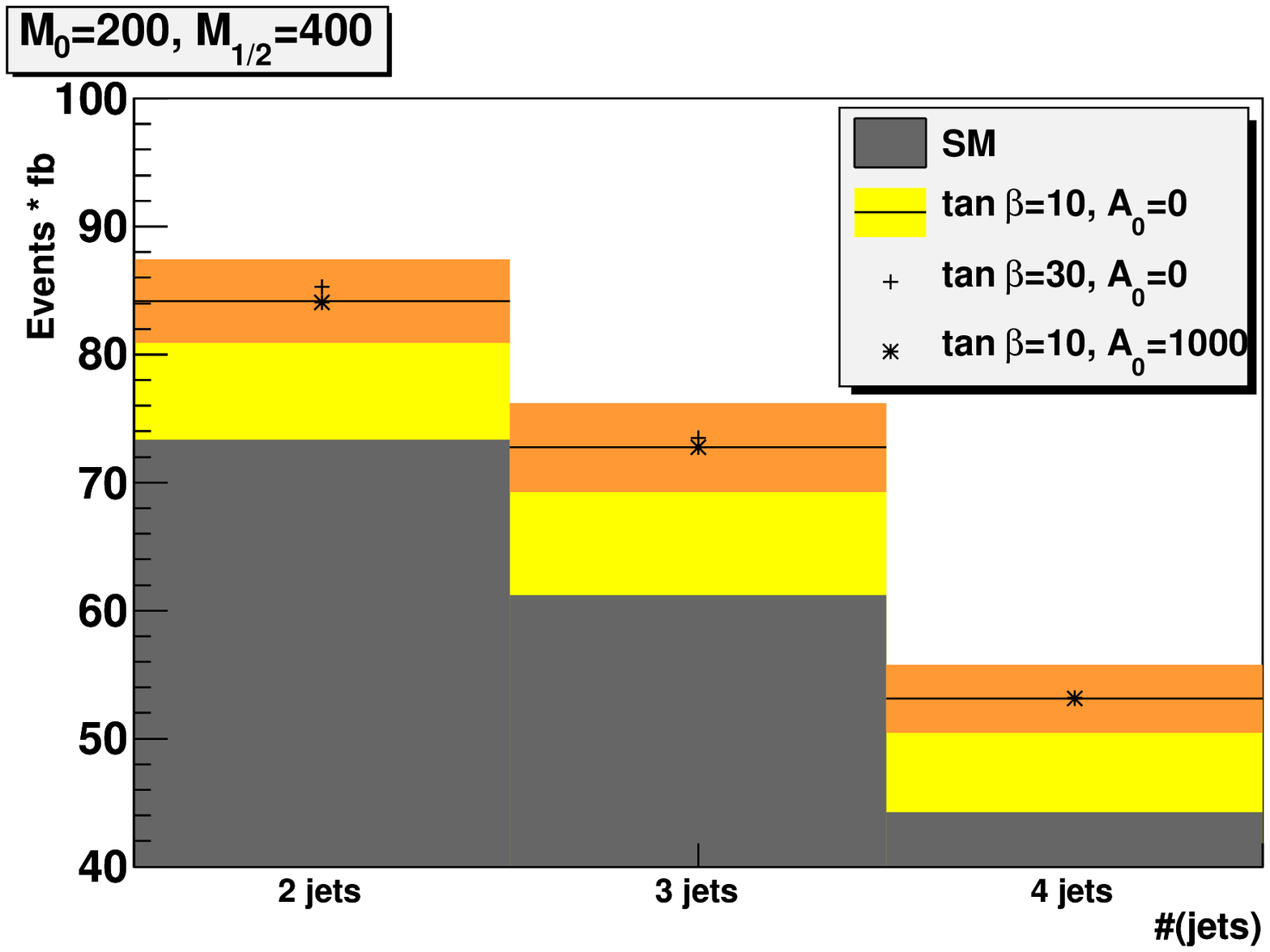}
  \label{fig:tanbetadep_200_400} } \subfigure[]{
  \includegraphics[width=0.49\textwidth,clip=]{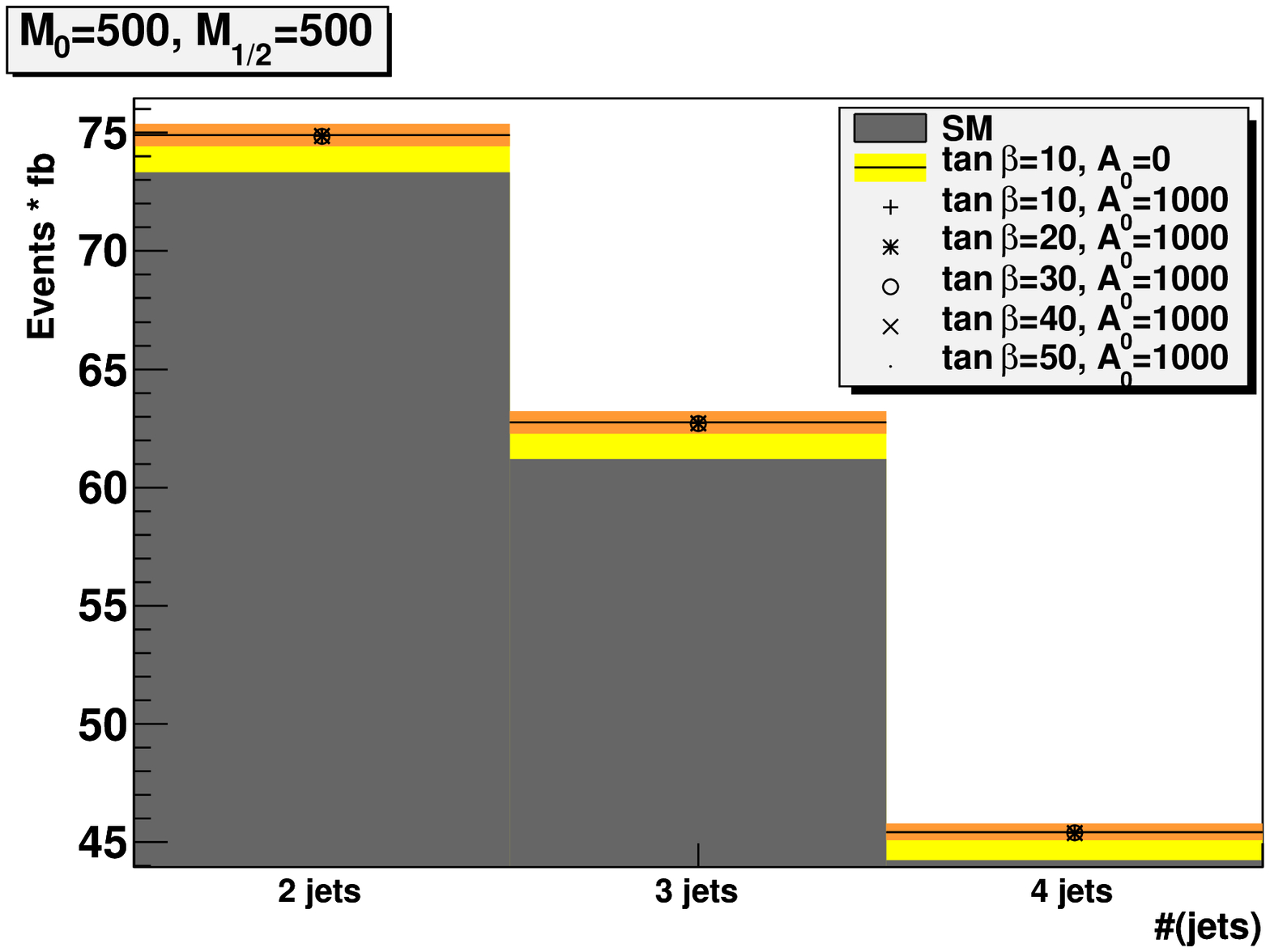}
  \label{fig:tanbetadep_500_500} } \caption{Simulated signal yield for
  events with 2, 3 or 4 jets for two points in the $(M_0, M_{1/2})$
  parameter space.  Shown are the SM background (gray), the CMSSM
  signal expectation for $\tan\beta =10$ and $A_0 = 0$ (yellow)
  together with the systematic uncertainty of 30\% (orange). Also
  shown through different markers are various other central value
  signal estimates based on different values of $\tan\beta$ and
  $A_0$.} \label{fig:tanbetadep}
\end{figure}

The signal estimate is normalized to the NLO+NLL SUSY-QCD prediction
for the inclusive squark and gluino
cross-sections~\cite{Beenakker:1996ch, Beenakker:1997ut,
  Kulesza:2008jb, Kulesza:2009kq, Beenakker:2009ha, Beenakker:2010nq,
  Beenakker:2011fu} and the NLO prediction for the production of
electroweak sparticles~\cite{Beenakker:1999xh}. While the theoretical
uncertainty of the signal cross-section varies along the parameter
space, we assign a constant 30\,\% error to the signal yield, which is
a conservative estimate in the vicinity of the current LHC sparticle
exclusions. The variation of the signal yield with the remaining CMSSM
parameters $\tan\beta$ and $A_0$ has been studied carefully. We found
that it is always well within the systematic uncertainty, as
exemplified in Fig.\,\ref{fig:tanbetadep} for two points in the $(M_0,
M_{1/2})$ parameter space.

\begin{figure}
  \begin{center}
    \includegraphics[width=0.75\textwidth,clip=]{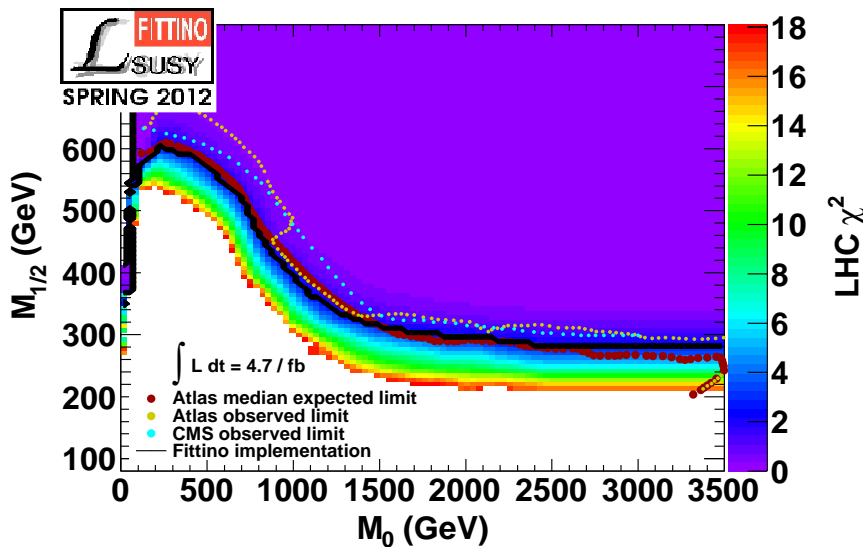}
  \end{center}
  \caption{$\chi^2$ contribution from the LHC SUSY search
    implementation compared to the published ATLAS and CMS
    limits~\cite{ATLAS-CMS-Moriond}. Good agreement of the
    estimated limit with the expected limit of the LHC collaborations
    is achieved.}\label{fig:LHCcomparison}
\end{figure}

Our estimate of the SM background contribution is based on the ATLAS
analysis\,\cite{LHC:SUSYsearch}, which reports the results and
statistical interpretation of the ATLAS fully hadronic inclusive SUSY
searches using 165 pb$^{-1}$ of integrated luminosity. A systematic
uncertainty of 23\% on the background has been estimated
in\,\cite{LHC:SUSYsearch}. Using the ATLAS background and background
systematic uncertainty, together with a systematic uncertainty of
$30\,\%$ on the signal cross-section, we can exactly reproduce the
published ATLAS limits at ${\cal
  L}^{\mathrm{int}}=165\,\mathrm{pb}^{-1}$ and ${\cal
  L}^{\mathrm{int}}=1\,\mathrm{fb}^{-1}$. The ATLAS analysis for
${\cal L}^{\rm int}=4.7\,\mathrm{fb}^{-1}$, however, uses a profile
likelihood based interpretation and six signal regions with tighter
cuts. Because of the large amount of background control regions, this
method cannot easily be implemented in an exact way within the fast
Monte Carlo analysis used here.  Most importantly, the background
systematics are controlled within the profile likelihood
technique. Therefore, we approximate the generic LHC results by
reducing the systematic uncertainties used in the likelihood ratio
analysis within realistic bounds and by slightly increasing the signal
efficiency. Fig.~\ref{fig:LHCcomparison} shows that this fast Monte
Carlo technique is able to describe the LHC results within the
1\,$\sigma$ variation of the expected limit, both in scale as well as
in shape. Thus, this method should provide a good description of
the $\chi^2_{\rm LHC}$ contribution at every point in SUSY parameter
space.

The exclusion limit is calculated in the parameter space for each
point with expected yields for signal $s$ and background $b$ and
observed yield $n$, calculating a test statistic $t=-2\ln Q$, with $Q$
being the likelihood ratio:
\begin{equation}
Q=\frac{L(s+b,n)}{L(b,n)},
\end{equation}
with the Poisson statistics $L(\mu,n)=\mu^n\exp^{-\mu}/n!$, the number
of expected and observed events, $\mu$ and $n$, respectively, for the
given point. The value of $s$ is a function of the SUSY parameters,
whilst $b$ is fixed. The additional uncertainties are taken into
account by a smearing of the Poisson distribution, and we consider a
signal excluded with 95~$\%\,$CL if
\begin{equation}
\mathrm{CL}_{s+b}=\displaystyle\int_{t_{\rm obs}}^{\infty}P_{s+b}(t)\,\mathrm{d}t<0.05,
\end{equation}
with $P_{s+b}$ the probability density function of $t$ assuming the
presence of a signal, and $t_{\rm obs}$ the actually observed value of
$t$. A given $\mathrm{CL}_{s+b}$ value can be approximately translated
into a $\chi^2$ contribution using~\cite{Ellis:2007fu}:
\begin{equation}
\chi^2=2[\mathrm{erf}^{-1}(1-2\mathrm{CL}_{s+b})]^2.
\end{equation}
To obtain the expected exclusion limits the Asimov data set $n=b$ was used.


\section{Results} 
\label{sec:results}

In this section, we present the results of our CMSSM and NUHM1 fits
for various sets of input observables. We first summarize our general
findings and then discuss the features of the different fits in more
detail in Sections \ref{sec:results_leo} to \ref{sec_results_ft}.

In Tables~\ref{tab:fitsummary} and \ref{tab:fitsummaryNUHM1}, we
display selected results of the fits within the CMSSM and the NUHM1
for various sets of input observables.  For all fits we require the
lightest neutralino to be the LSP, consistent radiative electroweak
symmetry breaking and the absence of tachyons.

\begin{table}[t]
  \renewcommand{\arraystretch}{1.5}
  \caption{Summary of the results for various CMSSM fits with
    different sets of input observables. The names 
    of the fits and the observables included are explained in the 
    text.
  }\label{tab:fitsummary}
  \begin{center}
    \begin{tabular}{cccccc}
      Fit & $M_0\, {\rm [GeV]}$ & $M_{1/2}\, {\rm [GeV]}$ & $\tan\beta$ & $A_0$ & $\chi^2$/$ndf$ \\ \hline\hline
      LEO & $84.4^{+144.6}_ {-28.1}$  & $375.4^{+174.5}_{ -87.5}$ & $14.9^{+16.5}_{-7.2}$ & 
      $186.3^{+831.4}_{-843.7} $ & 10.3/8 \\
      LHC & $304.4^{+373.7}_{-185.2}$ & $664.6^{+138.3}_{-70.9}$  & $34.4^{+10.9}_{-21.3}$ &
      $884.8^{+1178.0}_{-974.9}$ & 13.1/9 \\
      LHC+XENON1T &  $296.1^{+1366.8}_{-150}$  & $747.4^{+303.4}_{-143.5}$  &
      $28.3^{+21.2}_{-17.8}$  & $-518.7^{+5266.3}_{-2166.1}$ &
      15.0/9 \\
      LHC+$m_h=126$\,GeV & $1163.2^{+1185.3}_{-985.7}$  &
      $1167.4^{+594.0}_{-513.0}$  &  $39.3^{+16.7}_{-32.7}$  & $-2969.1^{+6297.8}_{-1234.9}$ & 18.4/9 \\
    \end{tabular}
  \end{center}
\end{table}
\begin{table}[t]
  \renewcommand{\arraystretch}{1.5}
  \caption{Summary of the results for the NUHM1 fit. 
  }\label{tab:fitsummaryNUHM1}
  {\small 
    \begin{center}
      \begin{tabular}{ccccccc}
        Fit & $M_0\, {\rm [GeV]}$ & $M_{1/2}\, {\rm [GeV]}$ & $M_H^2\,{\rm [GeV^2]}$ & $\tan\beta$ & $A_0$ & $\chi^2$/$ndf$ \\ \hline\hline
        NUHM1+$m_h$ & $124.3^{+95.2}_{-16.8}$  &
        $655.5^{+218.0}_{-65.0}$  & $(-1.7^{+0.5}_{-2.7})\times10^{6}$ & $29.4^{+3.3}_{-7.8}$  & $-511.2^{+574.7}_{-988.6}$ & 15.3/8 \\
      \end{tabular}
    \end{center}
  }
\end{table}
In total, 7 different input parameter sets have been tested for the
CMSSM. The fits based on the CoGeNT\footnote{The fit involving
  DAMA/LIBRA data has been omitted entirely since the CoGeNT regions
  lie closer to the accessible space in the CMSSM.}~\cite{cogent}
direct dark matter detection claim yield a prohibitively large minimal
$\chi^2$, implying that these potential DM observations cannot be
accommodated within the CMSSM. We thus do not discuss them
further. For all fits we require the lightest neutralino to be the
LSP, consistent radiative electroweak symmetry breaking and the
absence of tachyons. Note that a positive sign of $\mu$ is preferred
to describe the anomalous magnetic moment of the muon, so we have
fixed sgn$(\mu) = +$ in our fits.

\begin{figure}[t]
  \begin{center}
  \includegraphics[width=0.65\textwidth,clip=]{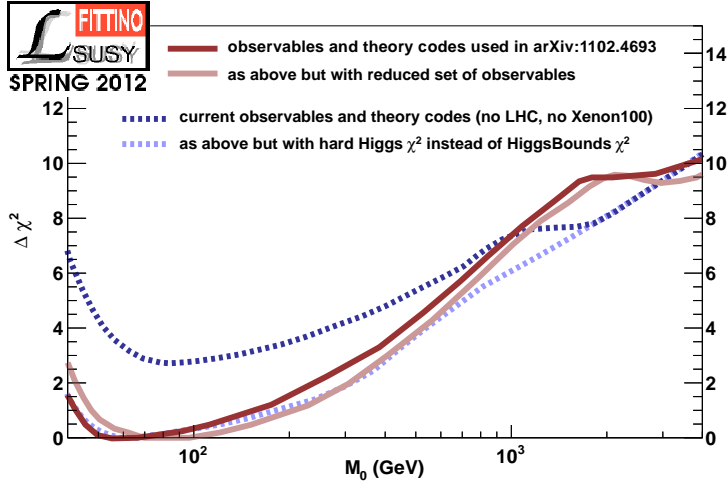}
  \end{center} 
  \caption{Comparison of fits with small/large set of observables and
  with and without exact \texttt{HiggsBounds} $\chi^2_h$. The minimum
  of all fits apart from the one using \texttt{HiggsBounds} for the
  calculation of the full Higgs $\chi^2$ contribution is set to
  $\Delta\chi^2=0$ for better comparison. We show here only the
  dependence on $M_0$. $M_{1/2}$, $A_{0}$ and $\tan\beta$ are profiled
  out.}\label{fig:noHB}
\end{figure}

The fit labeled ``LEO'' (``low energy observables'') in
Tab.~\ref{tab:fitsummary} is performed excluding the direct search for
SUSY at the LHC, but including all constraints from supersymmetric
loop corrections (Section~\ref{subsec:leos}), astrophysical
observations (\ref{sec:Astro}) and sparticle and Higgs boson search
limits from LEP and the Tevatron (\ref{sec:LEPchargino} and
\ref{sec:HiggsBounds}).  The best fit point and the
1-dimensional~1$\sigma$ parameter uncertainties are in excellent
agreement with earlier studies~\cite{Bechtle:2011dm}, despite the fact
that only a fraction of the observable set from~\cite{Bechtle:2011dm}
is used here. As mentioned in Sections~\ref{sec:rareBdecays} and
\ref{sec:EW} various flavor and electroweak precision observables
have been omitted from the fit, as they do not add significant
information for constrained SUSY models. Instead they add a constant
offset to the $\chi^2$, which is nearly independent of the SUSY
parameter point or model under study.  The offset only depends on the
SM parameters and the SM precision observables themselves. In
Ref.~\cite{Bechtle:2009ty} it has been shown that the SM parameters
decouple completely from the CMSSM parameters for the observables used
here. Therefore, a reduced observable set results in a clearer message
on the actual agreement of the SUSY prediction with the
\textit{relevant} data for a given model point. This is shown in
Fig.~\ref{fig:noHB}. The red lines show the profile of the $\chi^2$
minimum of the fit as a function of $M_0$, once with the full
observable set from~\cite{Bechtle:2011dm} (dark red), and once with
only the set of observables described in Section~\ref{sec:input}
(light red), but using the theory codes and observable vales
from~\cite{Bechtle:2011dm}. No significant difference can be observed.

An interesting effect can be observed for the observable ${\cal
  B}(B\to\tau\nu)$: While SUSY points with strong disagreement to the
measurement exist, the majority of the allowed region exhibits a
prediction of ${\cal B}(B\to\tau\nu)$ near its SM value, since the SUSY
contribution tends to decrease the branching fraction, leading to
stronger disagreement than in the SM. Thus, also ${\cal
  B}(B\to\tau\nu)$ contributes to a constant offset of the $\chi^2$
with respect to previous results with lower observed values of the
branching fraction (\textit{e.g.}~\cite{Bechtle:2009ty}), while having no
significant effect on the allowed region above the minimum.

In determining the number of degrees of freedom, $ndf$, we include
one-sided bounds as measurements, also if the minimum of the fit
fulfills the bound. This is justified because the model exhibits
parameter points violating the bounds. Removing the bound from the
$ndf$ count as soon as the minimum or the regions directly around it
satisfy it, would mean that the model is penalized for being
consistent with the data on the ground of constraints from other
observables. Therefore, all bounds with potential contribution to the
$\chi^2$ are counted in $ndf$.

Using the new observable set, and the new set of codes for the
prediction of the observables as outlined in
Sections~\ref{sec:ParameterDetermination} and \ref{sec:input}, we
obtain a $\chi^2/ndf$ of 10.3/8, hinting at a worse agreement than
using the full observable set~\cite{Bechtle:2011dm}. However, as
outlined in Section~\ref{sec:statistics}, no exact ${\cal P}$-value
can be calculated due to computational limitations and the
non-Gaussian nature of the fit. Clearly the overall fit quality of the
CMSSM is reduced when using the reduced observable set of
Tab.~\ref{tab:leobserables}, and taking the improvements in the
sensitivity of the measurements between the old and new observable set
into account. Additional small effects stem from the update to newer
versions of theory codes as outlined in Section~\ref{sec:fittino}.

Another significant difference between the fit results
from~\cite{Bechtle:2011dm} and the results presented here is the more
precise treatment of the LEP Higgs limits. In \cite{Bechtle:2011dm},
\texttt{HiggsBounds} was used to calculate the exact position of the
$95\,\%$\,CL on $m_h$ for each model point. Then, the left half of a
parabola corresponding to a theoretical uncertainty of $\Delta
m_h=3$~GeV and with the minimum of $\chi^2_h=0$ at the exact
$95\,\%$\,CL point was used to calculate $\chi^2_h$. Model points
above the actual limit received a contribution of $\chi^2_h=0$, while
model points below received a $\chi^2_h$ contribution corresponding to
the theoretical uncertainty.

\begin{figure}[t]
  \subfigure[]{
    \includegraphics[width=0.47\textwidth,clip=]{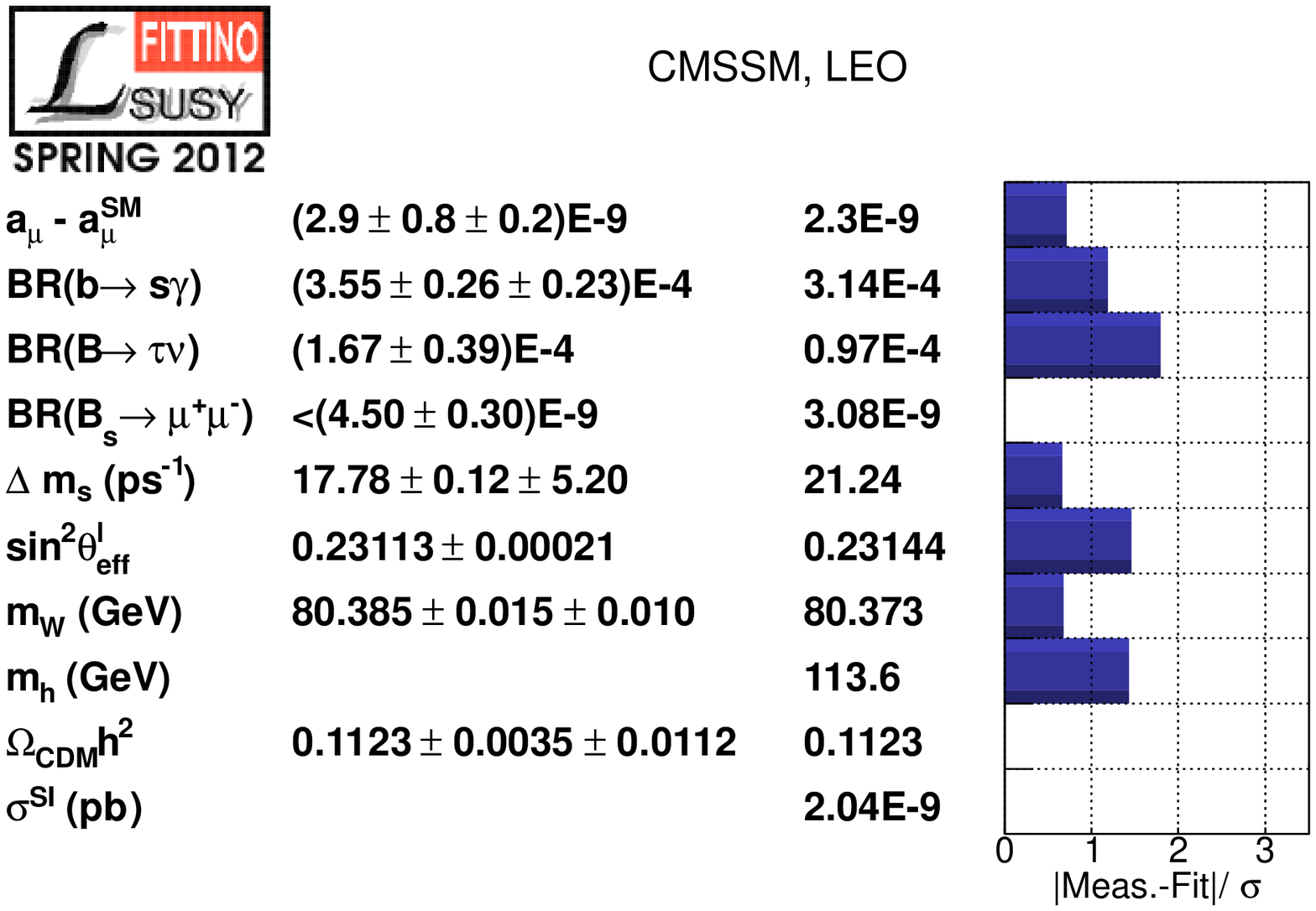}
    \label{fig:IndividualvariablePullPlots_0fbHBAFXn}
  }
  \subfigure[]{
    \includegraphics[width=0.47\textwidth,clip=]{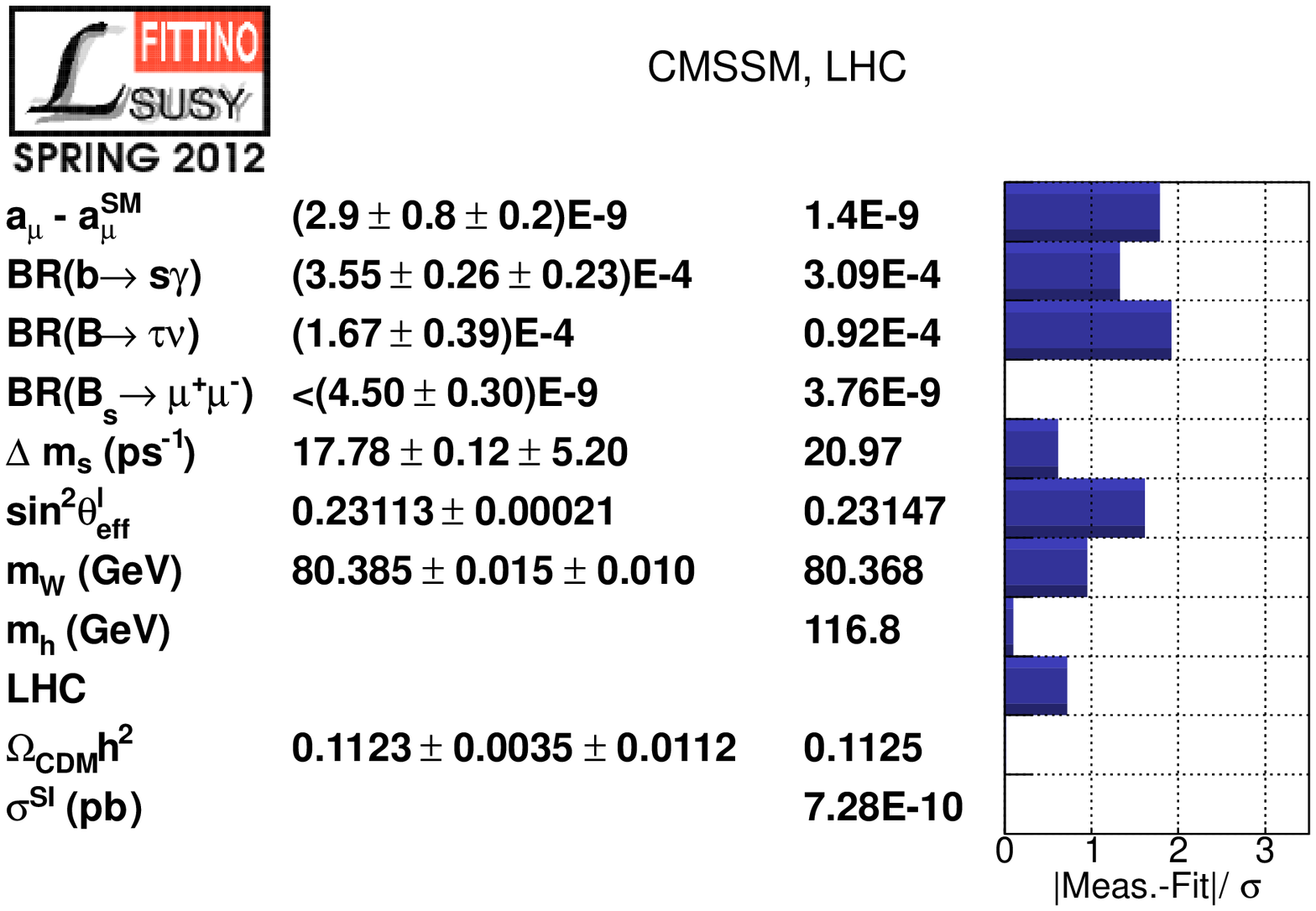}
    \label{fig:IndividualvariablePullPlots_2fbHBAFXn}
  }
  \subfigure[]{
    \includegraphics[width=0.47\textwidth,clip=]{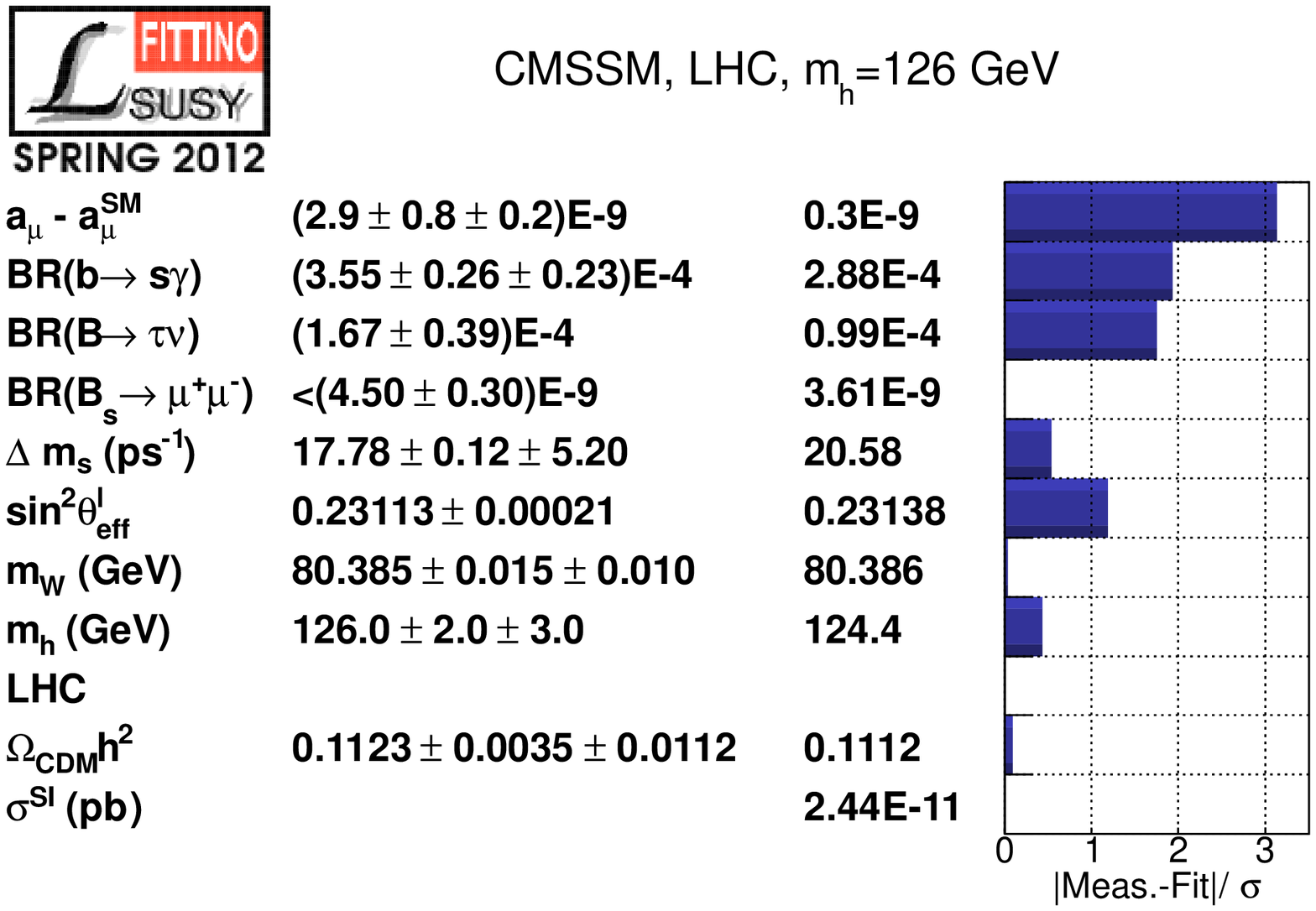}
    \label{fig:IndividualvariablePullPlots_2fbXnMh126}
  }
  \subfigure[]{
   \includegraphics[width=0.47\textwidth,clip=]{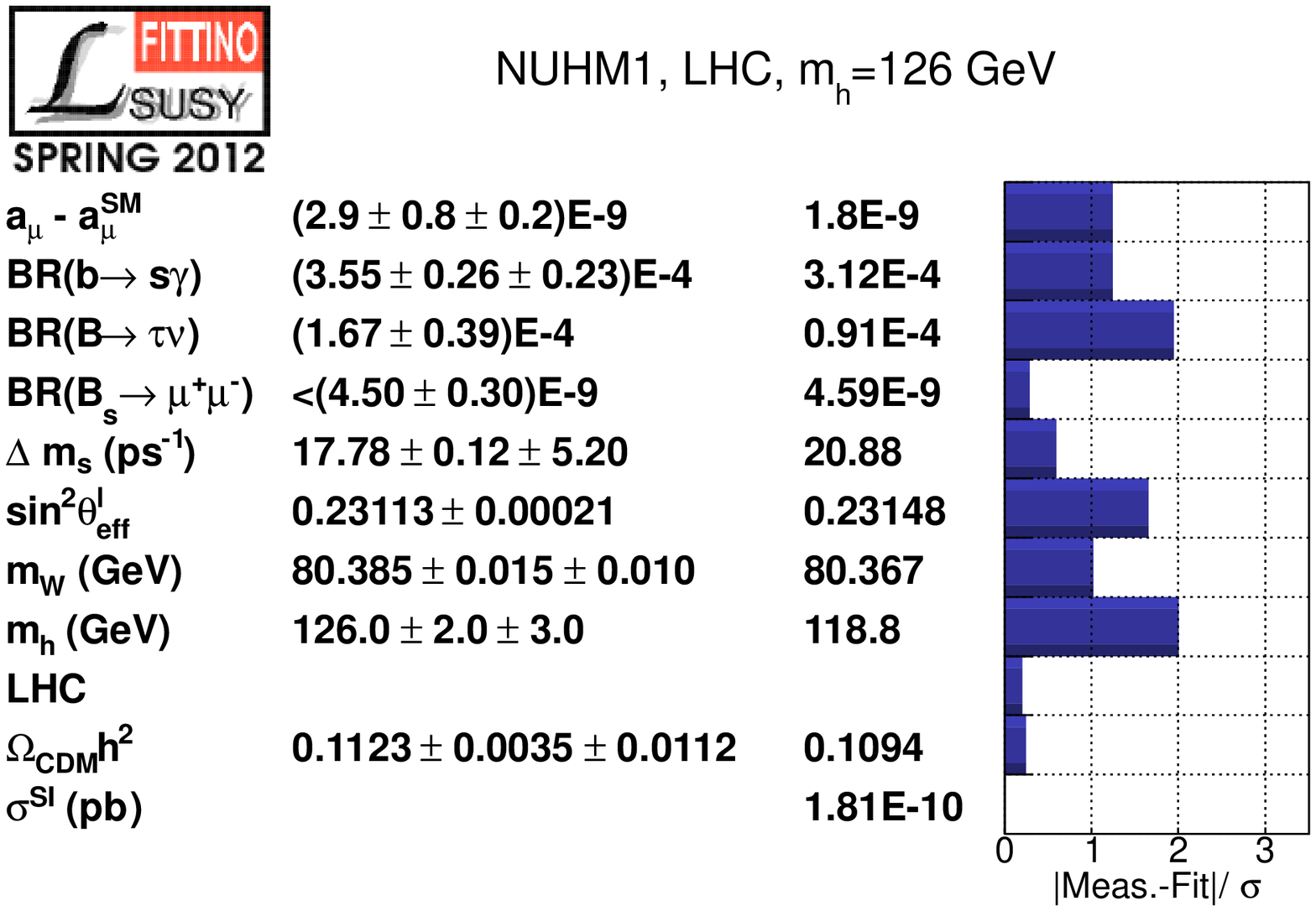}
    \label{fig:IndividualvariablePullPlots_2fbXnMh126NUHM1}
  }
  \caption{Contribution of the individual observables listed in
    Tab.~\ref{tab:leobserables} to selected fits.
    \subref{fig:IndividualvariablePullPlots_0fbHBAFXn} the LEO fit;
    \subref{fig:IndividualvariablePullPlots_2fbHBAFXn} the LHC fit;
    \subref{fig:IndividualvariablePullPlots_2fbXnMh126} the
    LHC+$m_h\!=$126 CMSSM fit; and
    \subref{fig:IndividualvariablePullPlots_2fbXnMh126NUHM1} the
    LHC+$m_h\!=$126 NUHM1 fit. In addition the limits from indirect
    WIMP detection and the chargino limit at LEP count for the
    calculation of the number of d.o.f.}
  \label{fig:IndividualvariablePullPlots}
\end{figure}

For the results presented here, \texttt{HiggsBounds} is used to
calculate the exact $\chi^2_h$ for every $m_h$ below and above the
$95\,\%$\,CL bound, folded with the theoretical uncertainty (see
Section~\ref{sec:HiggsBounds}). Since for the best fit point of the
LEO input set we obtain $m_h\sim(113\pm3)$\,GeV, adding the actual
$\chi^2_h$ contribution to the minimum makes a significant difference
in the overall $\chi^2$ of the minimum. This can be seen from the blue
lines of Fig.~\ref{fig:noHB}. The minimum in the dark blue curve is
lifted significantly with respect to the fits without
\texttt{HiggsBounds} (light blue). Since larger values of $M_0>2$\,TeV
yield $m_h>118$\,GeV, $\chi^2_h$ does not contribute to the fit for
large $M_0$ and the two lines are in perfect agreement for large
$M_0$. Thus, a thin stripe of the focus point region at large $M_0$
and very small $M_{1/2}$ is included in the allowed region, see
Sec.~\ref{sec:results_leo}. Its width in $M_{1/2}$ is mostly
constrained by requiring $m_{\chi^{\pm}_1}>102.5$\,GeV. The pull plot
in Fig.~\ref{fig:IndividualvariablePullPlots_0fbHBAFXn} shows the
contribution of the individual observables to the best fit point of
the LEO fit.


The fit labeled ``LHC'' in Tab.~\ref{tab:fitsummary} uses the same
input as the LEO fit, plus the results from the LHC SUSY search
described in Section~\ref{subsec:lhc}. The effect of the LHC limit is
a shift of the minimal $\chi^2$ and an increase of the number of
degrees of freedom to $\chi^2/ndf=13.1/9$. This clearly shows a
further reduced agreement of the fit with the data, and the minimum is
found at higher $M_0$, $M_{1/2}$ and $\tan\beta$.

From the pull plot in
Fig.~\ref{fig:IndividualvariablePullPlots_2fbHBAFXn} one can deduce
that the LHC limit pushes the fit into a region where, in particular,
$a_{\mu}$ is not described very well anymore.  In comparison, the
direct contribution of $\chi^2_{\rm LHC}$ is small at the best fit
point. The reason is that the $\chi^2$ contribution from the low
energy observables is almost constant for large enough $M_0$ and
$M_{1/2}$. Thus, if the LHC pushes the fit above a certain threshold
in the mass scale, the minimum can also move higher until $\chi^2_
{\rm LHC}$ is not dominant anymore. In addition, $\chi^2_h$ also
prefers higher sparticle masses, balancing the push to lower mass
scales from the precision observables. A further consequence of the
LHC exclusions is that the preferred Higgs mass is shifted upwards to
$m_h \approx 117$\,GeV.


The fit labeled ``LHC+XENON1T'' in Tab.~\ref{tab:fitsummary} uses all
the results from Tab.~\ref{tab:leobserables} and the direct LHC SUSY
search, but the current XENON100 limit replaced by the projection for
XENON\,1T as outlined in Section~\ref{sec:directDM}, assuming no
observation. The XENON\,1T projection increases the minimal $\chi^2$
to 15.0. In order to achieve lower direct detection cross-sections,
the fit is pushed to somewhat lower values of $\tan\beta$ and excludes
large positive values of $A_0$, resulting in an $A_0$ dependence which
is almost symmetric around $A_0=0$. The detailed reasons are given in
Section~\ref{sec:results_af}.


We have also considered the possibility that ${\cal B}(B_s\to\mu\mu)$
is measured at its SM value of ${\cal
  B}(B_s\to\mu\mu)=(3.2\pm0.3)\times 10^{-9}$ at
LHCb~\cite{LHCb:BsmumuProj} (instead of the otherwise employed limit
of ${\cal B}(B_s\to\mu\mu)<4.5\times 10^{-9}$). Since the minimum of
the LHC fit predicts ${\cal B}(B_s\to\mu\mu)$ near its SM value, no
significant change in fit probability or best fit point is observed,
see the discussion in Section~\ref{sec:results_lhc}.

The new limits on the SM Higgs boson mass of $117.5<m_{h}<127.5$\,GeV
at $95\,\%$\,CL~\cite{ATLAS:2012ae,Collaboration:2012tx} from the
ATLAS and CMS experiments are not included in the detailed $\chi^2$
calculation in \texttt{HiggsBounds}. Therefore, no detailed
calculation of the $\chi^2$ contribution from that constraint is
performed. However, applying a simple cut of $114.5<m_h<130.5$\,GeV
(including a theoretical uncertainty of $\Delta m_h=\pm3$\,GeV) on the
LHC fit does not lead to a significant effect on the allowed parameter
space, since for the largest part of the parameter space in the LHC
fit $m_h$ is predicted in the allowed range due to the heavy sparticle
mass scales.

Finally, we explore a possible Higgs observation at $m_h=(126\pm2({\rm
  exp}) \pm3({\rm theo}))$\,GeV, as outlined in
Section~\ref{sec:HiggsBounds}. The corresponding fit is called
``LHC+$m_h=$126'' in Table~\ref{tab:fitsummary}. Apart from exchanging
the \texttt{HiggsBounds} result for $\chi^2_h$ for the assumption of a
direct measurement, the inputs to the fit are identical to the LHC
fit. The high sparticle masses required to yield such a high Higgs
mass push the best fit point into the focus point region at $M_0>4\,
$TeV. The $\chi^2/ndf=18.4/9$ shows that $m_h\gtrsim125$\,GeV is
hardly compatible with a highly constrained SUSY model such as the
CMSSM. This is also obvious from the pull contributions in
Fig.~\ref{fig:IndividualvariablePullPlots_2fbXnMh126}. While a
sufficiently high value of $m_h=124.4$\,GeV can be reached for the
best fit point, the necessary high sparticle mass scales push the fit
into a region which is incompatible with $(g-2)_{\mu}$ and $B$-physics
observables at the 2 to 3~$\sigma$ level.


By decoupling the Higgs sector from the squark and slepton sector in
the NUHM1 model, this problem can be partly remedied. As shown in
Tab.~\ref{tab:fitsummaryNUHM1} and Fig.~\ref{fig:IndividualvariablePullPlots_2fbXnMh126NUHM1}, a slightly better fit is possible for
the NUHM1 ($\chi^2/ndf=15.3/8$) than for the CMSSM fit with
$m_h=(126\pm2\pm3)$\,GeV, albeit with reduced $ndf$. Also, the added
degree of freedom deepens the $\chi^2$ profile for $M_0$ and
$M_{1/2}$ around the minimum, since the Higgs limit does not push the
fit into an area with limited agreement between predictions and
indirect observables. For a more detailed description of the NUHM1
fit, see Section~\ref{sec:higgs}.

In the following sections, the results for individual input observable
sets are discussed in detail.

\subsection{Fit without LHC exclusions}
\label{sec:results_leo}

\begin{figure}[t]
  \subfigure[]{
    \includegraphics[width=0.49\textwidth,clip=]{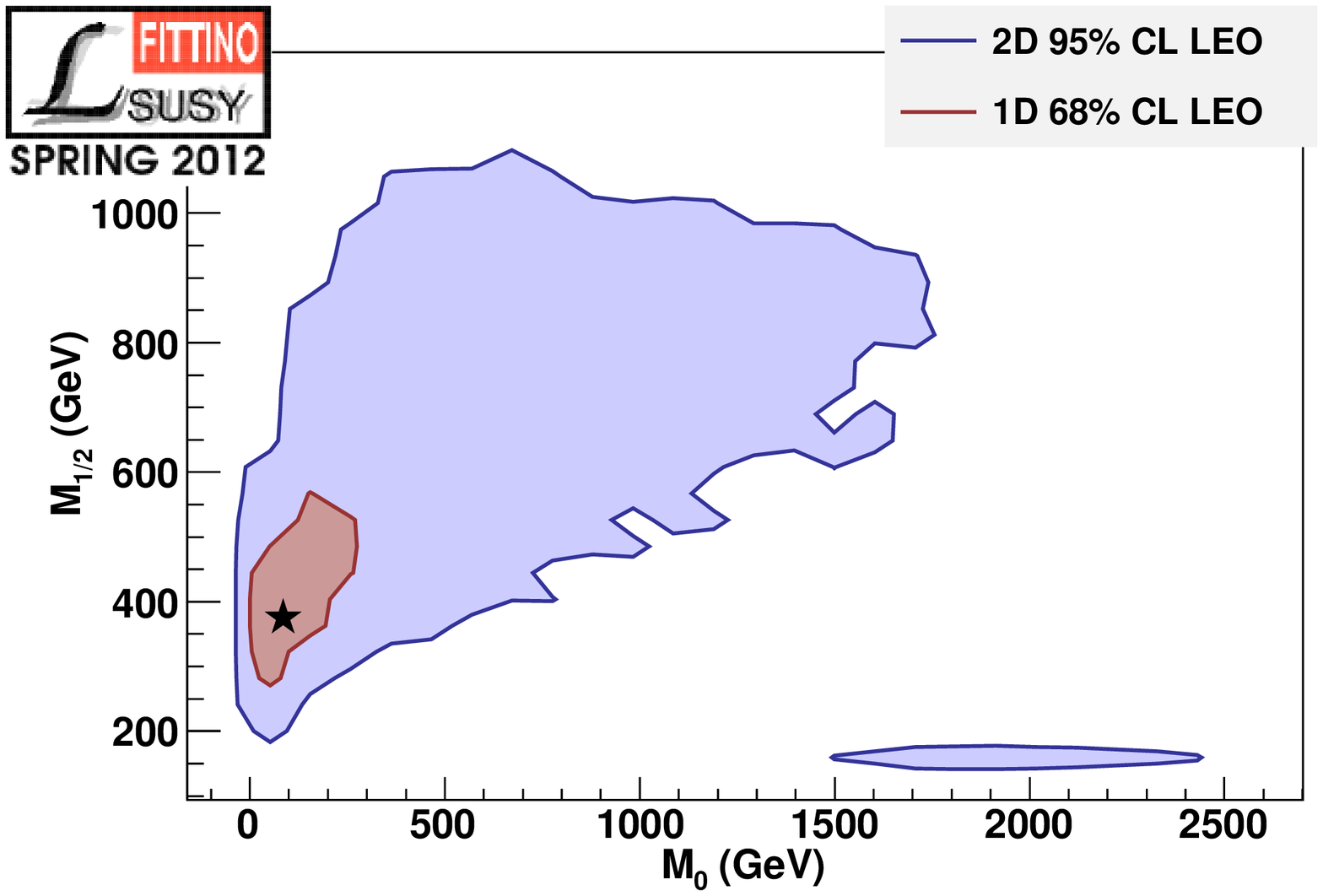}
    \label{fig:noLHC_M0M12}
  }
  \subfigure[]{
    \includegraphics[width=0.49\textwidth,clip=]{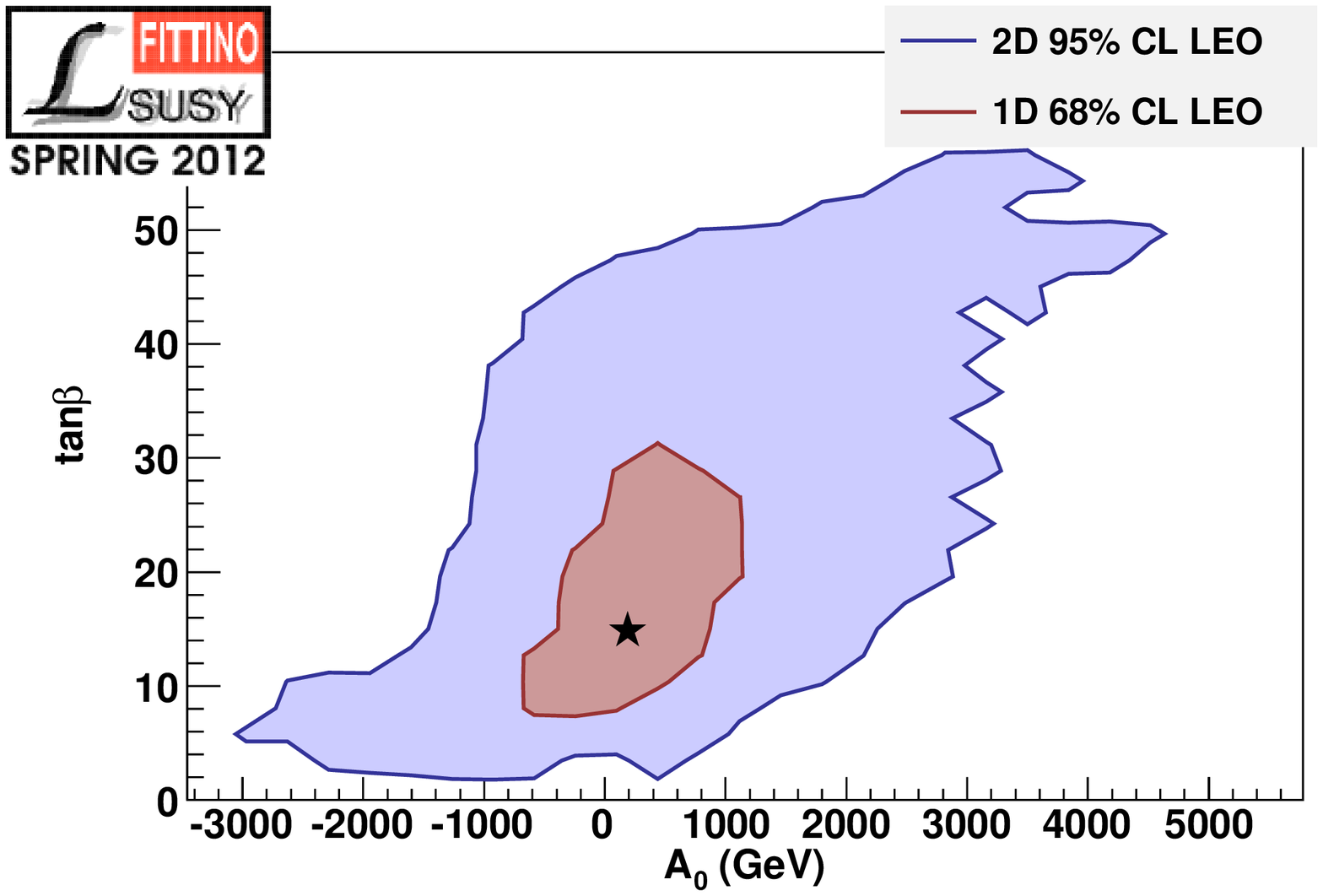}
    \label{fig:noLHC_A0Tanb}
  }
    \caption{Parameter distributions for the LEO fit. Details about
    the best-fit point, marked by a star, are given in
    Table~\ref{tab:fitsummary}. In \subref{fig:noLHC_M0M12}, the
    profile likelihood projection of the $\Delta\chi^2<1$ (1-dimensional 1$\sigma$) [red] and $\Delta\chi^2<5.99$ (2-dimensional 2$\sigma$) [blue] regions of the full 4-dimensional
    fit around the minimum are shown into the ($M_0,M_{1/2}$)
    plane. In \subref{fig:noLHC_A0Tanb}, the same is shown in the
    ($A_0,\tan\beta$) plane.
      }\label{fig:noLHC}
\end{figure}

In Fig.~\ref{fig:noLHC} the resulting parameter regions are shown for
the LEO fit. The best-fit point is marked by a star.  All hidden
dimensions are profiled. As in all further plots of the allowed
parameter space, the allowed regions display the full $\Delta\chi^2<1$
(1-dimensional 1$\sigma$) range in red and the $\Delta\chi^2<5.99$ (2-dimensional 2$\sigma$) range in blue. The 1-dimensional 1$\sigma$
range is shown in order to allow reading off the single parameter
uncertainties directly from the projection of the red areas onto the
horizontal and vertical axes. 

The resulting plot does not represent a parameter scan in the shown
dimensions with fixed hidden dimensions, but the full uncertainty
regions for the global fit of the four CMSSM parameters. As shown
in~\cite{Bechtle:2011dm}, including the SM parameters in the fit does
not change the result as long as the presently available observables
are used. This may change in the future, \textit{e.g.}  once very
precise measurements of $m_h$ or sparticle masses are available.

In contrast to the results from~\cite{Bechtle:2011dm}, and in contrast
to the profile likelihood result from Fig.~\ref{fig:BayesVsFreq} which
uses the observable set and prediction codes
from~\cite{Bechtle:2011dm}, the focus point region at low $M_{1/2}$
and high $M_0$ is allowed within the 2-dimensional $2\sigma$
uncertainty range. This is explained by the cut through the $\chi^2$
profile already shown in Fig.~\ref{fig:noHB}. The focus point region
is constrained to $M_{1/2}<200$\,GeV by the presence of the cut on
$m_{\chi^{\pm}}>102.5$\,GeV described in
Section~\ref{sec:LEPchargino}. The minimal $\chi^2$ shape in
Fig.~\ref{fig:noHB} also explains why the $2\sigma$ allowed region is
so much larger than what would be expected from the $1\sigma$
region. The fit exhibits a rather narrow minimum, where a good
agreement with $(g-2)_{\mu}$ and $\Omega_{\rm DM}$ can be achieved.  For
larger values of $M_0$ and $M_{1/2}$, there is an almost constant
disagreement with both measurements at the $2-3\sigma$ level. Hence,
no further discrimination is achieved and the $\chi^2$ profile becomes
almost flat, still within the $\Delta\chi^2<5.99$ range above the
minimum. 

\subsection{Fits with LHC exclusions}
\label{sec:results_lhc}

\begin{figure}[t]
  \subfigure[]{
    \includegraphics[width=0.49\textwidth,clip=]{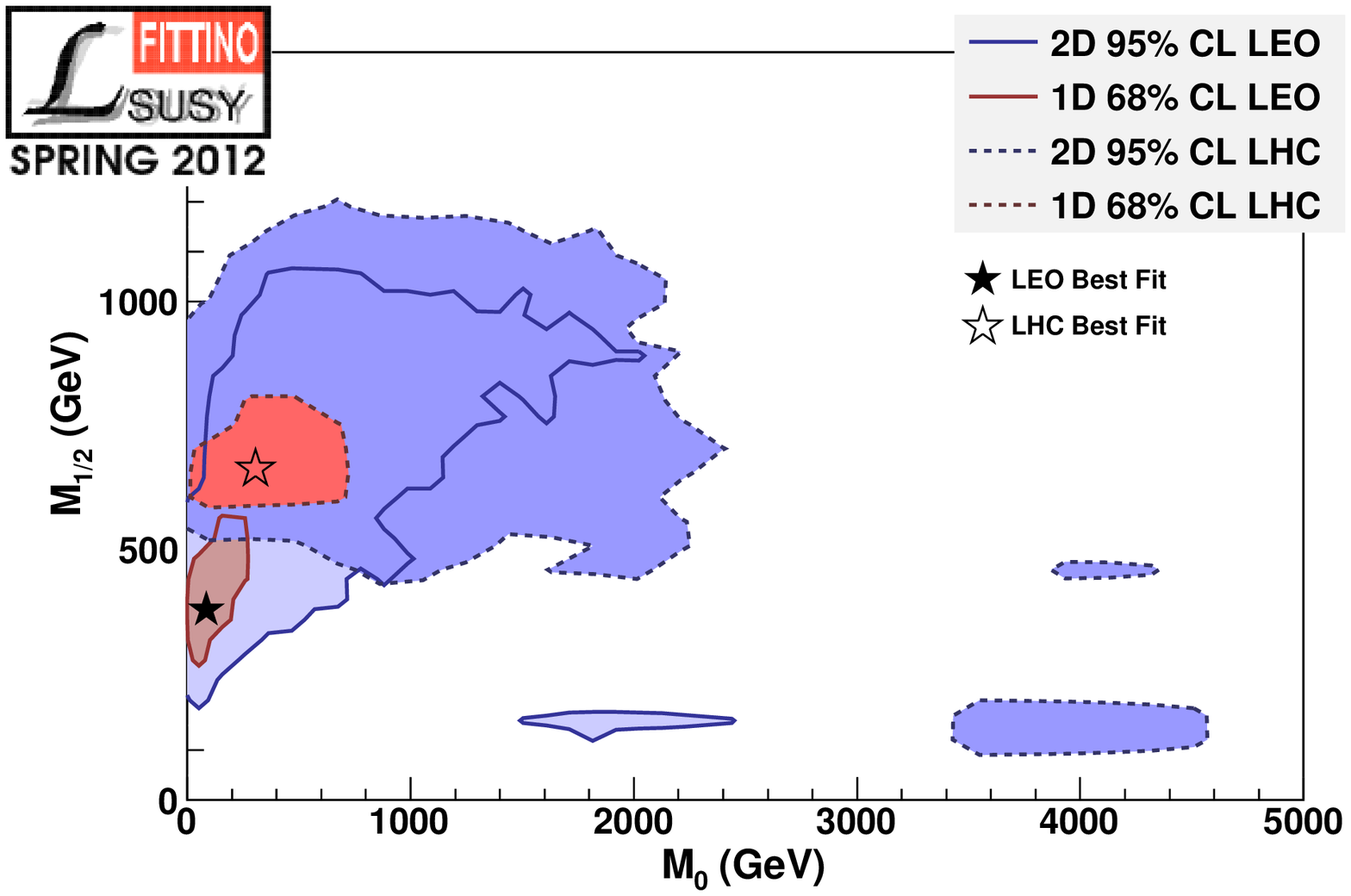}
    \label{fig:LHC_M0M12}
  }
  \subfigure[]{
    \includegraphics[width=0.49\textwidth,clip=]{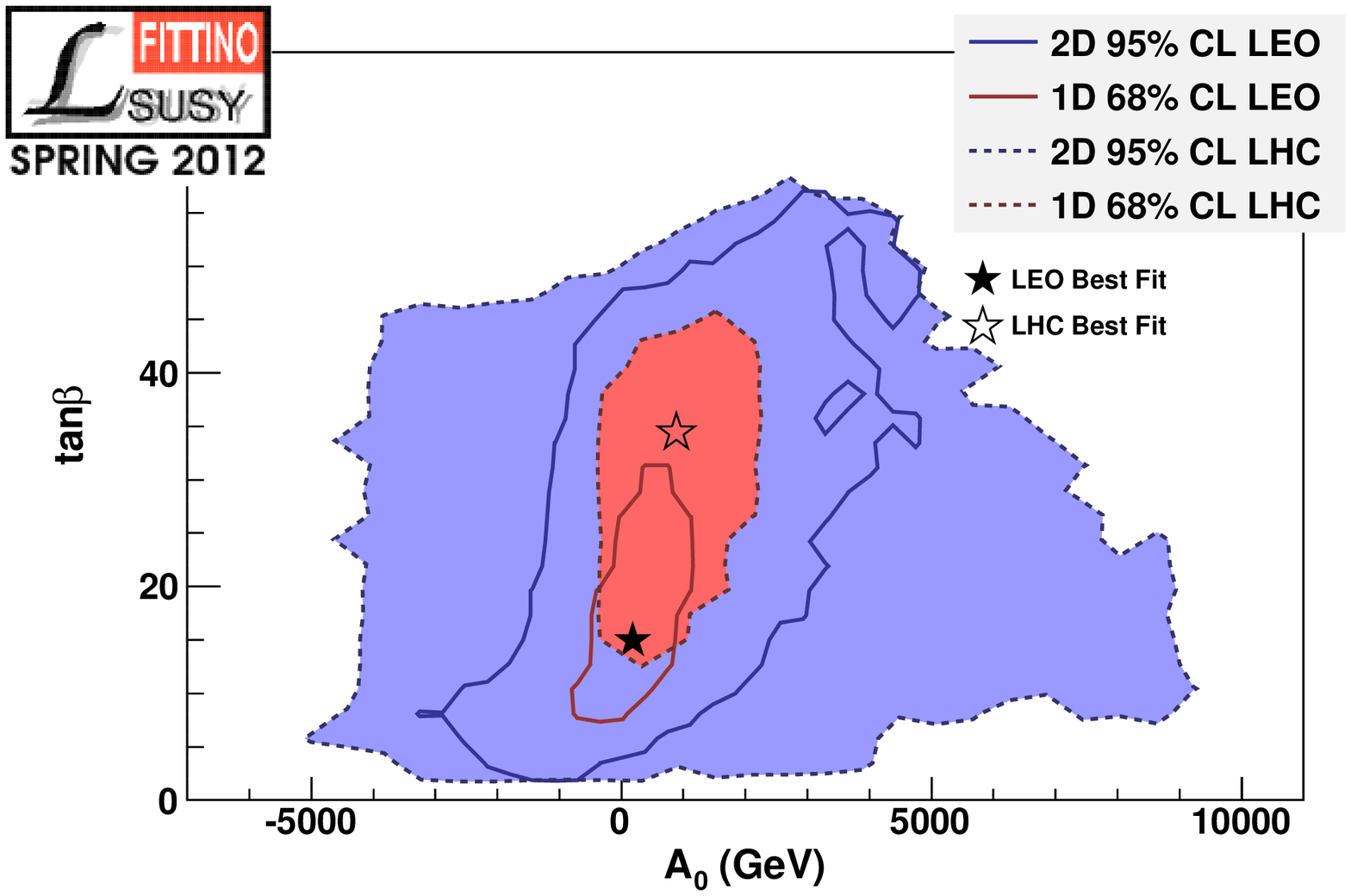}
    \label{fig:LHC_A0Tanb}
  }

  \caption{Best-fit regions in the CMSSM including LHC
    results. Details about the LHC best-fit points are given in
    Table~\ref{tab:fitsummary}. Results for the $1\sigma$ and
    $2\sigma$ regions are shown in comparison between the LEO and LHC
    fits in \subref{fig:LHC_M0M12} the ($M_0,M_{1/2}$) plane and in
    \subref{fig:LHC_A0Tanb} the ($A_0,\tan \beta$) plane.  The
    parameter projections and uncertainties are as in
    Fig.~\ref{fig:noLHC}.}\label{fig:LHC}
\end{figure}

In this section, we discuss the allowed CMSSM parameter space for
different inputs from the LHC. The inputs with the strongest impact are
the inclusive direct searches for SUSY at the ATLAS experiment
described in detail in Section~\ref{subsec:lhc} and the search for
$B_s\to\mu\mu$ described in Section~\ref{sec:rareBdecays}. The
additional very strong constraint stemming from a possible measurement
of the lightest Higgs boson mass is studied separately in
Section~\ref{sec:higgs}. Fig.~\ref{fig:LHC} shows the allowed
parameter range of the LHC fit introduced in
Table~\ref{tab:fitsummary}. The difference between the LHC and the LEO
fit is significant. The position of the best-fit point is lifted
outside of the directly accessible range of sparticle searches at low
$M_0$ and $M_{1/2}\approx650$\,GeV.
The focus point region is excluded by the LHC SUSY search up to our
maximum value of $M_0=3.5$\,TeV of the LHC limit implementation. Also
the LHC collaborations do not publish search results for larger values
of $M_0$, but it can be assumed that a dedicated interpretation at LHC
would exclude large parts of the rest of the focus point region from
the fit (\textit{i.e.} the two islands above $M_0\gtrsim3.5$\,TeV in
Fig.~\ref{fig:LHC_M0M12}). Due to the diminishing relevance of large
SUSY mass scales for the solution of the hierarchy problem and EWSB,
this is not further pursued here.

It can be observed in Fig.~\ref{fig:LHC_A0Tanb} that the best fit
point of $\tan\beta$ is shifted upwards significantly due to the
direct LHC SUSY search limits, which do not affect $\tan\beta$
directly, as shown in detail in Section~\ref{LHCSUSYanalysis}. This is
the result of the correlation between $\tan\beta$ on the one hand and
$M_{1/2}$ on the other hand due to $\Omega_{\rm DM}$ and
$(g-2)_{\mu}$. A decent agreement in those variables for larger
$M_{1/2}$ implies higher scales of $\tan\beta$.

\begin{figure}[t]
  \subfigure[]{
    \includegraphics[width=0.49\textwidth,clip=]{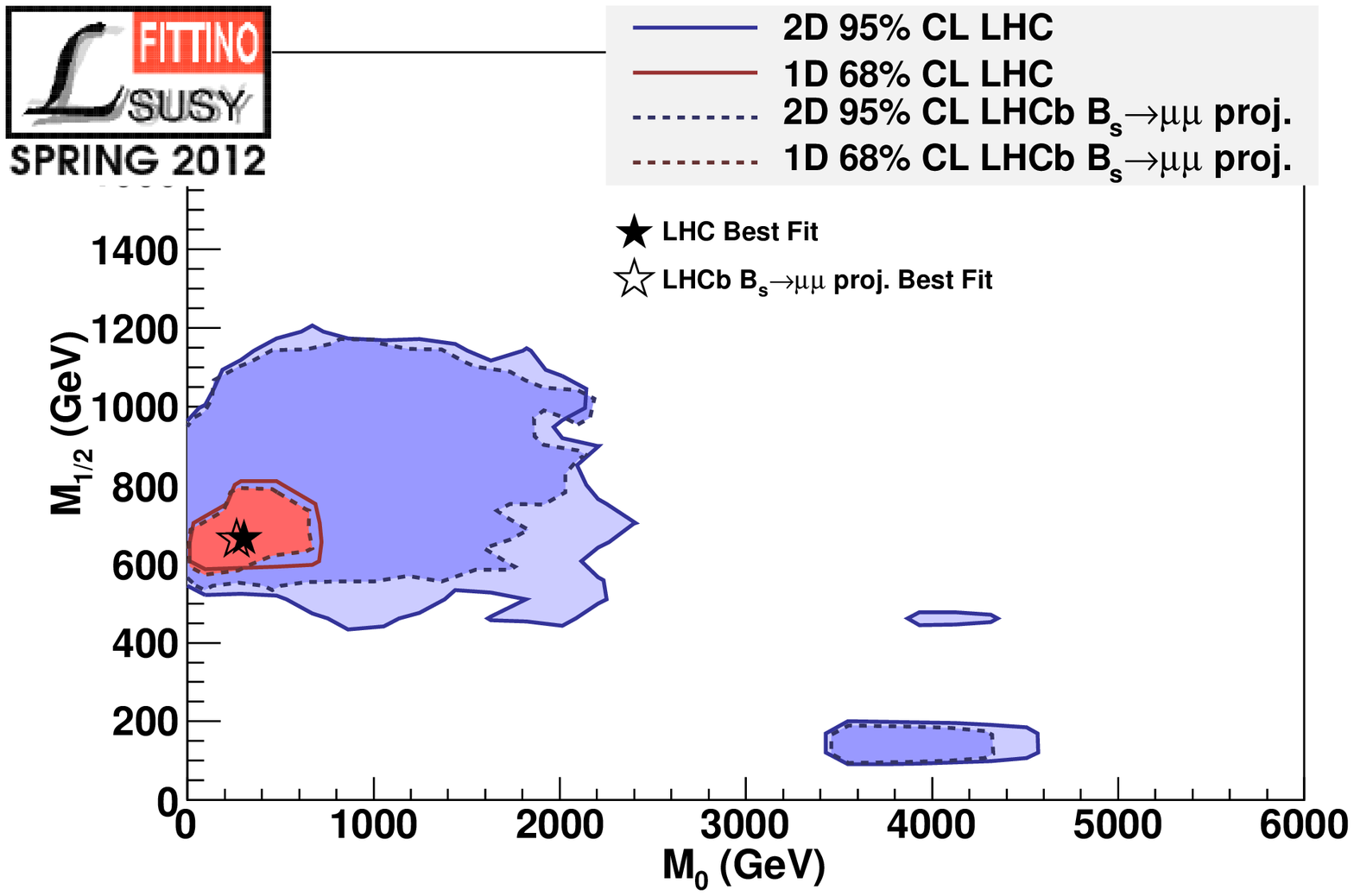}
    \label{fig:LHC_BsmumuProj_M0M12}
  }
  \subfigure[]{
    \includegraphics[width=0.49\textwidth,clip=]{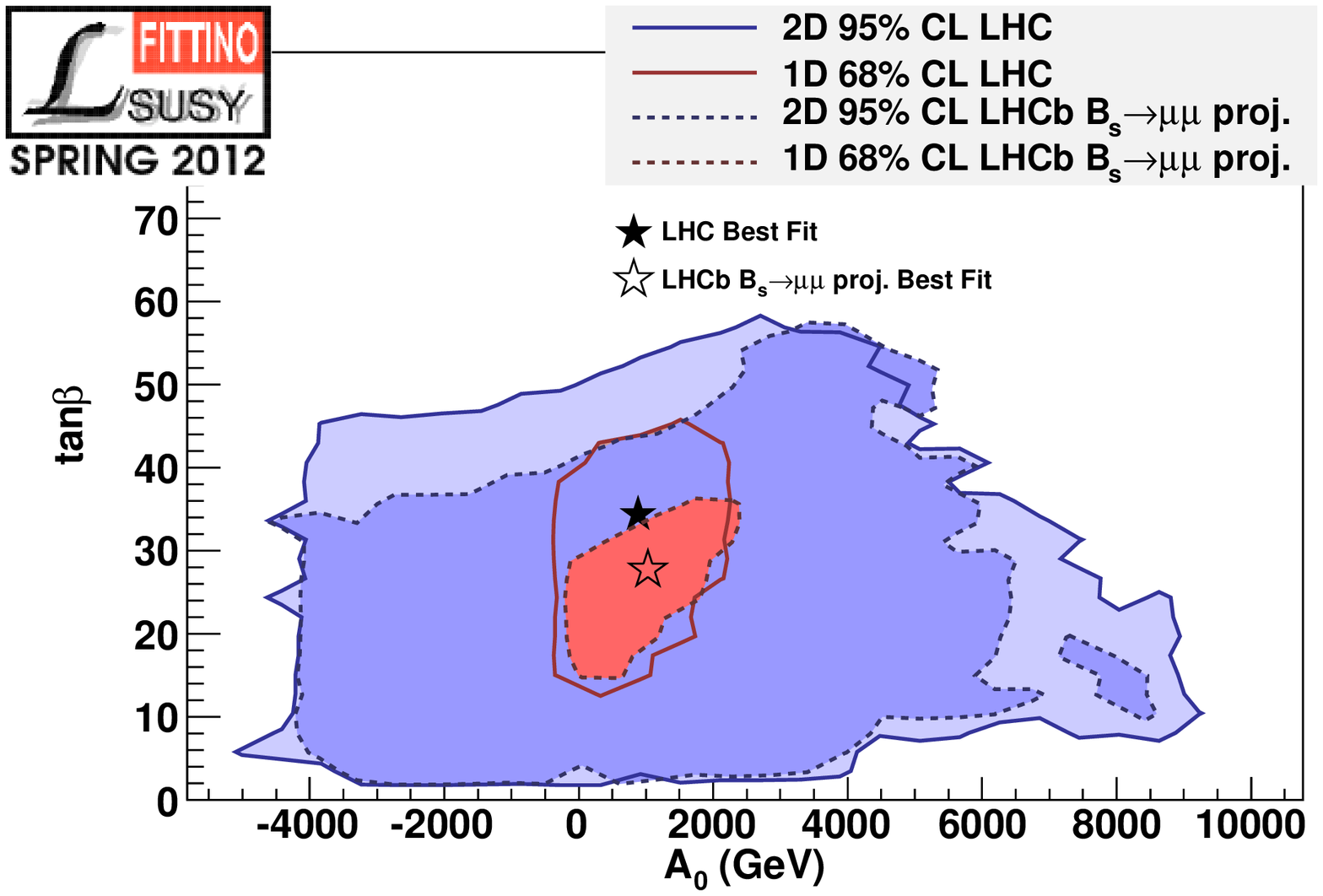}
    \label{fig:LHC_BsmumuProj_A0Tanb}
  }

  \caption{Parameter distributions for the LHC fit (${\cal
      B}(B_s\to\mu\mu)<4.5\times10^{-9}$) and the LHC$^{{\cal
        B}_{SM}(B_s\to\mu\mu)}$ fit in the ($M_0,M_{1/2}$) in
    \subref{fig:LHC_BsmumuProj_M0M12} and in the ($A_0,\tan\beta$)
    plane in \subref{fig:LHC_BsmumuProj_A0Tanb}.  Details about the
    best-fit points are given in Table~\ref{tab:fitsummary}. 
  }\label{fig:LHC_Bsmumu}
\end{figure}

In Fig.~\ref{fig:LHC_Bsmumu} we use the LHC limits as outlined above
and in addition study variations of the limit on ${\cal
  B}(B_s\to\mu\mu)$. In Fig.~\ref{fig:LHC_BsmumuProj_M0M12} and
\ref{fig:LHC_BsmumuProj_A0Tanb}, the results of the fit including a
potential measurement at the SM value ${\cal B}(B_s\to\mu\mu)=(3.2\pm
0.3)\times10^{-9}$ (assuming a small theoretical and experimental
error of $\pm0.3\times10^{-9}$ for the sake of studying the strongest
possible impact) at LHCb~\cite{LHCb:BsmumuProj} and possibly ATLAS and
CMS is shown. Since the best-fit point of the LHC fit predicts ${\cal
  B}(B_s\to\mu\mu)$ close to its SM value, this has only a marginal
effect on the minimal $\chi^2$ of the fit and the allowed parameter
space. Note that the situation would have been somewhat different for
a potential observation of ${\cal B}(B_s\to\mu\mu)$ at
CDF~\cite{CDF:Bsmumu}, which would have clearly disfavored the focus
point region.

\begin{figure}[t]
    \subfigure[]{
      \includegraphics[width=0.49\textwidth,clip=]{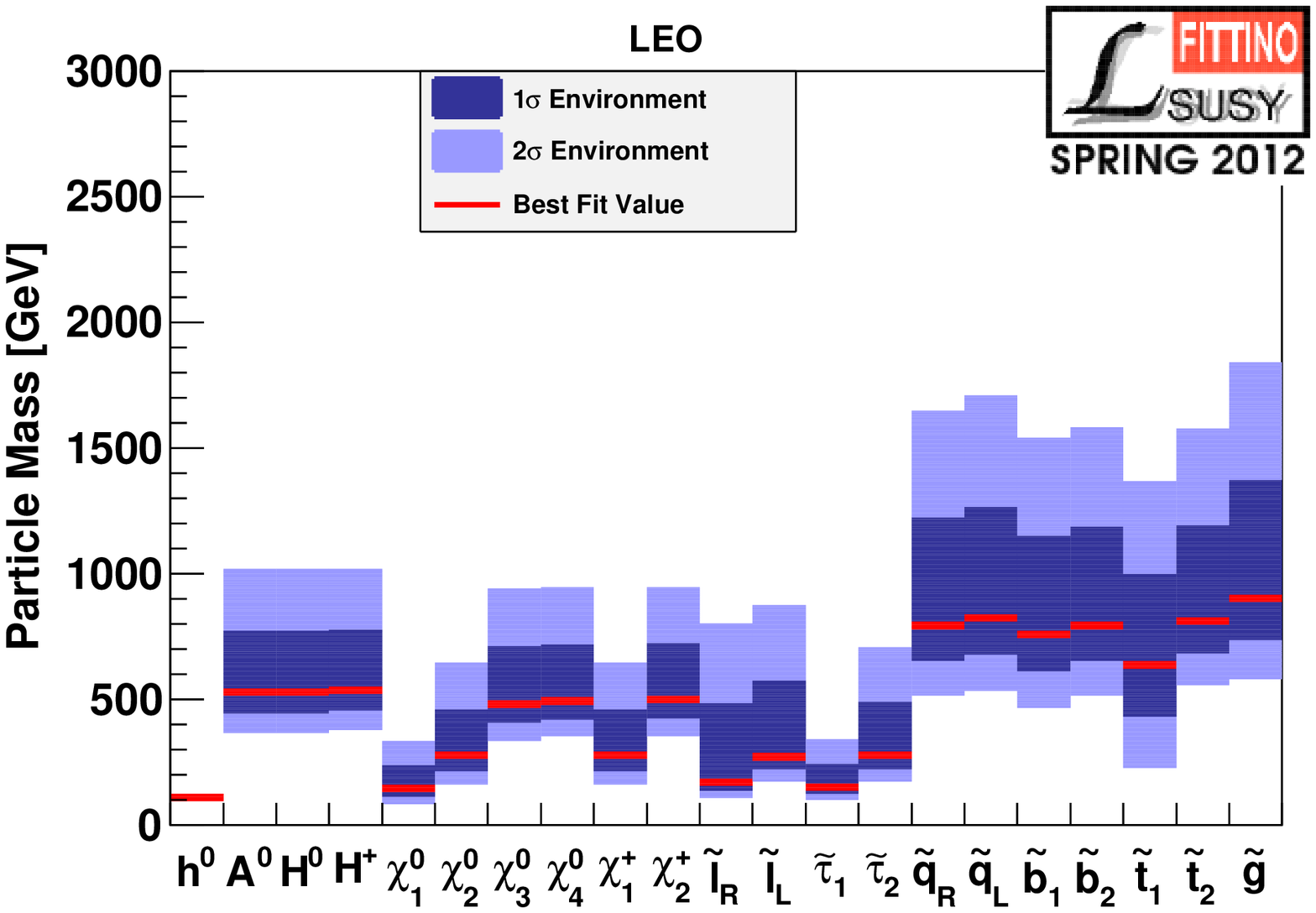} 
      \label{fig:mdp_lhc_0fbXn}
    }
    \subfigure[]{
      \includegraphics[width=0.49\textwidth,clip=]{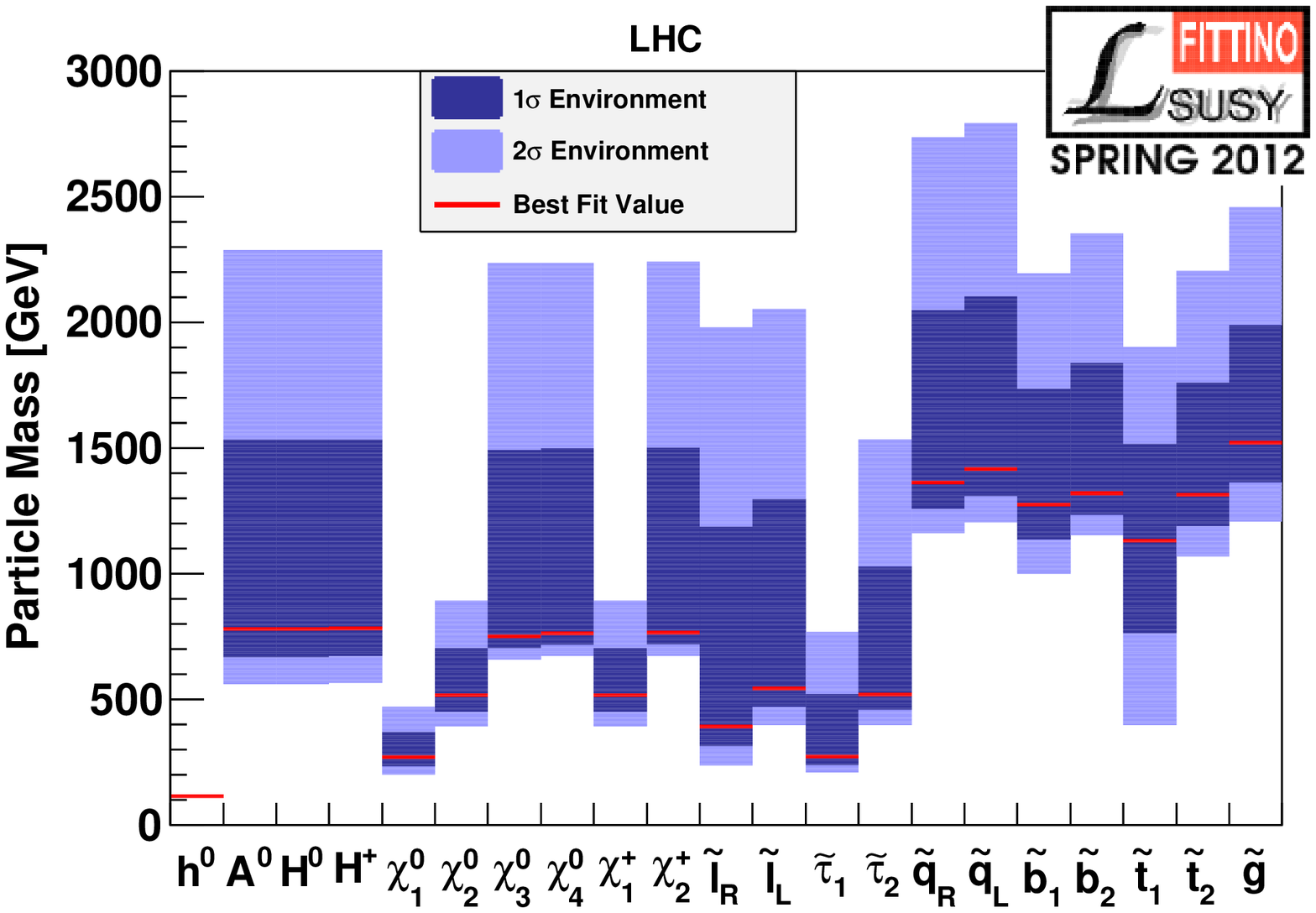} 
      \label{fig:mdp_lhc_02fbXn}
    }
    \caption{Predicted distribution of sparticle and Higgs boson
    masses from \subref{fig:mdp_lhc_0fbXn} the LEO fit and
    \subref{fig:mdp_lhc_02fbXn} the LHC fit. The full uncertainty band
    gives the 1-dimensional 2$\sigma$ uncertainty of each mass defined
    by the region $\Delta\chi^2<4$ after profiling over all hidden
    dimensions.}  \label{fig:mdp_lhc}
\end{figure}

We show the predicted range of sparticle and Higgs boson masses for
the LHC fit in Fig.~\ref{fig:mdp_lhc_02fbXn} compared to the same
prediction for the LEO fit in Fig.~\ref{fig:mdp_lhc_0fbXn}.  As
discussed above and in Fig.~\ref{fig:noHB}, the flat $\chi^2$ profile
for large $M_0$ and $M_{1/2}$ predicts sparticle masses at $m\approx
1$\,TeV for both the LEO and LHC fits. The only particle with a very
strong upper bound on its mass is the lightest Higgs boson. The
prediction of $m_h$ changes from $m_h=(113.5\pm3)$\,GeV for the LEO
fit to $m_h=(117 \pm1.5)$\,GeV for the LHC fit. The 1-dimensional
$1\sigma$ range quoted here is clearly incompatible with the potential
signal at $m_h=126\, $GeV. However the $\chi^2$ profile is flatter for
higher $m_h$, making it marginally possible to reconcile a heavier
Higgs mass with the CMSSM, as outlined in Section~\ref{sec:higgs}. In
addition, the lightest neutralino is bound from above at
$m_{\chi^0_1}\lesssim500$\,GeV, as a result of the combined constraints
on $\Omega_{\rm DM}$ and $(g-2)_{\mu}$.

We note that also in the LHC fit there is still enough room for
slepton and gaugino masses below $m\lesssim500 $\,GeV, which would be
observable at an $e^+e^-$ linear collider with
$\sqrt{s}\lesssim1$\,TeV, despite the stringent search limits from the
LHC. This is partly due to the fact that the LHC allowed $2\sigma$
region extends slightly below the 95\,\%\,CL exclusion. The increasing
LHC $\chi^2$ contribution for smaller sparticle mass scales is partly
canceled by the decreasing contribution from the LEO observables. The
stronger effect however is the fact that even highly constrained
models as the CMSSM still exhibit enough freedom in the parameter
space to reconcile light gauginos and sleptons with heavy squarks and
gluinos, as shown in Fig.~\ref{fig:mdp_lhc_02fbXn}. This feature would
be even more pronounced in more general SUSY models, in which the
colored and the non-colored sectors explicitly decouple.

\subsection{Adding a potential Higgs boson mass measurement}
\label{sec:higgs}


\begin{figure}[t]
  \begin{center}
    \includegraphics[width=0.65\textwidth,clip=]{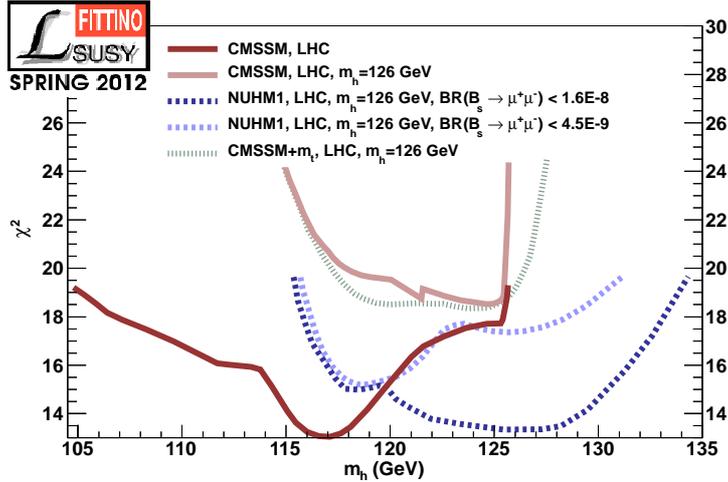}
  \end{center}
  \caption{The dependence of the minimal $\chi^2$ of the fit on
    $m_{h}$ for different input observable sets and for the CMSSM and
    NUHM1.  CMSSM and NUHM1 fits with and without
    $m_h=(126\pm2\pm3)$\,GeV are shown.  For the CMSSM, whose
    best-fit-point coincides with a prediction of the SM value of
    ${\cal B}(B_s\to\mu\mu)$, the strongest impact comes from
    including $m_h=(126\pm2\pm3)$\,GeV into the fit (compare dark and
    light red lines).  Adding $m_t$ as a fit parameter has limited
    effect (compare green and light red line).  For the NUHM1, a
    heavier Higgs boson is in principle easy to accommodate, however a
    heavy Higgs boson is disfavored in the NUHM1 by the present bound on
    ${\cal B}(B_s\to\mu\mu)$~\cite{LHCb-Moriond} compared to a
    previous looser limit~\cite{Serrano:2011px} (compare light and
    dark blue lines).  }\label{fig:chi2_1D_higgs}
\end{figure}

A very strong constraint on the CMSSM is expected from a direct
measurement of the lightest Higgs boson mass $m_h$. This is shown
explicitly in Fig.~\ref{fig:chi2_1D_higgs}, where we display the $\chi^2$
profile of $m_h$ for different fits: the LHC fit, the LHC+$m_h=$126
fit with fixed $m_t$, the same with $m_t=173.2\pm1.34 $\,GeV floating,
and the NUHM1 fit with the same observables. It can be seen that the
CMSSM fits (red and green curves) can barely incorporate
$m_h=126$\,GeV. At this edge the $\chi^2$ rises dramatically excluding
any higher Higgs boson mass values. This limit is just slightly increased by floating
$m_t$ (green curve), but the overall $\chi^2/ndf$ of this fit is still
at the same rather inconsistent level of 18/9 as for the fit with
fixed $m_t$, for which 18.4/9 is observed. It is obvious that the
NUHM1 fit, which decouples the Higgs boson mass parameter squared
$M_{H}^2=M_{H_u}^2= M_{H_d}^2$ from $M_0^2$ at the GUT scale, shows
both a significantly better agreement of $\chi^2/ndf=15.3/8$ and the
ability to accommodate higher values of $m_h$. 

In Ref.~\cite{{Dedes:2001fv}} a strong correlation between ${\cal
  B}(B_s\to\mu\mu)$ and $a_\mu$ was observed for fixed $m_h$. It was 
shown that the correlated region for the current experimental 
values of ${\cal
  B}(B_s\to\mu\mu)$ and $a_\mu$ is inconsistent with a Higgs boson mass at
$m_h=(126\pm2({\rm exp})\pm3 ({\rm theo}))$\,GeV. Accordingly the pull plot
Fig.~\ref{fig:IndividualvariablePullPlots_2fbXnMh126} has a poor fit
for $a_\mu$. Performing a fit just for the three observables ${\cal
  B}(B_s\to\mu\mu)$, $a_\mu$, and $m_h=(126\pm2({\rm exp})\pm3 ({\rm theo}))$\,GeV
results in a $\Delta\chi^2$ penalty of 7 with respect to just fitting
${\cal B}(B_s\to\mu\mu)$ and $a_\mu$. This explains the strong
difference between the fits with and without $m_h=(126\pm2({\rm exp})\pm3
({\rm theo}))$\,GeV.  To further exhibit the combined effect of the
limit on ${\cal
  B}(B_s\to\mu\mu)$ and a potential Higgs boson observation, we show in
Fig.~\ref{fig:chi2_1D_higgs} 
the profile of $\chi^2$ against $m_h$ for different options of
the measurement of ${\cal B}(B_s\to\mu\mu)$. While a measurement at
the level of the SM of ${\cal
  B}(B_s\to\mu\mu)=(3.2\pm0.3)\times10^{-9}$ has no strong effect on
the maximal achievable value of $m_h$ for the CMSSM fit, there is a
significant difference for the NUHM1 fit, where large $m_h$ is 
disfavored for a low ${\cal B}(B_s\to\mu\mu)$. Even the current LHCb
limit severely impacts the fit quality with respect to a looser bound on
${\cal B}(B_s\to\mu\mu)$, since the NUHM1 fit with
$m_h=(126\pm2({\rm exp})\pm3 ({\rm theo}))$\,GeV prefers ${\cal
  B}(B_s\to\mu\mu)\gtrsim6.5\times10^{-9}$. Of course this is not
a general observation, since the $B$-physics and the Higgs
sector observables are not directly governed by the same parameters in
the MSSM, while they are connected via GUT
scale assumptions in the CMSSM and NUHM1.  Therefore, also in this
example, testing more general SUSY models than the CMSSM and the
NUHM1 would be beneficial. However, the current limitations on the
correct implementation of the LHC results, as outlined in
Section~\ref{subsec:lhc}, has to be overcome.

\begin{figure}[t]
  \subfigure[]{
    \includegraphics[width=0.49\textwidth,clip=]{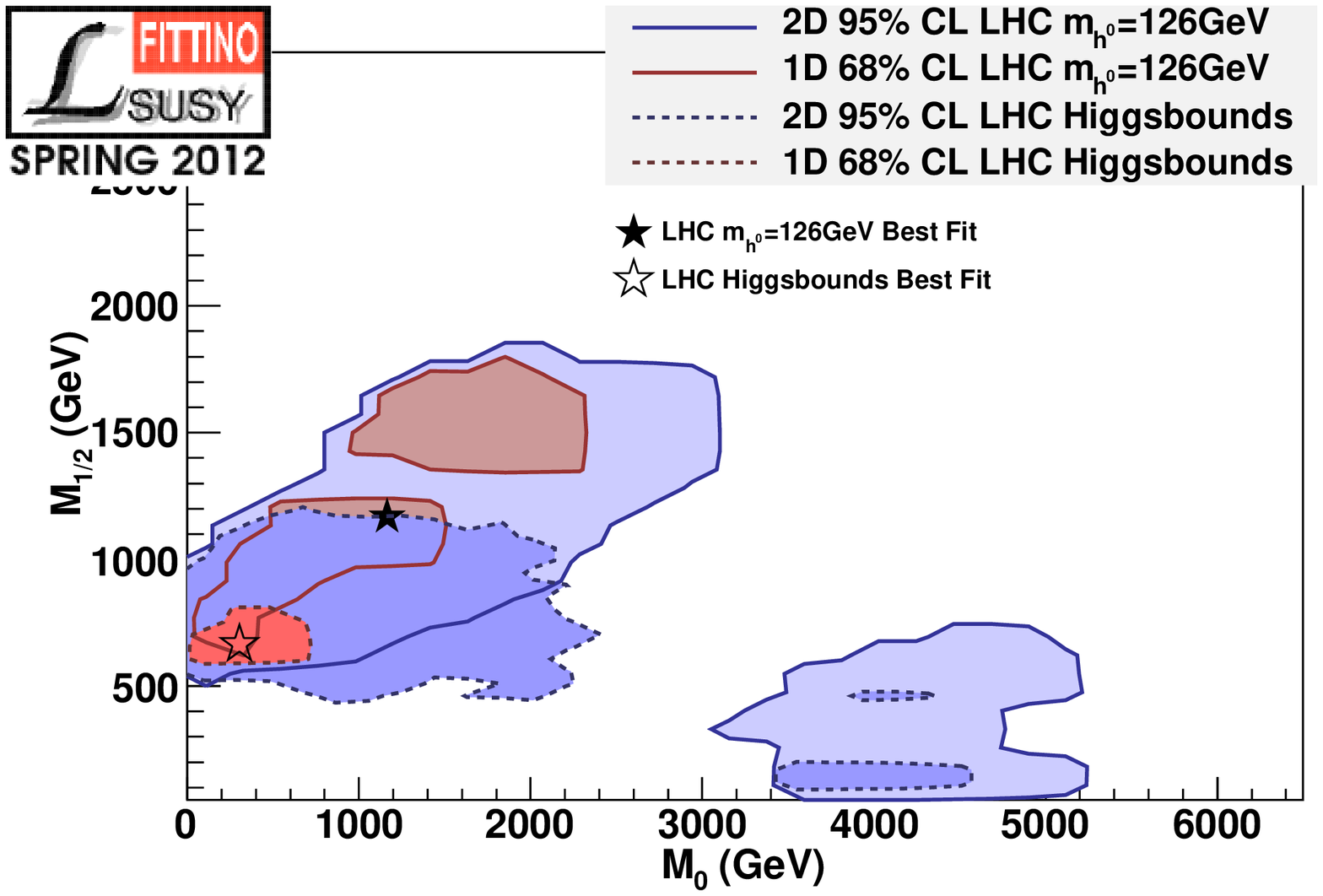}
    \label{fig:msugra_parameter_dist_mtopfix_M0M12}
  }
  \subfigure[]{
    \includegraphics[width=0.49\textwidth,clip=]{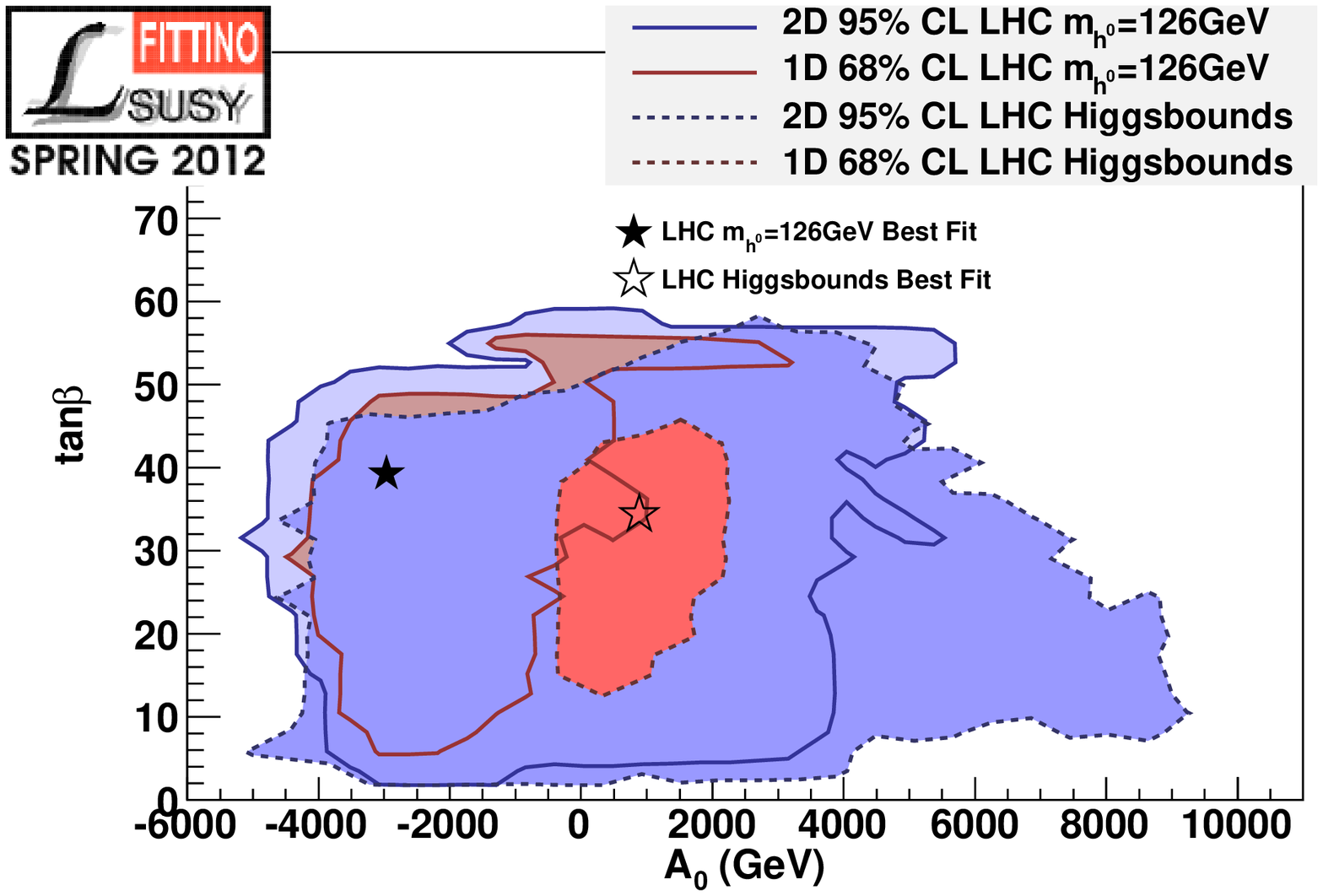}
    \label{fig:msugra_parameter_dist_mtopfix_A0Tanb}
  }
  \subfigure[]{
    \includegraphics[width=0.49\textwidth,clip=]{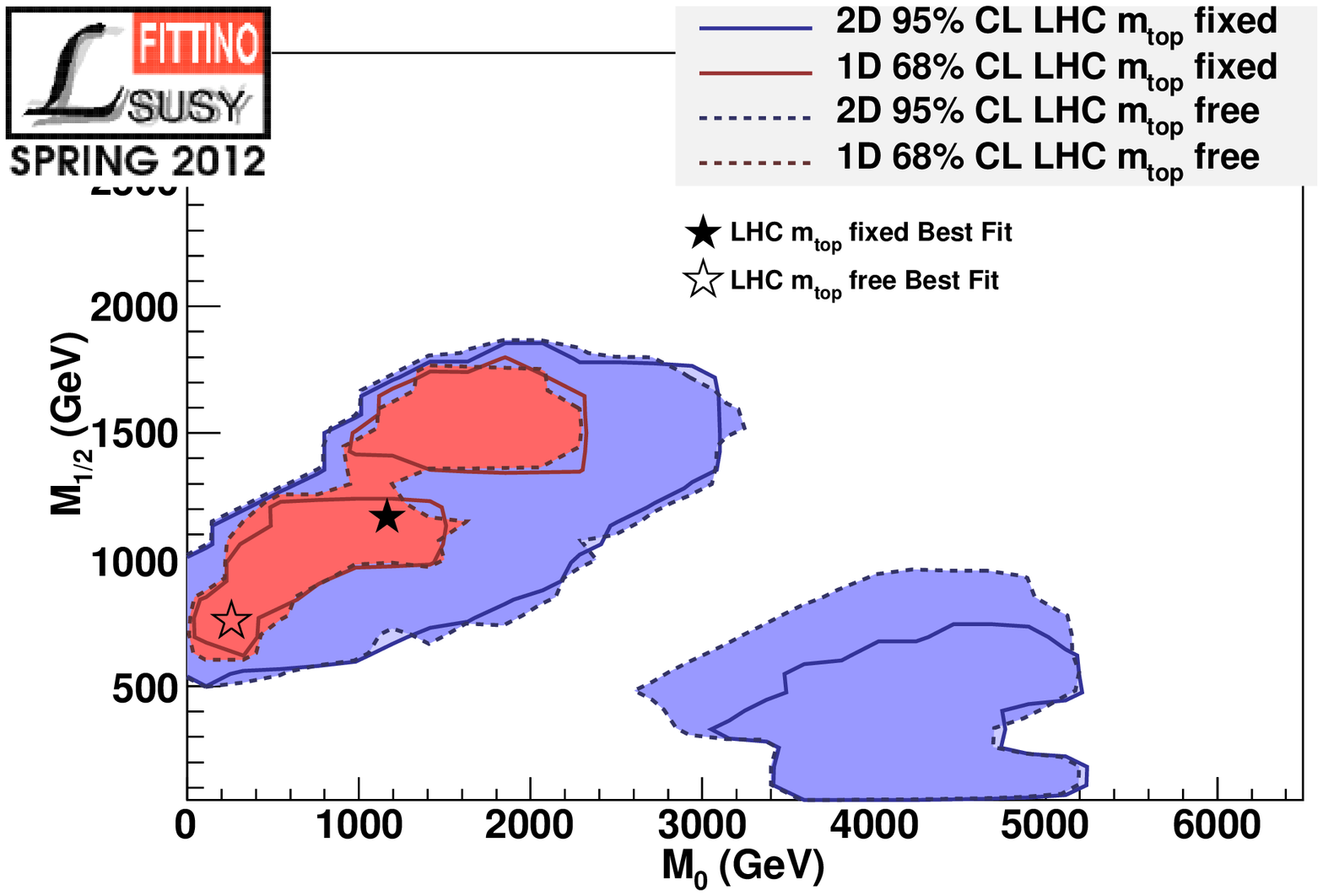}
    \label{fig:msugra_parameter_dist_mtopfree_M0M12}
  }
  \subfigure[]{
    \includegraphics[width=0.49\textwidth,clip=]{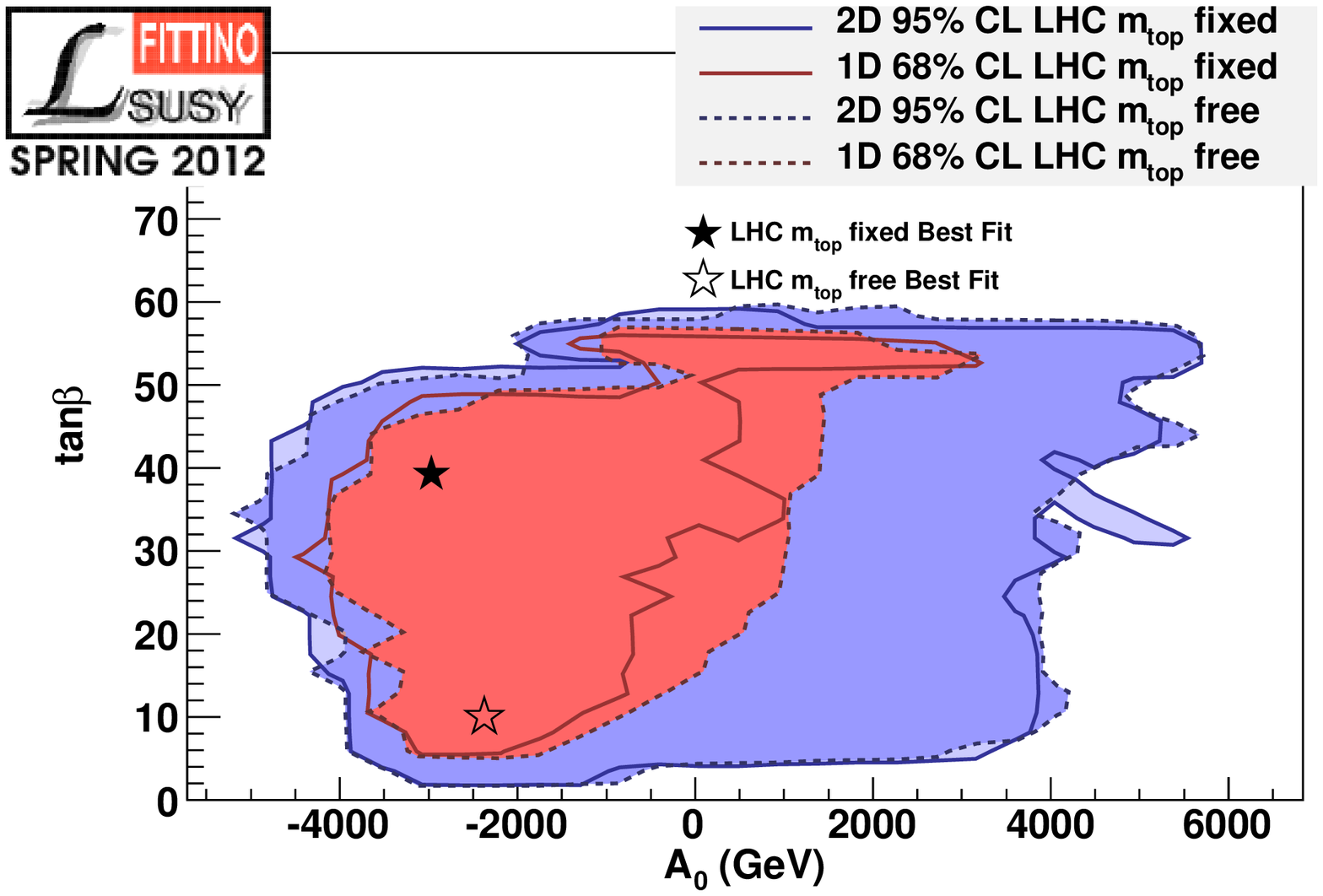}
    \label{fig:msugra_parameter_dist_mtopfree_A0Tanb}
  }
  \caption{CMSSM parameter distributions in ($M_0,M_{12}$) and
    ($\tan\beta,A_0$).
    \subref{fig:msugra_parameter_dist_mtopfix_M0M12} and
    \subref{fig:msugra_parameter_dist_mtopfix_A0Tanb} show the fit
    results for $m_h=(126\pm2\pm3)$\,GeV and fixed $m_t=173.2$\,GeV,
    compared to the LHC
    fit. \subref{fig:msugra_parameter_dist_mtopfree_M0M12} and
    \subref{fig:msugra_parameter_dist_mtopfree_A0Tanb} show the fit
    with the same input observable set, but with
    $m_t=173.2\pm1.34$\,GeV floating free in the fit, in comparison to
    the fit with fixed $m_t$.  }\label{fig:msugra_parameter_dist}
\end{figure}

Figure~\ref{fig:msugra_parameter_dist} shows the allowed parameter
range in the ($M_0,M_{12}$) and ($\tan\beta,A_0$) planes for the CMSSM
fits with $m_t$ fixed in
Fig.~\ref{fig:msugra_parameter_dist_mtopfix_M0M12} and
\ref{fig:msugra_parameter_dist_mtopfix_A0Tanb} and with $m_t=173.2
\pm1.34$\,GeV floating in
Fig.~\ref{fig:msugra_parameter_dist_mtopfree_M0M12} and
\ref{fig:msugra_parameter_dist_mtopfree_A0Tanb}. The possible signal
at $m_h\approx126$\,GeV shifts the allowed region strongly into
regions of higher $M_0$ and $M_{1/2}$, as compared to the LHC
fit. This is due to the larger squark masses necessary to lift $m_h$
so strongly above the tree level bound of $m_h\leq m_Z$. Also, large
$\tan\beta$ is clearly preferred, while again showing a flat profile
in $\tan\beta$ outside the $1\sigma$ region. As expected from the
small effect which floating $m_t$ had on the fit in
Fig.~\ref{fig:chi2_1D_higgs}, there is no significant difference
between the allowed parameter ranges for $m_t$ fixed and $m_t$
floating, albeit there is a significant jump in the best fit point due
to the flatness of the $\chi^2$ profile. Since $m_t$ is expected to
have the strongest direct effect on the prediction of all SM
parameters, this confirms that with the current observable set the SM
parameters can be fixed in the fit, since their uncertainties decouple
completely from the SUSY parameter uncertainties.

\begin{figure}[t]
  \begin{center}
    \subfigure[]{
      \includegraphics[width=0.47\textwidth,clip=]{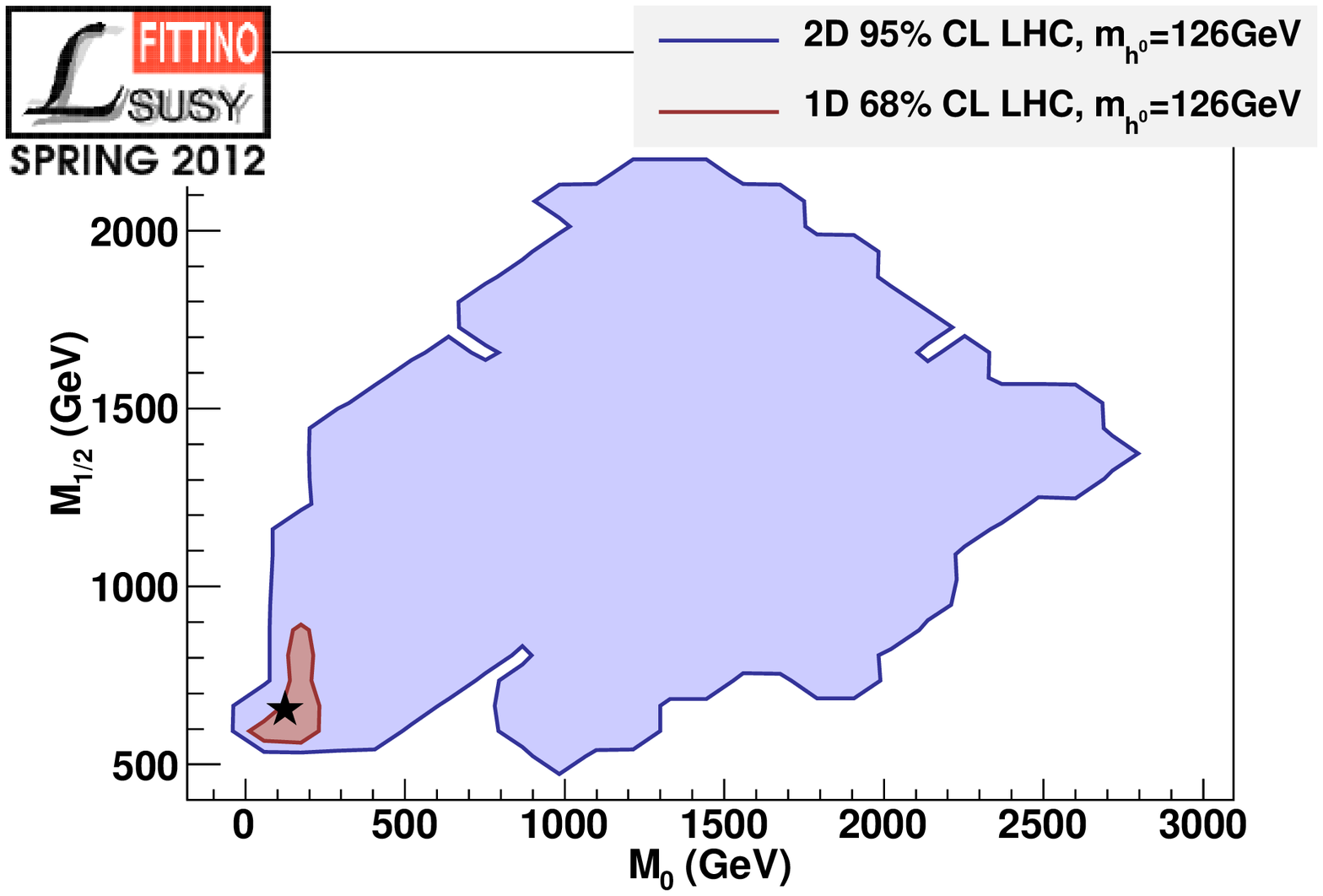}
      \label{fig:NUHM1_parameter_dist_M0M12}
    }
    \subfigure[]{
      \includegraphics[width=0.47\textwidth,clip=]{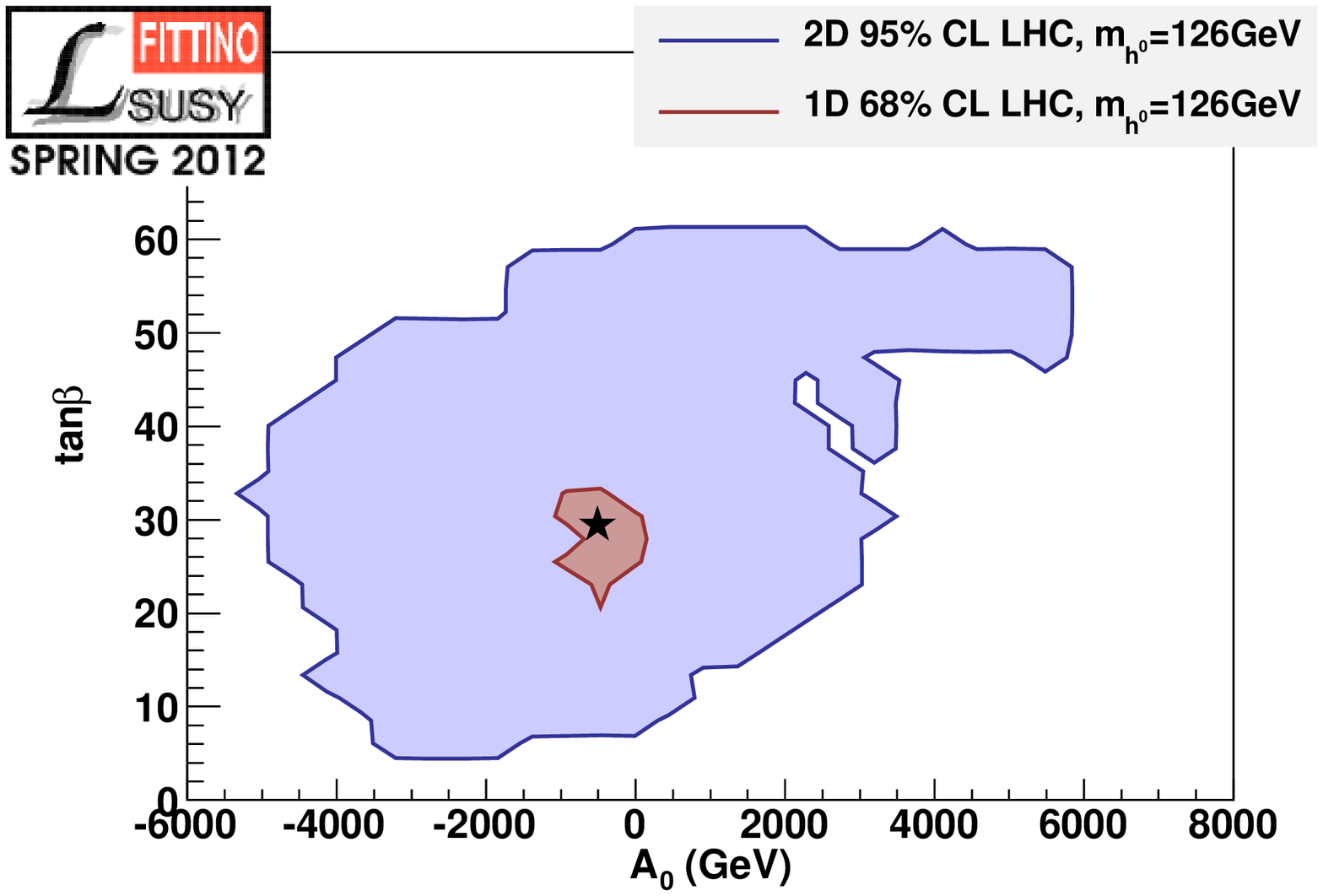}
      \label{fig:NUHM1_parameter_dist_A0Tanb}
    }
    \subfigure[]{
      \includegraphics[width=0.47\textwidth,clip=]{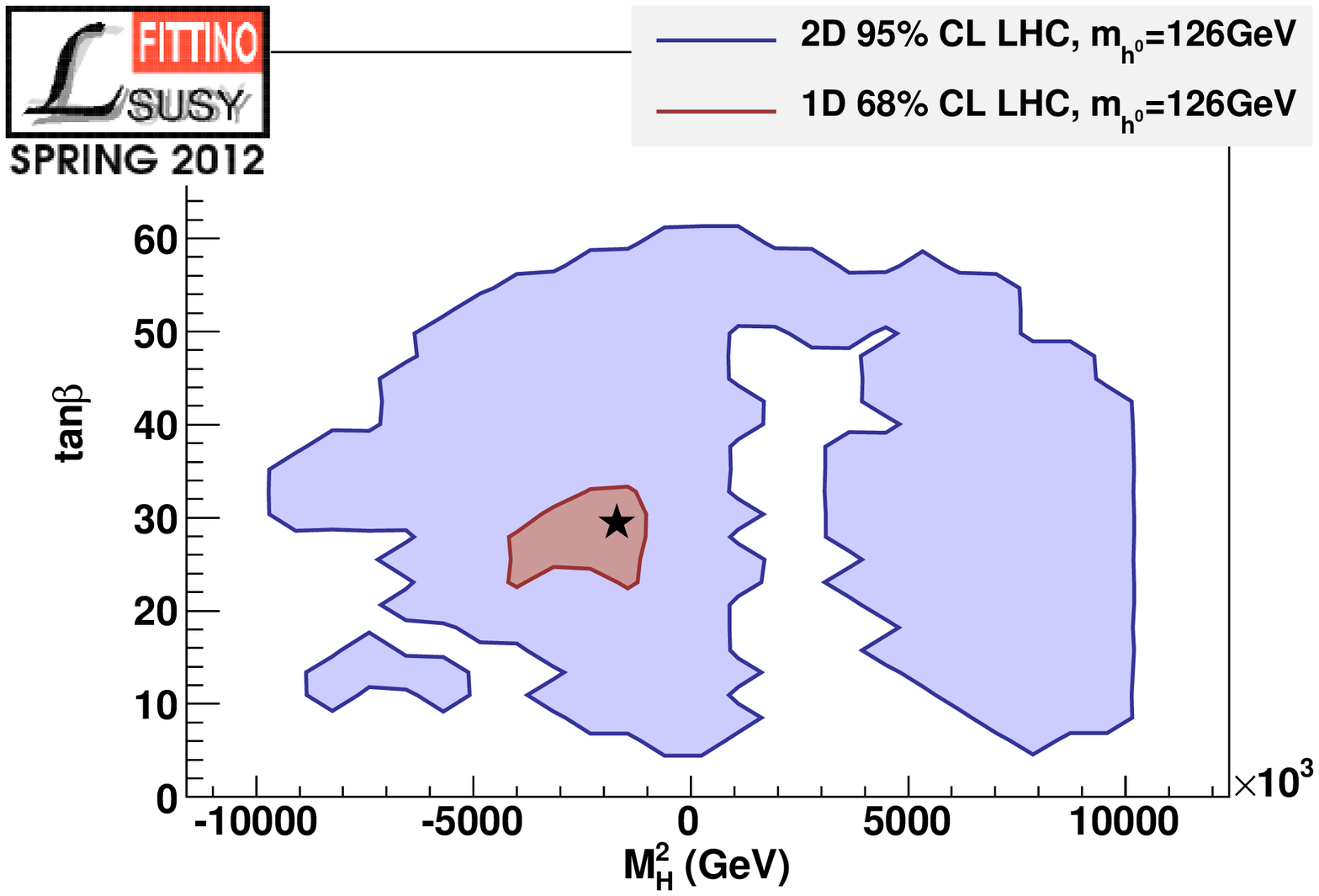}
      \label{fig:NUHM1_parameter_dist_M0HTanb}
    }
  \end{center}
  \caption{NUHM1 parameter distributions in ($M_0,M_{12}$) in
    \subref{fig:NUHM1_parameter_dist_M0M12}, ($\tan\beta,A_0$) in
    \subref{fig:NUHM1_parameter_dist_A0Tanb} and ($\tan\beta,M_H^2$)
    in \subref{fig:NUHM1_parameter_dist_M0HTanb} for the fit to the
    full observable set including
    $m_h=(126\pm2\pm3)$\,GeV.}\label{fig:NUHM1_parameter_dist}
\end{figure}

Figure~\ref{fig:NUHM1_parameter_dist} shows the allowed parameter
space of the NUHM1 model. Negative values for $M^2_H$ can be
considered because the relevant parameter combination for EWSB is
$|\mu|^2 + M^2_H$, which we checked to be positive above the
electroweak scale.  Since this model reaches lower $\chi^2$ for the
same observable set as the LHC+$m_h=$126 fit, its area of low $\chi^2$
is deep enough with respect to the surrounding flat profile (see
Section~\ref{sec:results_leo} and \ref{sec:results_lhc}) to exclude
the focus point region from the 2-dimensional $2\sigma$ region. Other
than that, no significant deviations from the CMSSM can be observed
for $M_0$ and $M_{1/2}$. However, the NUHM1 may also be sensitive to
exclusions from the search for $A^0\to\tau\tau$ at the
LHC. Unfortunately, these searches are not yet implemented in
\texttt{HiggsBounds} in a model independent way, and hence not
available for this fit.

\begin{figure}[t]
  \subfigure[]{
    \includegraphics[width=0.49\textwidth,clip=]{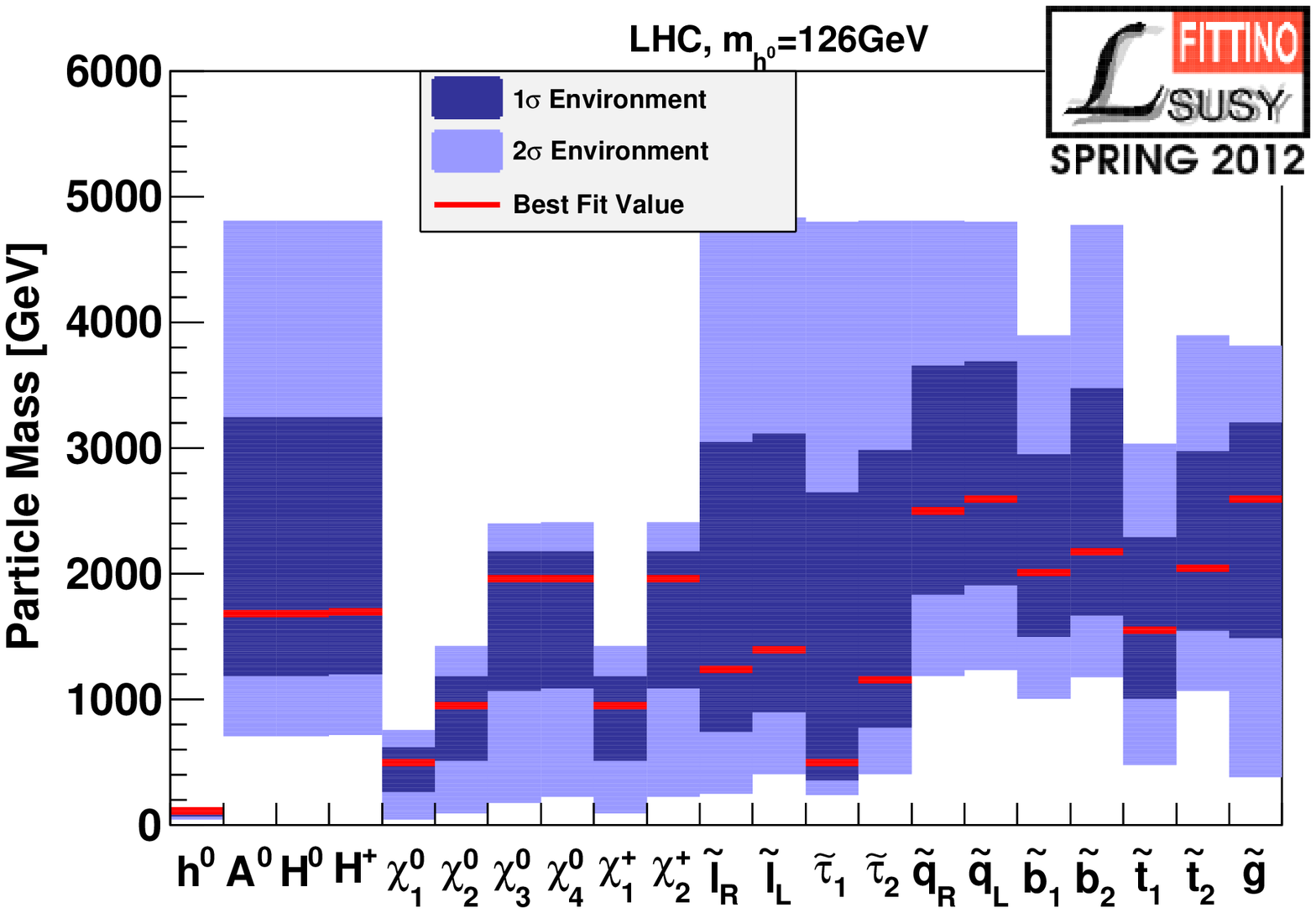} 
    \label{fig:mdp_higgs_5fbH126Xn}
  }
  \subfigure[]{
    \includegraphics[width=0.49\textwidth,clip=]{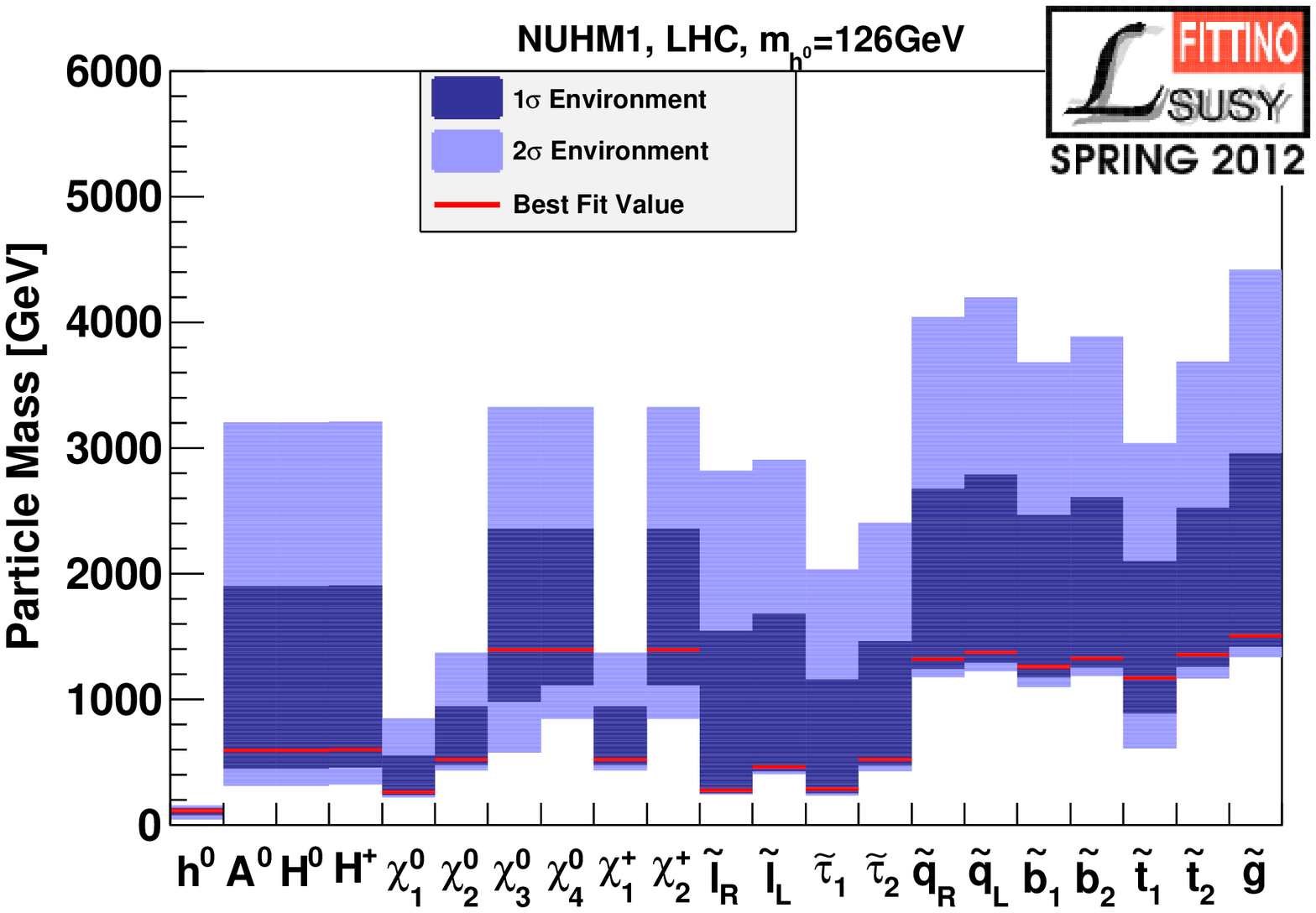} 
    \label{fig:mdp_higgs_nuhm5fbH126Xn}
  }  
  \caption{Predicted distribution of sparticle and Higgs boson masses
    from the CMSSM fit with $m_h=(126\pm2\pm3)$\,GeV in
    \subref{fig:mdp_higgs_5fbH126Xn} and the NUHM1 fit to the same
    observable set in \subref{fig:mdp_higgs_nuhm5fbH126Xn}.}
  \label{fig:mdp_higgs}
\end{figure}

In Fig.~\ref{fig:mdp_higgs_5fbH126Xn}, the predicted sparticle and
Higgs boson mass range for the CMSSM with $m_h=(126\pm2\pm3)$\,GeV is
shown. As discussed above, the high value of $m_h$ can best be reached
for high sparticle masses, shifting up the expected squark mass scale
significantly with respect to Fig.~\ref{fig:mdp_lhc_02fbXn}. The NUHM1
fit is not pushed into the same area of a flat $\chi^2$ profile for
heavy $M_0$ and $M_{1/2}$ as the CMSSM LHC+$m_h=$126 fit. Therefore,
and because it can decouple the Higgs sector via the separate
parameter $M_H^2$, it prefers significantly lower sparticle and heavy
Higgs boson mass scales than the CMSSM with the same observables.

\begin{figure}[t]
    \subfigure[]{
      \includegraphics[width=0.49\textwidth,clip=]{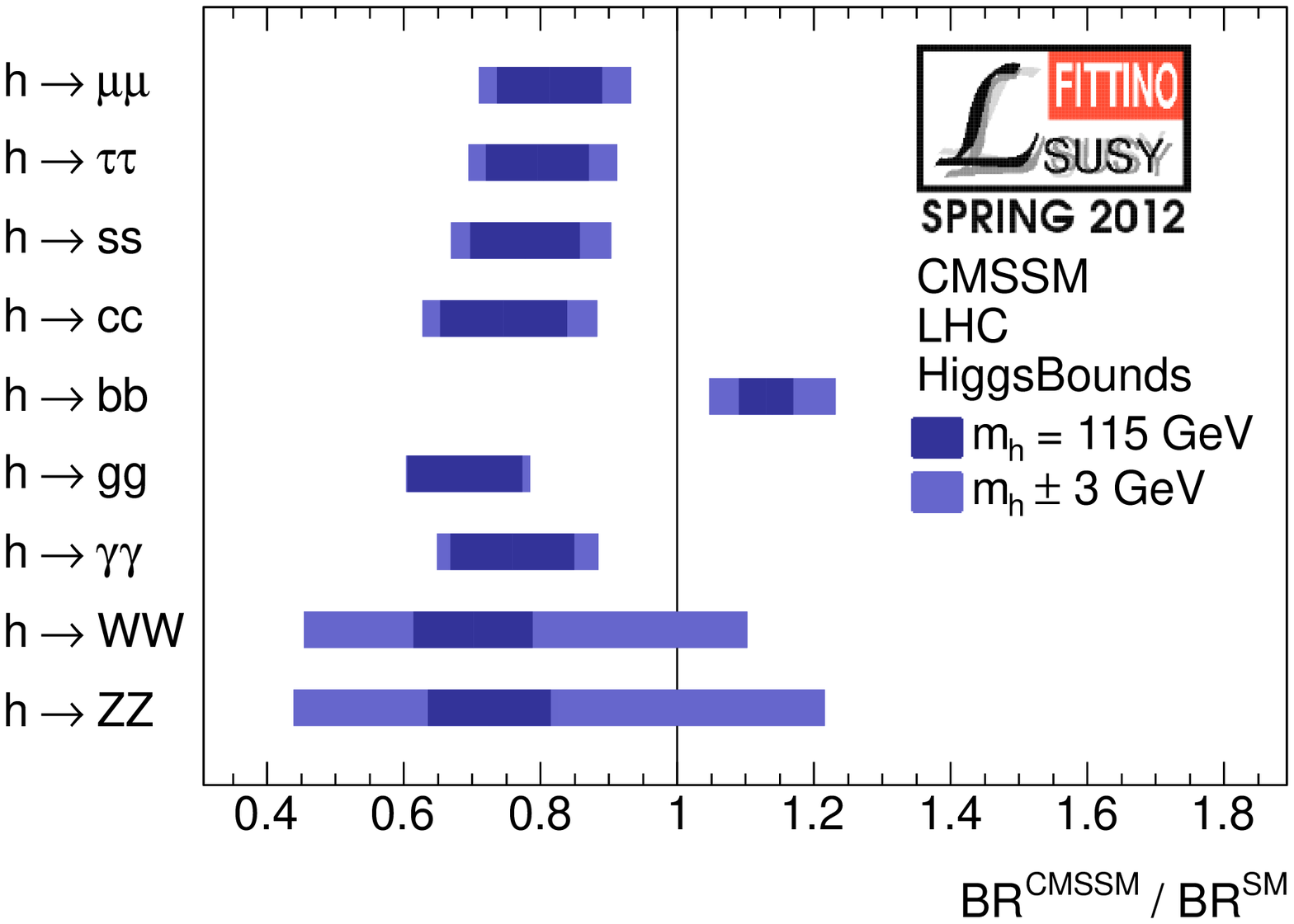}
      \label{fig:HiggsBrAllowedRanges_LHC2}
    }
    \subfigure[]{
      \includegraphics[width=0.49\textwidth,clip=]{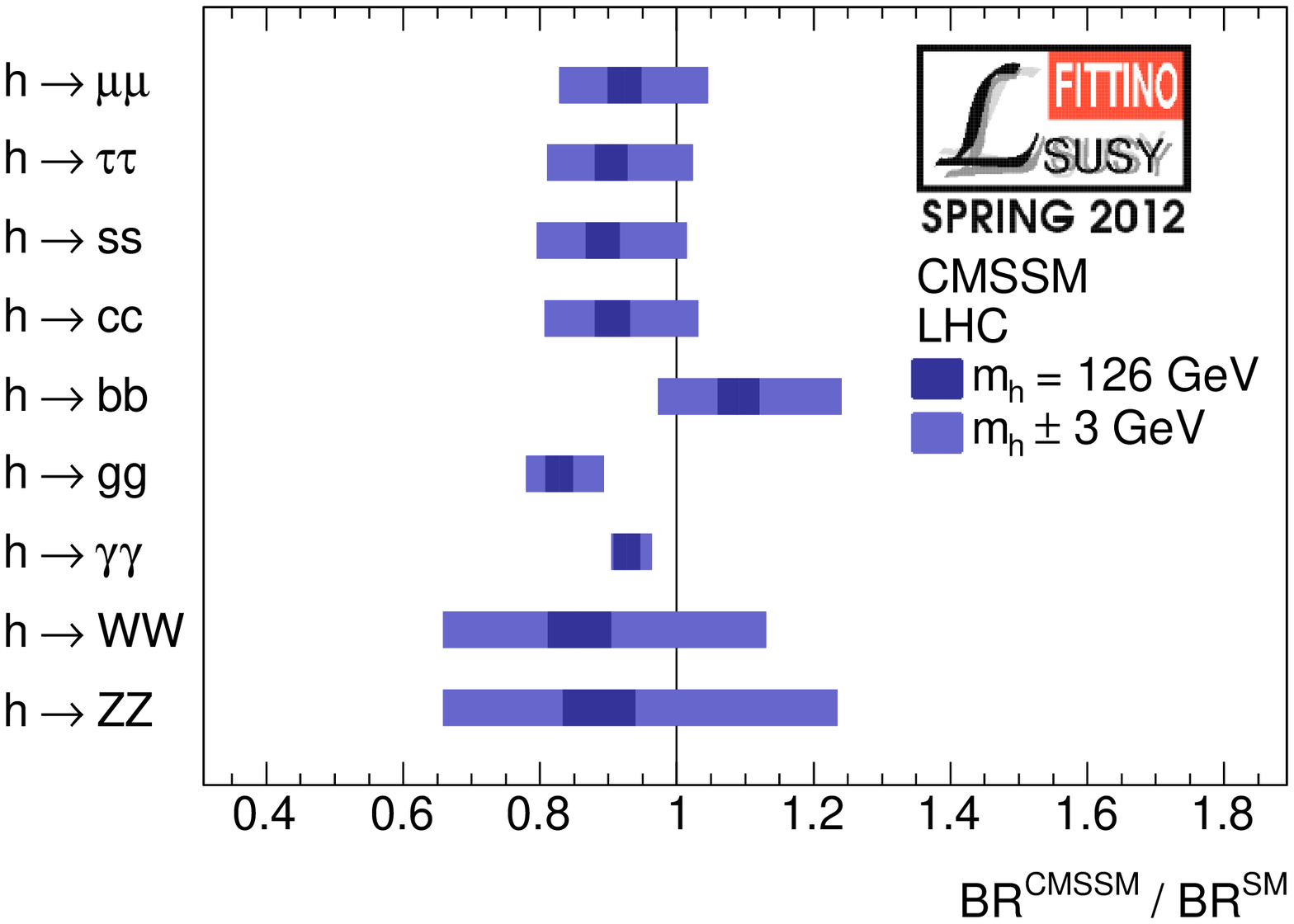}
      \label{fig:HiggsBrAllowedRanges_LHC2H126}
    }
    \subfigure[]{
      \includegraphics[width=0.49\textwidth,clip=]{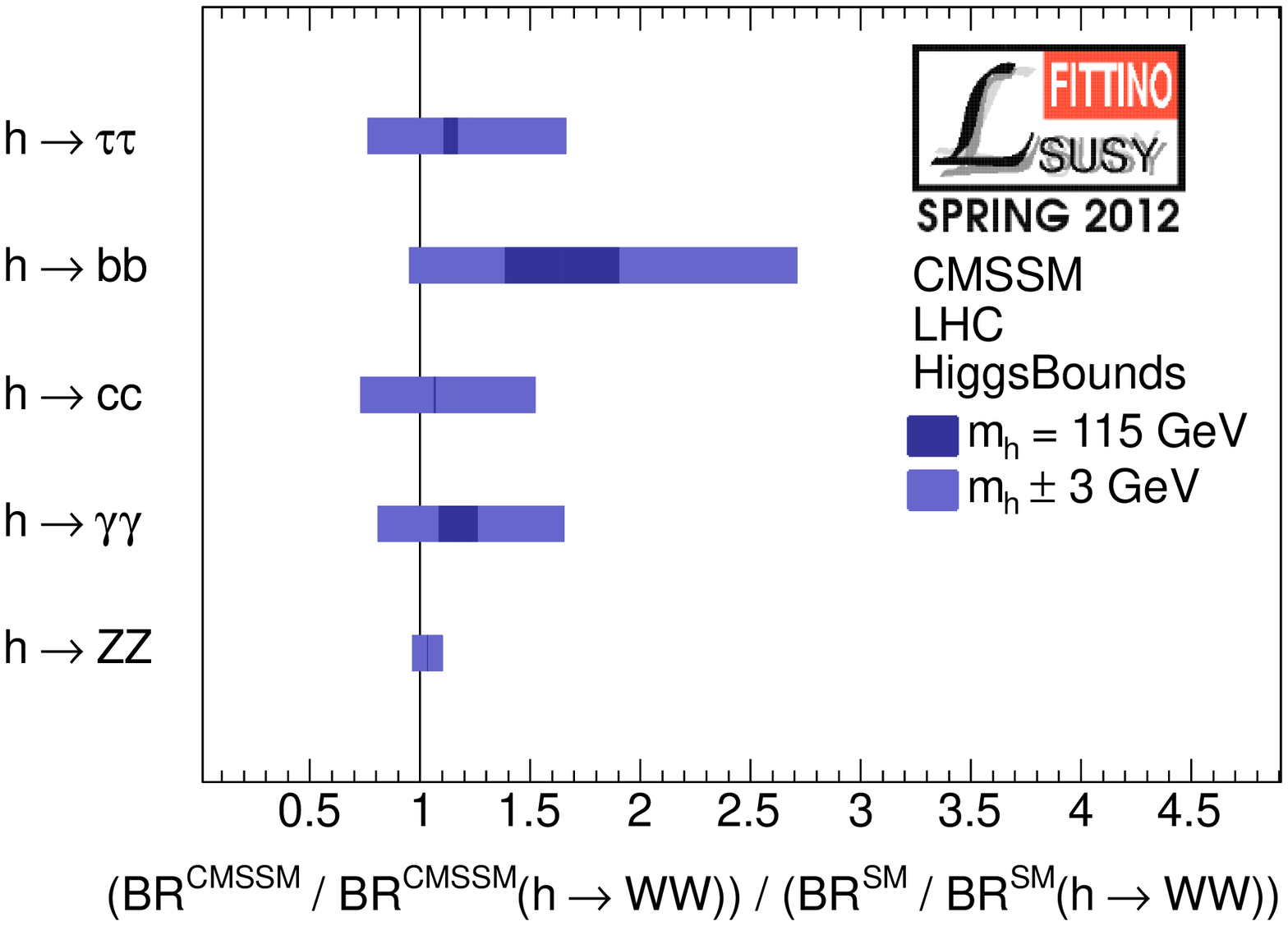}
      \label{fig:HiggsBrAllowedRatios_LHC2}
    }
    \subfigure[]{
      \includegraphics[width=0.49\textwidth,clip=]{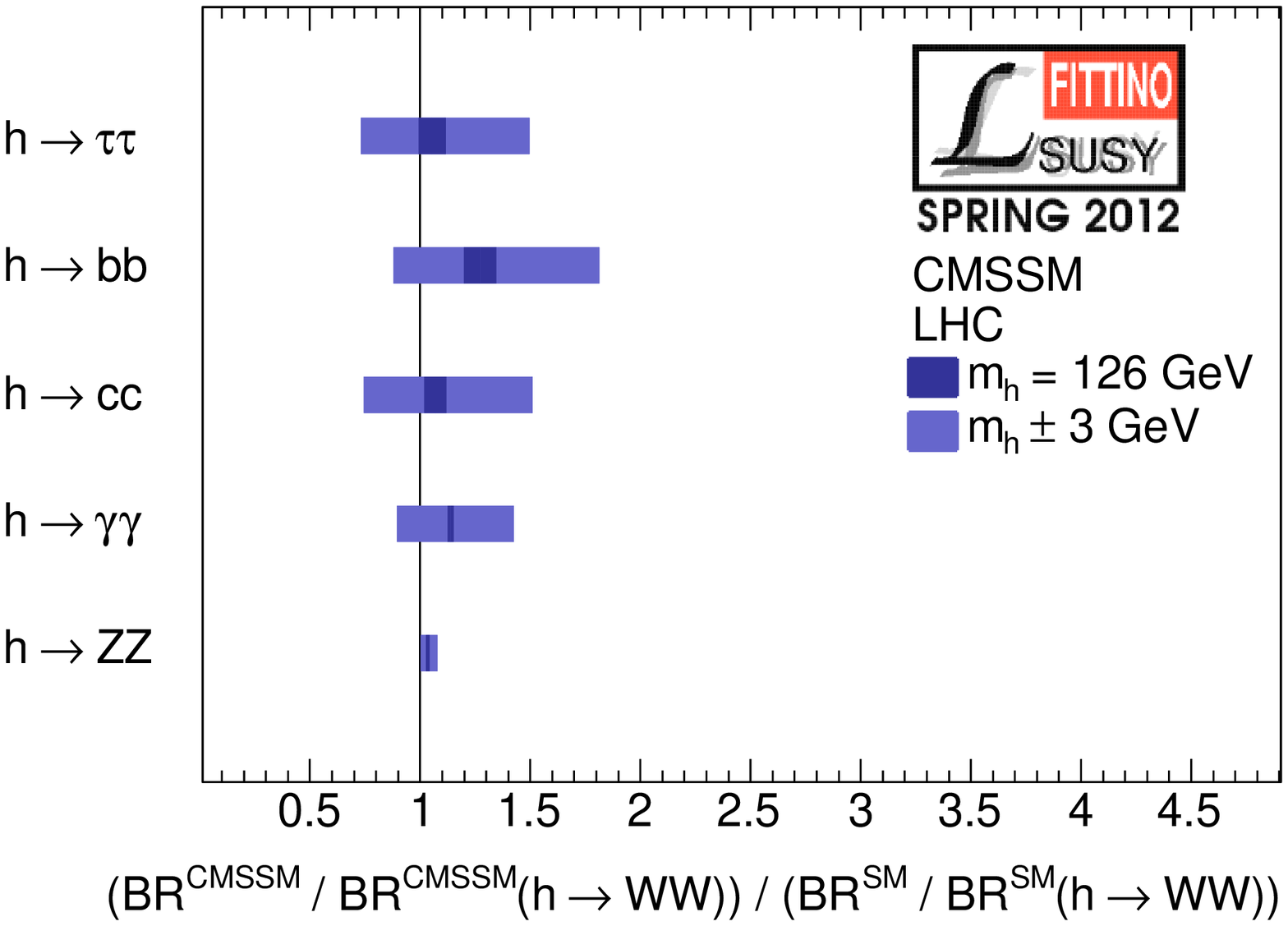}
      \label{fig:HiggsBrAllowedRatios_LHC2H126}
    }
    \caption{Predicted $2\sigma$ range of Higgs branching fractions for the LHC
      fit of the CMSSM in \subref{fig:HiggsBrAllowedRanges_LHC2} and
      the CMSSM fit with $m_h=(126\pm2\pm3)$\,GeV in
      \subref{fig:HiggsBrAllowedRanges_LHC2H126}. The results do
      include the theoretical uncertainty of the Higgs boson mass of
      $\pm3$\,GeV. Ratios of the potentially experimentally accessible
      branching fractions at LHC are given in in
      \subref{fig:HiggsBrAllowedRatios_LHC2} and
      \subref{fig:HiggsBrAllowedRatios_LHC2H126} for the same fits as
      above.}\label{fig:HiggsBrAllowedRanges}
\end{figure}

The observed fit results can be used to look at the range of allowed
Higgs branching fractions in the CMSSM. These measurements are not
used in the fit, and the range of predicted branching fraction values
can be plotted within the allowed parameter range. Since the program
\texttt{HDECAY}~\cite{Djouadi:1997yw} is not included in the fitting
process, it is used after the determination of the $2\sigma$ allowed
parameter range to predict the ratios of lightest Higgs branching
fractions between the CMSSM and the SM. In order to do this, the
parameters are scanned in a 4-dimensional grid around the best fit
point within their 1-dimensional $2\sigma$ uncertainties. We assume
that $m_h$ can be measured with an experimental precision of better
than 2\,GeV, once the branching fractions or ratios of branching
fractions are measured with sufficient precision.  The theoretical
uncertainty of the prediction for $m_h$ must be taken into account for
a prediction of ${\cal B}_{\rm CMSSM}/{\cal B}_{\rm SM}$. In
Fig.~\ref{fig:HiggsBrAllowedRanges_LHC2} the allowed range of ${\cal
  B}_{\rm CMSSM}/{\cal B}_{\rm SM}$ for various branching fractions is shown
for the LHC fit at $m_h=115$\,GeV.
Fig.~\ref{fig:HiggsBrAllowedRanges_LHC2H126} shows the same for the
CMSSM LHC+$m_h=$126 fit. In both cases one observes an enhancement of
the $b\bar{b}$ final state. $\tau^+\tau^-$ decreases due to different
contributions to the $\tan\beta$ enhanced terms which need to be
resummed~\cite{Carena:1999py,Dedes:2002er}.  The main difference is in
particular due to the gluino-sbottom and chargino-stop contributions
in case of the $b\bar{b}$ final state. A significant sensitivity
beyond the SM values can be observed in both cases, making potential
measurements of the branching fractions~\cite{Lafaye:2009vr} very
attractive to determine the model parameters further and to discover a
deviation from the SM even for SUSY mass scales beyond the LHC reach
at $\sqrt{s}=7$ or $8$\,TeV. In
Fig.~\ref{fig:HiggsBrAllowedRatios_LHC2} and
\ref{fig:HiggsBrAllowedRatios_LHC2H126} it can be seen that even the
(at the LHC) experimentally potentially accessible ratios of branching
fractions still exhibit significant potential for discovering
differences from the expected behavior of a SM Higgs boson. The
effect of such measurements on the fit can be expected to be even
stronger assuming the foreseen precision at an $e^+e^-$ linear
collider (see \textit{e.g.}~\cite{Djouadi:2007ik}).

\subsection{Implications for dark matter searches}
\label{sec:results_af}

\begin{figure}[t]
  \begin{center}
    \subfigure[]{
      \includegraphics[width=0.47\textwidth,clip=]{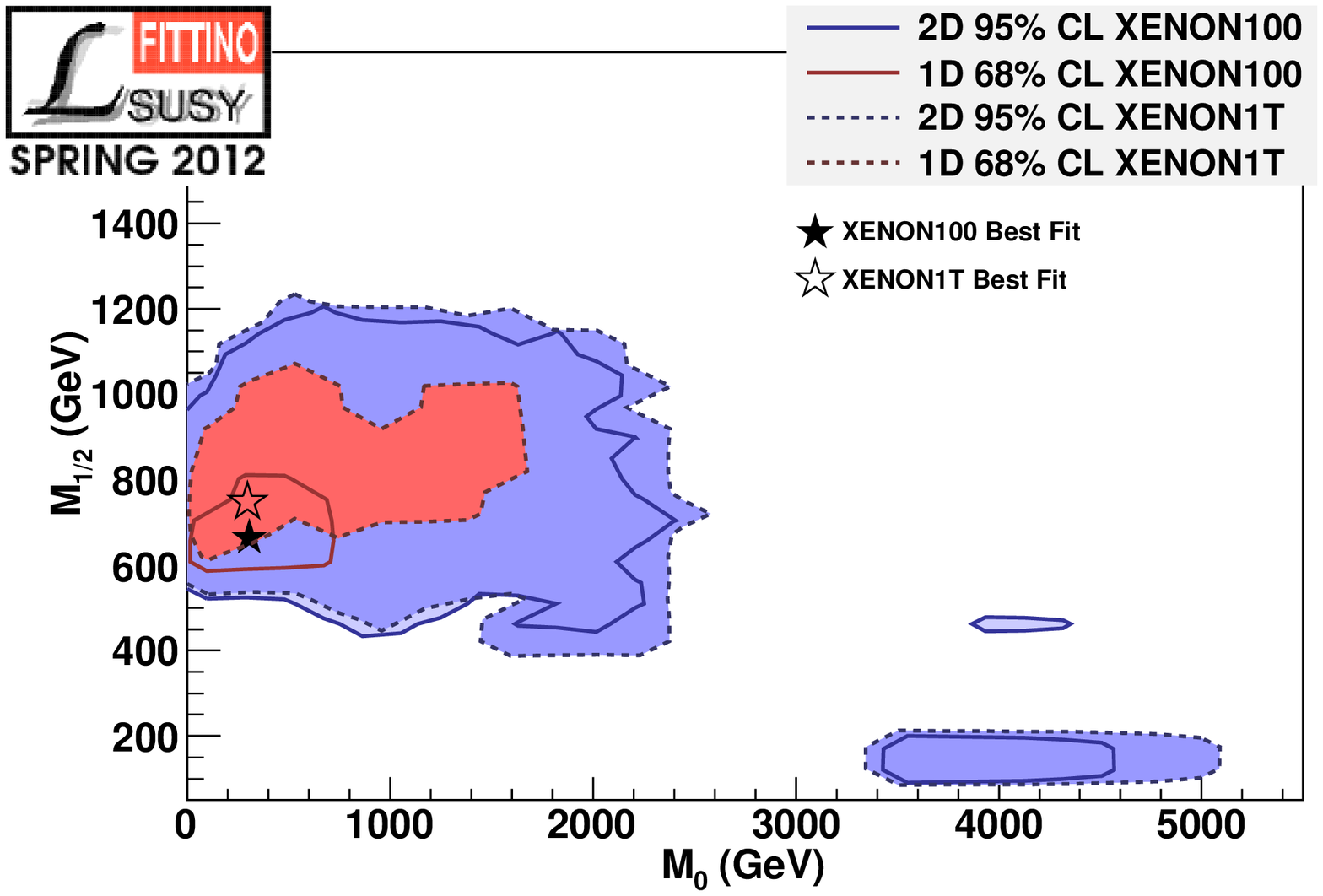} 
      \label{fig:af_CMSSMparameters_M0_M12}
    }
    \subfigure[]{
      \includegraphics[width=0.47\textwidth,clip=]{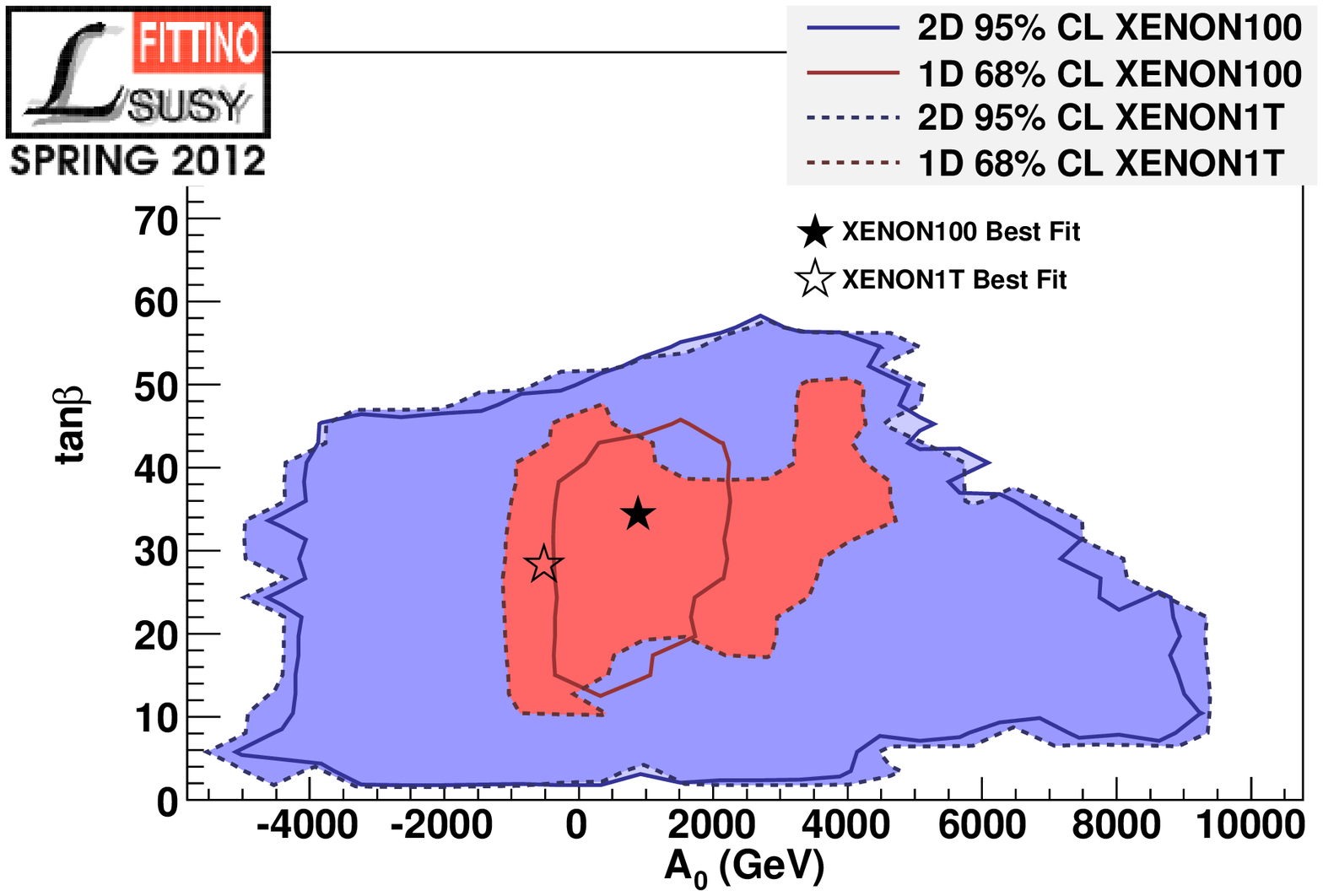} 
      \label{fig:af_CMSSMparameters_A0_tanb}
    } 
  \end{center}
  \caption{Solid lines show the results of the LHC fit (including  direct detection constraints from Xenon100 
   \cite{Aprile:2011hi}) on the  $(M_0,M_{1/2})$ (a) and $(A_0,\tan\beta)$ 
   plane (b). Increasing the runtime of Xenon100 (as aimed at with XENON100Goal) does not 
   have a significant impact on these contour plots. Even the considerably stronger constraints
    that would result from increasing the target mass (as planned with XENON1T, shown with 
    dashed lines) hardly  affects the 95\% CL regions -- at the expense, however, of making the 
    overall fit  quality worse; the 68\% CL regions, on the other hand,  become larger in that case 
    and the best fit point is shifted to a higher scale.}
  \label{fig:af_CMSSMparameters}
\end{figure}

In the following we discuss the impact of the astrophysical input
parameters on our fit results. Recall that adding the claimed signals
in \textit{direct detection} from the CoGeNT experiment
\cite{cogent,talk-collar} to our fit leads to an unacceptably high
$\Delta\chi^2_{\rm direct}$ contribution. We thus confirm earlier
observations~\cite{Baltz:2004aw} revealing that the large scattering
rates associated with these signals are incompatible with originating
solely from neutralino DM in the CMSSM. The same holds for the
DAMA/LIBRA \cite{dama} signal.

\begin{figure}[t]
  \begin{center}
    \subfigure[]{
      \includegraphics[width=0.47\textwidth,clip=]{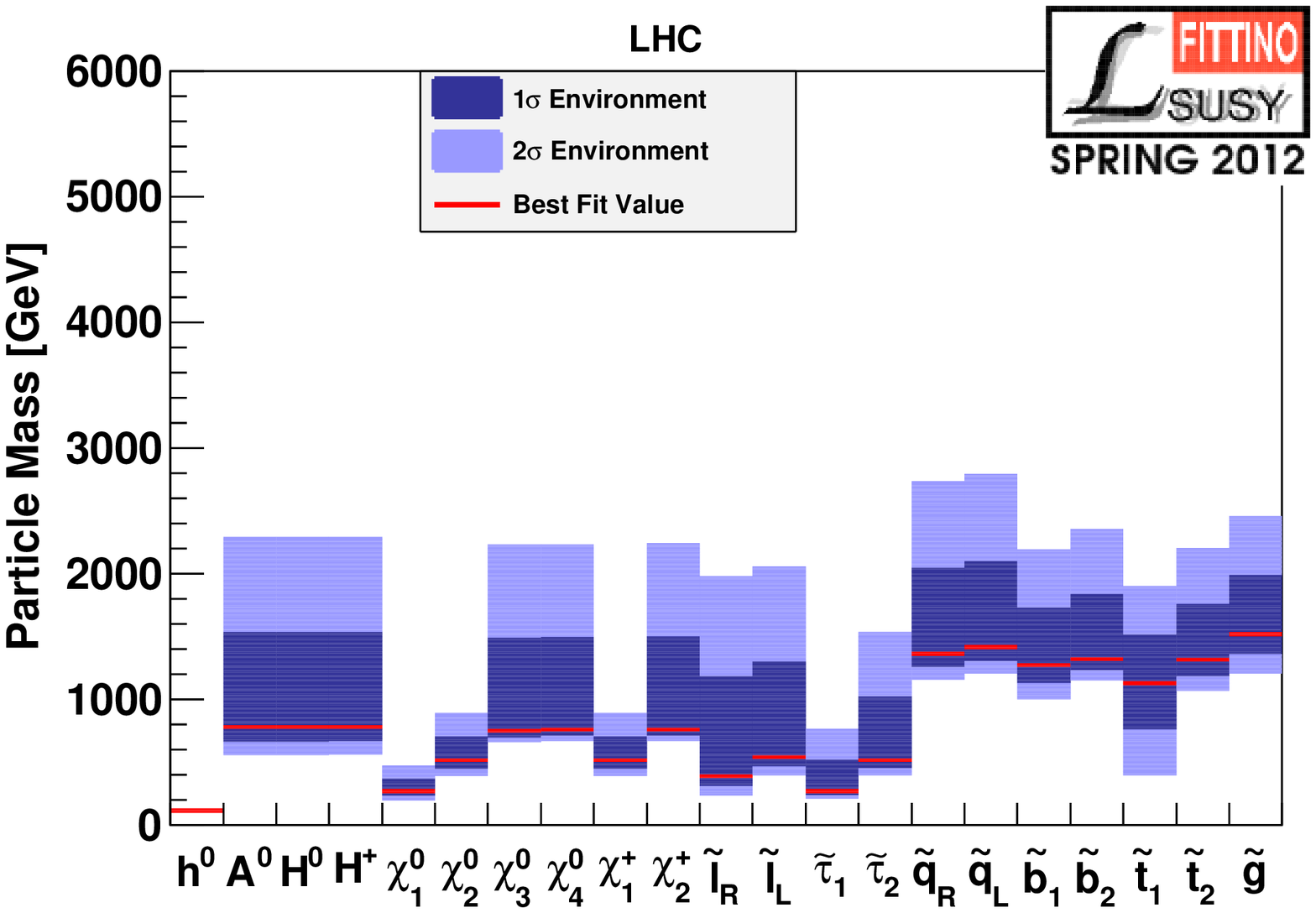} 
      \label{fig:mdp_lhc_XENON1T_02fbXn}
    }
    \subfigure[]{
      \includegraphics[width=0.47\textwidth,clip=]{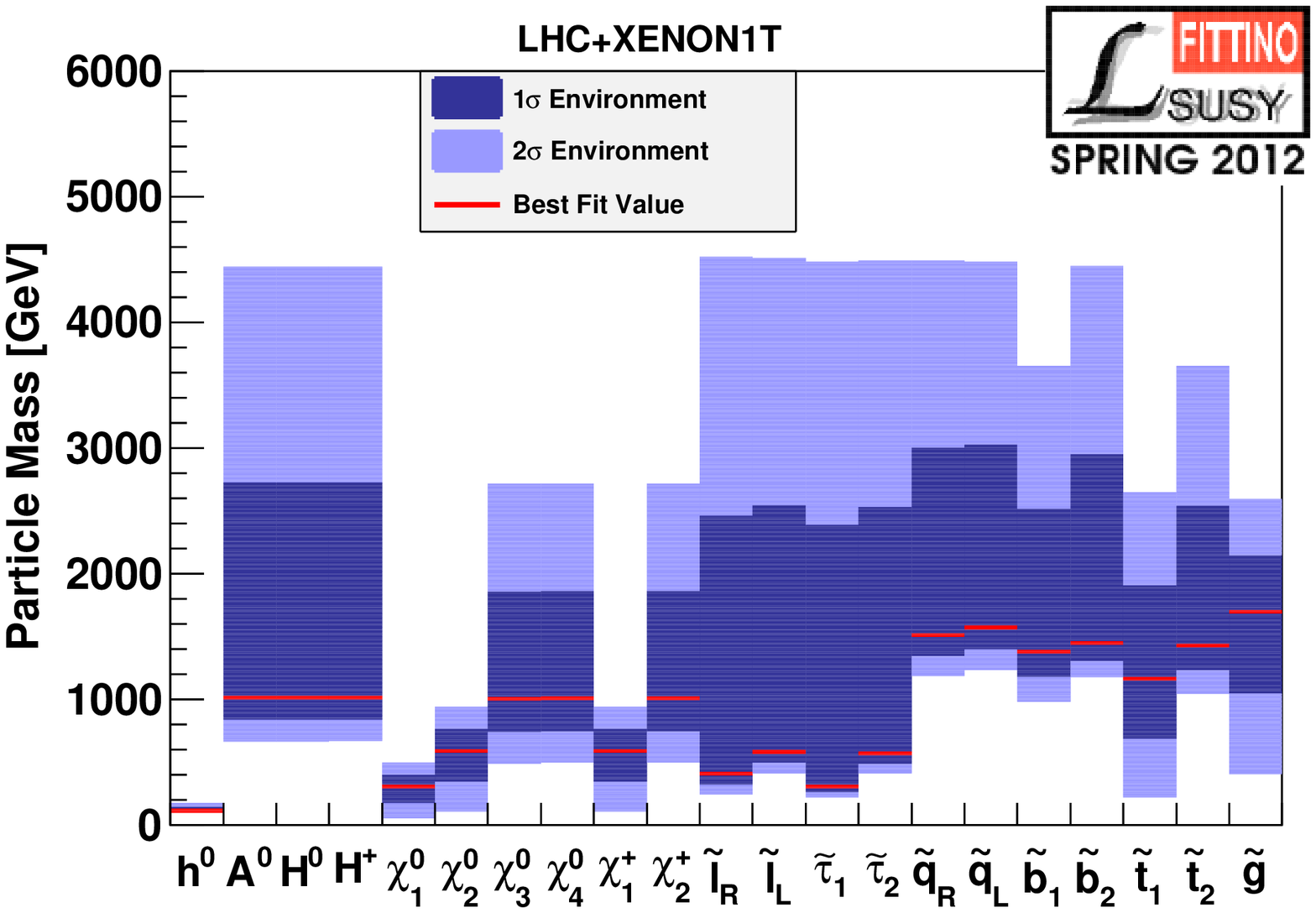} 
      \label{fig:mdp_lhc_XENON1T_02fbXn1T}
    }
  \end{center}
  \caption{Predicted distribution of sparticle and Higgs boson masses
    from \subref{fig:mdp_lhc_XENON1T_02fbXn} the LHC and
    \subref{fig:mdp_lhc_XENON1T_02fbXn1T} the LHC+XENON1T fit. The
    full uncertainty band gives the 1-dimensional 2$\sigma$
    uncertainty of each mass defined by the region $\Delta\chi^2<4$
    after profiling over all hidden dimensions. Note the different
    scales on the ordinate (mass) axis compared to
    Fig.~\ref{fig:mdp_lhc}.}  \label{fig:mdp_XENON1T}
\end{figure}

Upper limits from the Xenon100 \cite{Aprile:2011hi} experiment
presently do not constrain the CMSSM parameters beyond the LEO, LHC (5
fb$^{-1}$) data and the relic density of CDM alone. Concerning future
prospects \cite{Baudis:2012bc} for direct detection, we find that
stronger constraints on the spin-independent scattering cross-section
$\sigma_{\rm SI}$ per nucleon that would result from an increased
runtime (as planned with XENON100Goal) and a non-observation do not have
a large impact, either.  Increasing the target mass (XENON1T) and a
non-observation, on the other hand, would enlarge the allowed 68\% CL
region in both the ($M_0,M_{1/2}$) and the ($A_0,\tan\beta$) plane (see Fig.~\ref{fig:af_CMSSMparameters}) and move the best fit point to
somewhat higher scales.  At 95\% CL, the contours comprise only
slightly larger regions -- with the largest change visible in the
focus point region at $M_0\gtrsim4.5\,$TeV.  The overall fit quality,
however, becomes considerably worse in this case and the minimal
$\chi^2/ndf$ increases from 13.1/9 to 15.0/9, as already stated in
Table~\ref{tab:fitsummary}. In fact, Fig.~\ref{fig:mdp_XENON1T} shows
that implementing the projected XENON1T limits has the effect of
considerably raising the best-fit mass of essentially all particles
with respect to the current baseline expectation, in particular for
the heavy Higgs bosons and squarks.

\begin{figure}[t]
 \begin{center}
    \subfigure[]{
      \includegraphics[width=0.47\textwidth,clip=]{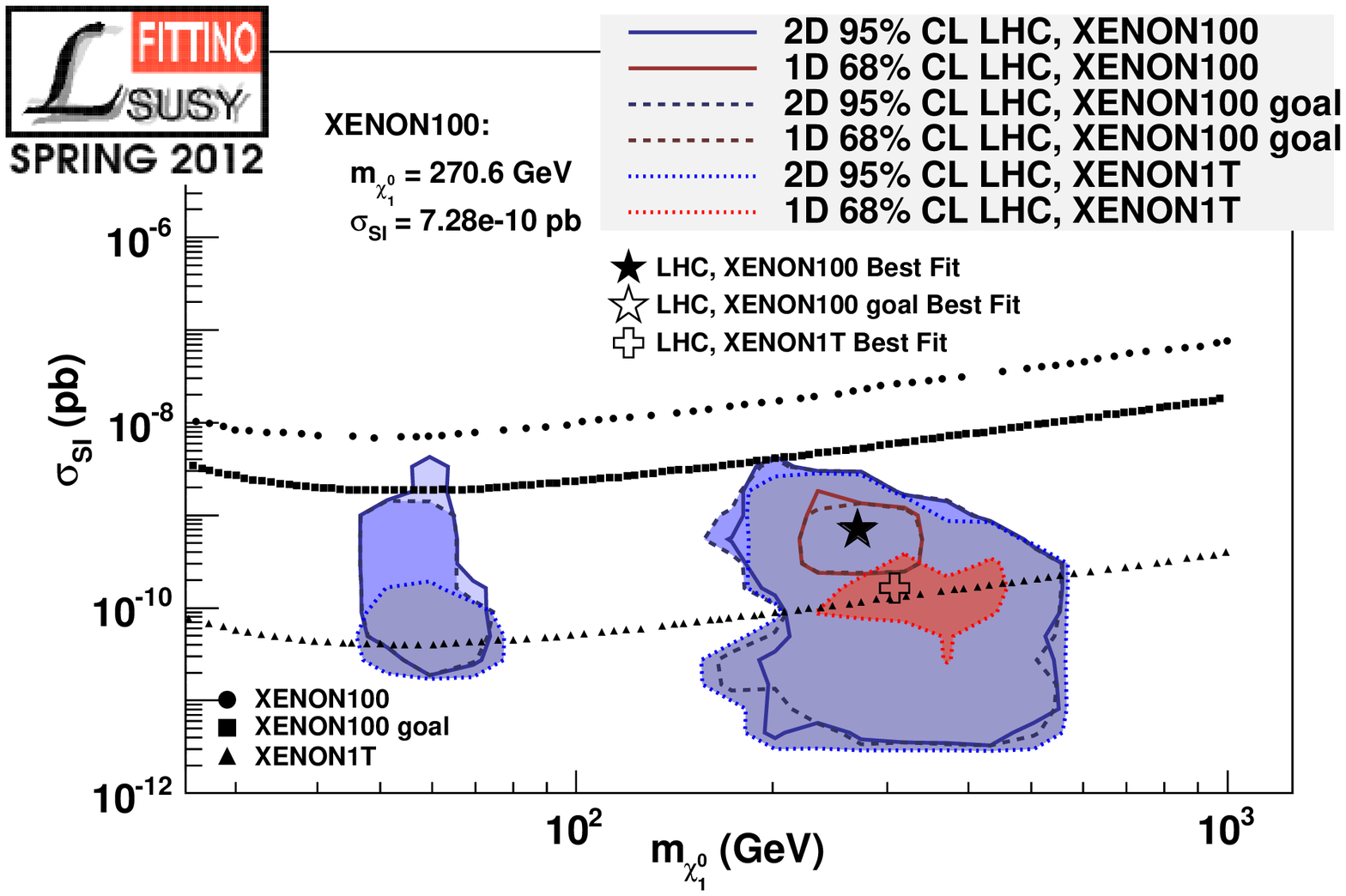} 
      \label{fig:afOverlay_XENON}
    }
    \subfigure[]{
      \includegraphics[width=0.47\textwidth,clip=]{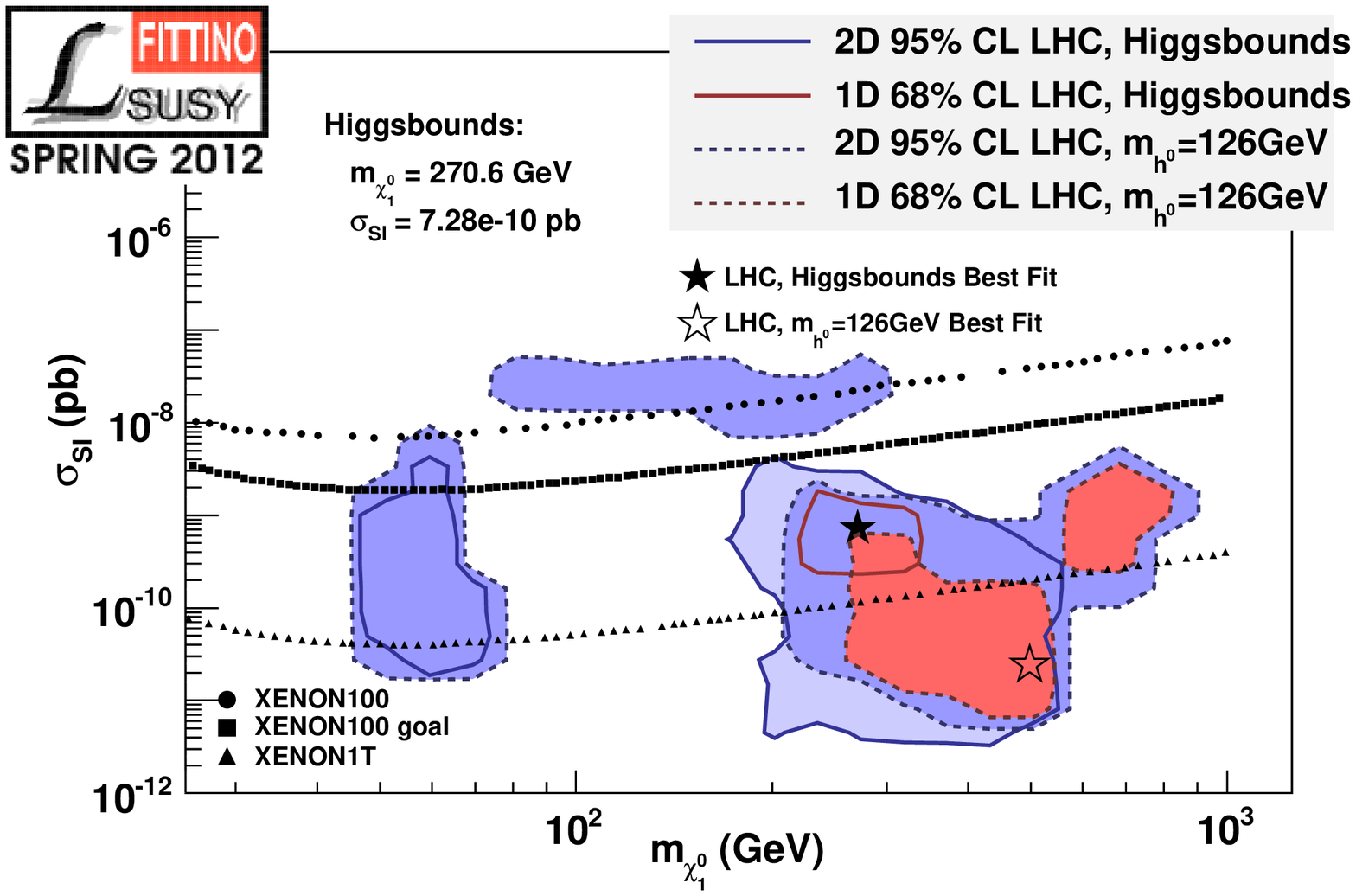} 
      \label{fig:afOverlay_Higgs}
    }
  \end{center}
  \caption{Current and future limits from the Xenon experiment, when
    \subref{fig:afOverlay_XENON} using \texttt{HiggsBounds} or
    \subref{fig:afOverlay_Higgs} assuming a Higgs boson mass of $m_{h}
    = [126\pm2\pm3]\,$GeV. In \subref{fig:afOverlay_XENON}, the minima
    for the first two fits are identical. See the text for further
    comments.  }
  \label{fig:afOverlay}
\end{figure}

We also considered the combined impact of direct detection limits and
a Higgs boson mass of $m_{h} \approx 126\,$GeV. In
Fig.~\ref{fig:afOverlay_XENON}, we only use limits from \texttt{HiggsBounds} to
constrain the Higgs boson mass: in an overlay plot, we show how constraints
from direct detection experiments will hardly improve with increased
runtime (XENON100Goal) but would do so with an increased target mass
(XENON1T) -- however at the cost of an increased $\chi^2_{\rm min}$,
see the comment above. Note that our XENON1T allowed region extends
into the regions nominally excluded by the experiment. The reason for
this is that we adopted the rather conservative choice of assigning a
theoretical uncertainty of $50\%$ to the calculation of the
spin-independent scattering cross-section per nucleon $\sigma_{\rm
  SI}$, see Section \ref{sec:directDM}. Restricting the Higgs boson mass to
$m_{h} = [126\pm2\pm3]\,$GeV instead, we see in
Fig.~\ref{fig:afOverlay_Higgs} that the preferred neutralino mass
moves from $270\,$GeV to $497\,$GeV; this trend to higher masses is of
course expected because a large Higgs boson mass generally requires rather
high values for the SUSY breaking scale (at least in the minimal SUSY
version considered here).  Concerning future prospects for direct
detection, we can see that a large Higgs boson mass worsens the situation as
it pushes the best-fit $\sigma_{\rm SI}$ to lower values.

\begin{figure}[t]
  \begin{center}
    \subfigure[]{
      \includegraphics[width=0.47\textwidth,clip=]{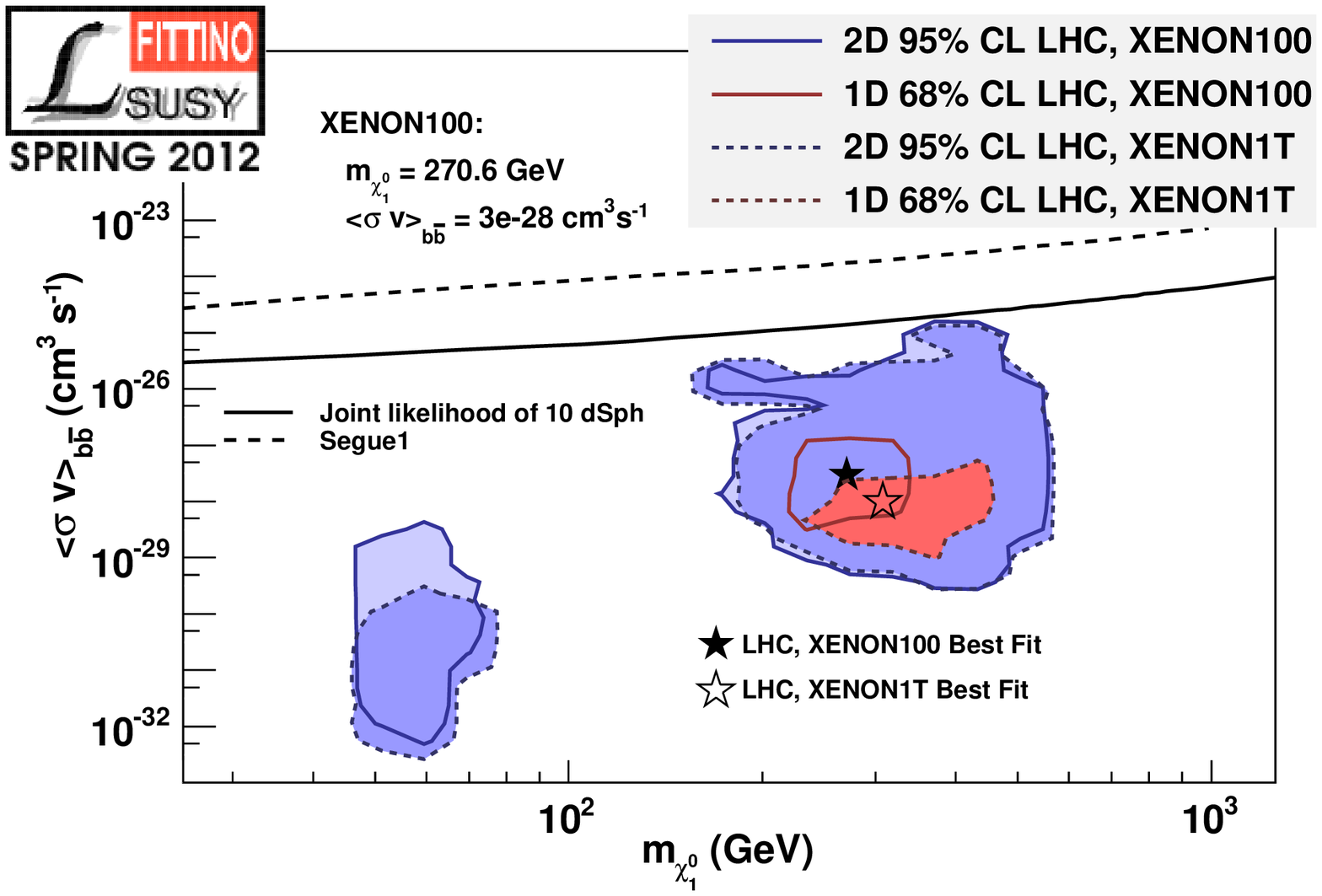}
      \label{fig:indirectDM}
    }
    \subfigure[]{
      \includegraphics[width=0.47\textwidth,clip=]{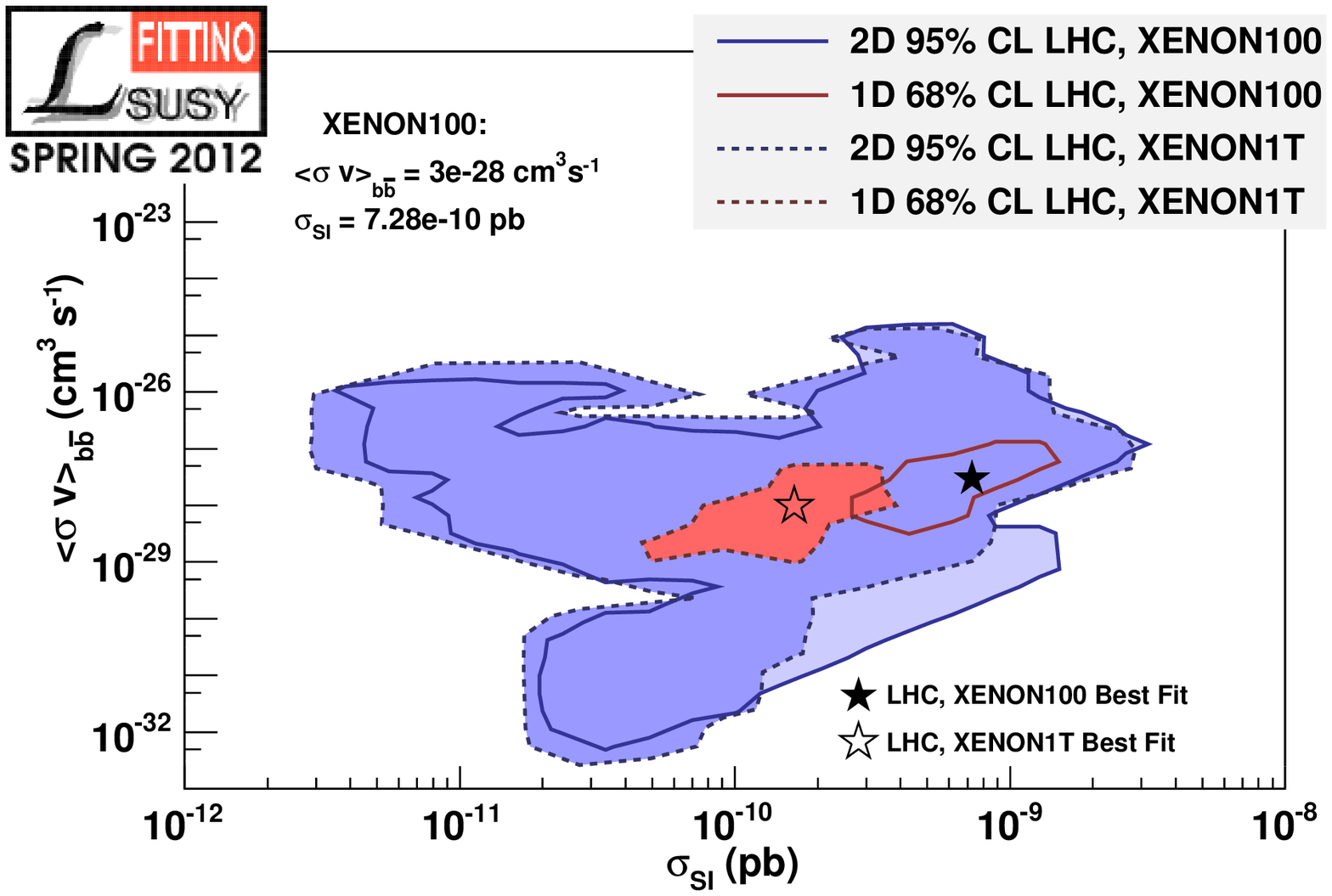} 
      \label{fig:compDM}
    }
      \end{center}
\caption{1$\sigma$ and 2$\sigma$ contours in the $\langle\sigma v\rangle_{\chi\chi\rightarrow\bar b b}$ vs.~$m_{\tilde \chi_1}$ plane (a), relevant for \textit{indirect} dark matter detection, slightly change when applying current and projected limits from \textit{direct} searches for dark matter. Also shown are the gamma-ray limits which we adopted here \cite{Abdo:2010ex} (dashed horizontal line) as well as the currently most stringent limits \cite{Ackermann:2011wa} (solid horizontal line) that will be used in an update of this study. In (b) the annihilation cross-section vs.~the spin-independent scattering cross-section is shown in order to demonstrate the complementarity \cite{Bergstrom:2010gh} of direct and indirect dark matter searches. }
\end{figure}

We find that our \textit{indirect detection} upper limits from dwarf
spheroidal galaxies, using gamma-ray observations by the Fermi
satellite, are still too weak to give a noticeable \(\Delta
\chi^2\)-contribution for neutralino DM in the CMSSM. This is not a
great surprise as the limits barely touch the annihilation
cross-section of $\sim3\times10^{-26}{\rm cm}^3/{\rm s}$, which is
naively expected for thermally produced DM. Concretely, we used the
photon flux upper limits from Ref.~\cite{Abdo:2010ex}, for
$E_\gamma>100\,$MeV, on neutralino pair annihilation into $\bar b b$
final states -- which very often gives the dominant contribution to
the total flux (mostly from photons with $E_\gamma\ll m_\chi$). An
improved treatment would also take into account the photons from other
final states. However we caution that this is not straightforward to
implement in those regions of the parameter space where the photon
spectra are very model-dependent \cite{Bringmann:2007nk}.  Further
improvement is possible by using updated limits from a combined
(`stacked') analysis of {\it all} dwarf spheroidal galaxy data taken
by Fermi \cite{Ackermann:2011wa}. We thus expect that a more accurate
treatment of the combined gamma-ray limits in \texttt{AstroFit}, which
is planned for future fits, would actually impact the CMSSM parameter
space (as also found in \textit{e.g.}~Ref.~\cite{Scott:2009jn}). This
expectation is reflected in Fig.~\ref{fig:indirectDM}, where we show
$\langle\sigma v\rangle_{\chi\chi\rightarrow\bar b b}$ vs.~$m_{\tilde
  \chi_1}$: While the limits that we have implemented indeed do not
touch the $2\sigma$ regions, the improved limits from the joined dwarf
spheroidal galaxy analysis \cite{Ackermann:2011wa} do. Those limits
were not available in \texttt{AstroFit} when the scans were set
up. Let us also stress that we plot here only the annihilation
cross-section into $\bar b b$ final states. Future prospects for
indirect dark matter detection are thus actually much better than what
is naively inferred from this figure -- especially when explicitly
taking into account gamma-ray spectral features in the analysis rather
than only counting the number of photons \cite{Bringmann:2012vr}.
 
In Fig.~\ref{fig:compDM}, we plot the neutralino annihilation
cross-section against the spin-in\-de\-pen\-dent scattering cross-section,
demonstrating that indirect and direct dark matter searches indeed
probe the parameter space from an orthogonal direction
\cite{Bergstrom:2010gh} and are highly complementary even for very
constrained scenarios like the CMSSM. In particular, improving current
gamma-ray limits by about one order of magnitude (as might be rather
straight-forward with future air \^Cerenkov telescopes
\cite{Bergstrom:2010gh}) would allow to probe models that are
completely out of reach even for XENON1T. Models in the upper right
corner of Fig.~\ref{fig:compDM}, on the other hand, would in principle
allow for a future {\it simultaneous} detection of dark matter with
both direct and indirect methods which evidently would make any claim
for a corresponding signal much more convincing. We checked that
adding the Higgs-mass constraint $m_{h} = [126\pm2\pm3]\,$GeV does not
have a major impact on the 2$\sigma$ region in this plane. The
1$\sigma$ region, on the other, hand blows up considerably.  The best
fit point moves to $\sigma_{\rm SI}\sim10^{-11}$pb and $\langle\sigma
v\rangle_{b\bar b}\sim10^{-29}$cm$^3$s$^{-1}$. This again just
reflects the overall worse quality of the fit.

\begin{figure}[t]
  \begin{center}
    \subfigure[]{
      \includegraphics[width=0.47\textwidth,clip=]{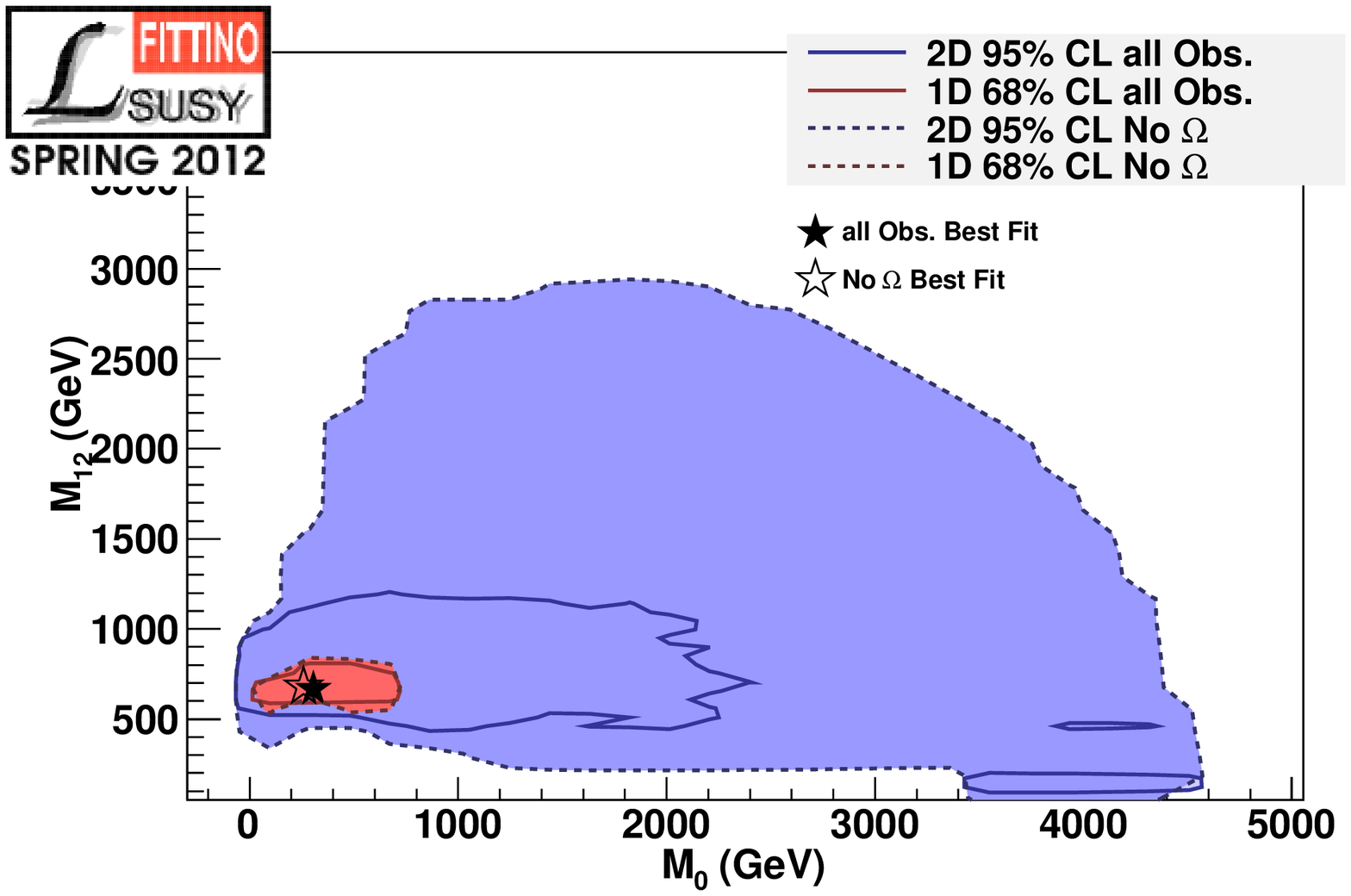}
      \label{fig:M0M12_NoOmega}
    }
    \subfigure[]{
      \includegraphics[width=0.47\textwidth,clip=]{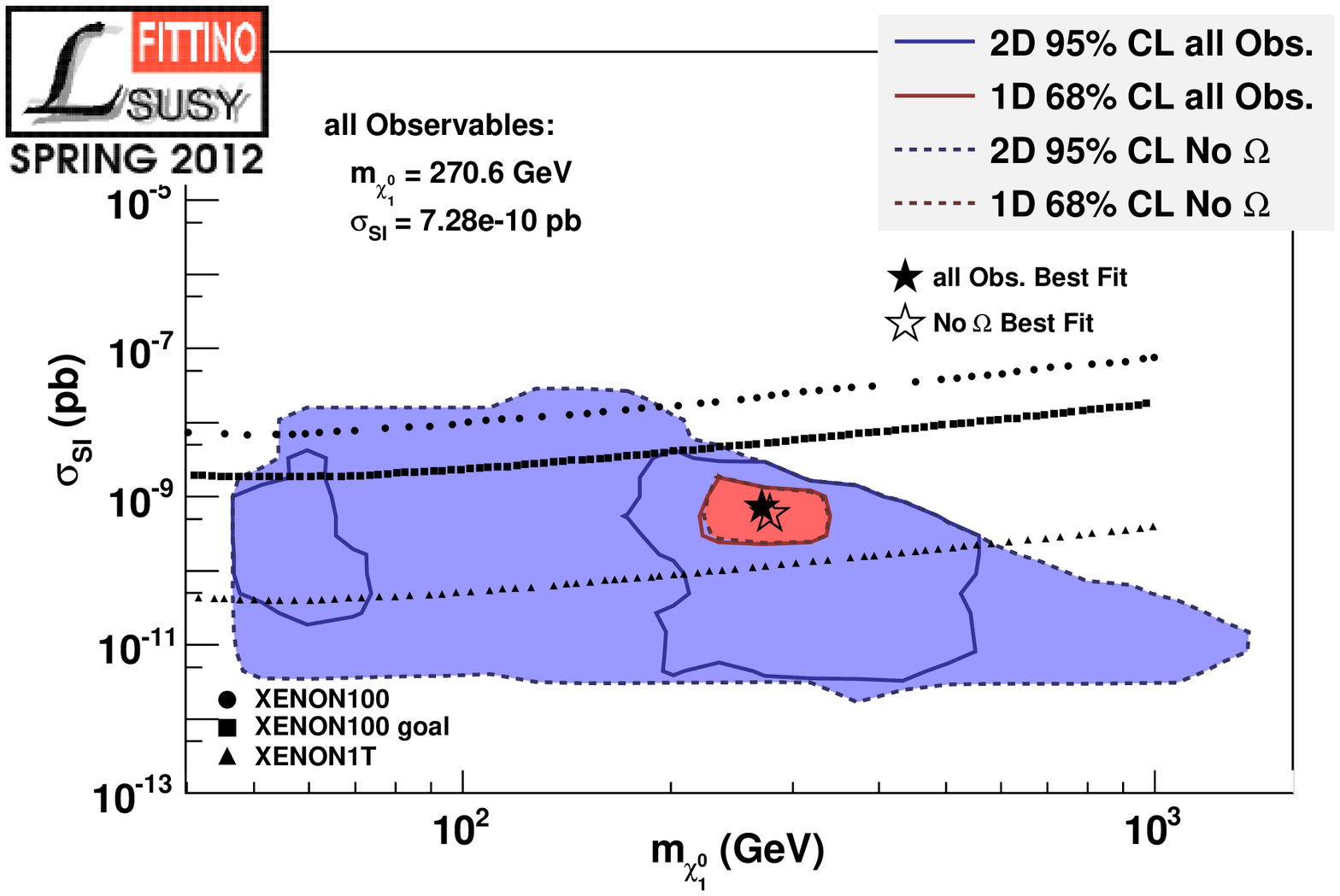} 
      \label{fig:afOverlay_NoOmega}
    }
      \end{center}
      \caption{Solid lines show the standard LHC fit, dashed lines the
        impact of the relic density constraint on
        \subref{fig:M0M12_NoOmega} the $(M_0,M_{1/2})$ plane and
        \subref{fig:afOverlay_NoOmega} $\sigma_{\rm SI}$
        vs. $m_{\tilde \chi^0_1}$. In \subref{fig:afOverlay_NoOmega}
        also the current (Xenon100) as well as near (XENON100Goal) and
        far (XENON1T) future reaches of direct detection experiments
        are shown (from top to bottom).}
  \label{fig:NoOmega}
\end{figure}

The relic density of cold dark matter remains a strong
constraint on the fit.  Indeed, it is well known that only relatively
small regions in the full parameter space of the CMSSM can account for
thermally produced neutralino dark matter with the correct relic
density: (i) the bulk region at low values of $M_0$ and $M_{1/2}$,
which is 
now essentially excluded by collider data, (ii) the funnel region
at intermediate values of $M_0$ and $M_{1/2}$ with $m_A \simeq
2m_\chi$ so that annihilation via the pseudo-scalar Higgs boson
becomes resonant, (iii) the focus point region at $M_0\gg M_{1/2}$
where the lightest neutralino has a sizable Higgsino component and
(iv)
 the stau 
co-annihilation region at large $M_{1/2}$
but small $M_0$ with $m_\chi\simeq m_{\tilde\tau}$.
Moreover, for $A_0\gg0$ there is the stop co-annihilation region
 at somewhat larger
values of $M_0$ and correspondingly very large values of
$M_{1/2}$. The impact of the relic density constraint can best be
demonstrated in the $(M_0,M_{1/2})$ plane, as shown in
Fig.˜\ref{fig:M0M12_NoOmega}. Note that for fixed values of $A_0$ and
$\tan\beta$ the allowed regions would only appear as thin strips in
this plane. While we find that the relic density calculated by 
{\tt DarkSUSY}~\cite{darksusy} and {\tt micrOMEGAs}~\cite{Belanger:2008sj}
can differ up to a factor two for some parameter combinations, 
this does not
have a significant impact on our best fit regions and 2D contour plots.
The most striking difference appears in Fig.˜\ref{fig:M0M12_NoOmega} 
where the co-annihilation region does not extend all the way out to 
$M_0\sim2.2\,$TeV but only to $M_0\sim1.8\,$TeV when using 
{\tt DarkSUSY} instead of {\tt micrOMEGAs}. 
 Figure~\ref{fig:afOverlay_NoOmega} shows that
the relic density  together with the particle physics input
favors neutralino masses in the range between 200~GeV and 500~GeV.
At 95\% CL an additional range between 50 and 70~GeV is allowed. This corresponds
to a resonance in the focus point region, see below.
However, the expected spin-independent scattering cross-section is hardly 
affected by the relic density constraint.
 
Note that the $2\sigma$ region with $m_{\chi^0_1}\sim50-70\,$GeV that
appears in several of the 2D plots discussed above does not show up in
the mass distribution plots (Fig.~\ref{fig:mdp_XENON1T}) because it is
not contained in the 1-dimensional $2\sigma$ environment.  These
neutralinos lie in the focus point region appearing in
Fig.~\ref{fig:M0M12_NoOmega} at very low values of $M_{1/2}$ and
$M_0\sim4\,$TeV. They have a Higgsino fraction of roughly 1\%. Their
relic density is almost exclusively set through $s$-channel
annihilation via the light scalar Higgs $h$.

Moving along the focus point `strip' to larger masses, the Higgsino
fraction increases and at $M_{1/2}\gtrsim2.5\,$TeV the neutralino
becomes an essentially pure Higgsino that acquires the correct relic
density for $m_{\chi^0_1}\sim1.1\,$TeV. 
In our fit, however, models with such large neutralino masses do not
appear because the sampling algorithm runs out of statistics at these
very large values of $M_0>4.5$\,TeV due to the excessive
inter-parameter fine-tuning (discussed in
Section~\ref{sec_results_newft}) for such high $M_0$\footnote{ See,
  however, Ref.~\cite{Baer:2005ky} for a detailed discussion of the
  focus point region arguing that, even for very large values of
  $M_0$, the actual fine-tuning is small if expressed in {\it
    physical} parameters.}. For the same reason, we do not sample the
tail of the co-annihilation strip which in principle admits TeV-scale
neutralinos co-annihilating with almost degenerate staus.  We mention
that the very small region at $M_0\sim4\,$TeV and slightly higher
values of $M_{1/2}$ also belongs to the focus point region. Albeit in
this case it contains models where the lightest neutralino has a
sizable Higgsino fraction of $\mathcal{O}(10\%)$ and a mass around
$\sim\,$200\,GeV. The relic density is thus mostly set through
(non-resonant) annihilation into $SU(2)$ gauge bosons, which is helped
by the presence of relatively light charginos and neutralinos, though
not light enough that co-annihilations would be important.

\subsection{Fine-tuning}
\label{sec_results_ft}

A major motivation for supersymmetry at or near the electroweak scale
is to solve the hierarchy problem
\cite{Gildener:1976ai,Veltman:1980mj}.  This is a fine-tuning problem
in the Higgs sector of the theory, where the low-energy observable
Higgs boson mass depends sensitively on the input parameters at a
postulated new scale of physics, such as the unification scale
$M_{\mathrm{GUT}}$, or the Planck scale. This fine-tuning is removed
and the hierarchy problem solved for unbroken supersymmetry. A
solution is retained after supersymmetry breaking, if the
supersymmetry masses are not too heavy. How heavy depends on how much
fine-tuning is perceived to be acceptable. This is the origin of the
expectation, that the supersymmetric masses should be
$M_{\mathrm{SUSY}}<\mathcal{O} (10\,\mathrm{TeV})$, or even less.

Beyond the Higgs boson mass, also other low-energy parameters can be
fine-tuned, such as the electroweak breaking scale, represented by the
$Z^0$ boson mass, $m_Z$. The fine-tuning can be quantified by various
measures. A popular one was introduced in \cite{Barbieri:1987fn}. It
is based on the logarithmic derivatives of any given observables with
respect to the parameters considered.  For $m_Z$ considered as a
function of variables $a_i$ the relevant numbers are
\begin{equation}
c(a_i)\equiv
\left|\frac{a_i}{m_Z^2}\frac{\partial m_Z^2(a_i)}{\partial a_i}\right|\,.
\label{eq:convFineTuningDefinition}
\end{equation}
The overall fine-tuning at a given parameter point is then given by
\begin{eqnarray}
\Delta \equiv \textrm{max}_i \left[ c(a_i) \right].
\label{eq:convDeltaParameter}
\end{eqnarray}
It has since been widely used to quantify the fine-tuning in the
(C)MSSM both before 
\cite{Chankowski:1997zh,Barbieri:1998uv,Chankowski:1998xv,Batra:2003nj,Giudice:2006sn,Cassel:2009ps} 
and more recently also including LHC data 
\cite{Cassel:2011tg,Strumia:2011dy,Ross:2011xv}.


For the CMSSM, \texttt{SOFTSUSY} \cite{Allanach:2001kg} can be
configured to calculate all of the derivatives in
Eq.~(\ref{eq:convFineTuningDefinition}). The set of parameters $\{a_i
\}$ consists of $M_0$, $M_{1/2}$, $A_{0}$, $m_3$, $\mu$, and $h_{t}
$.\footnote{\texttt{SOFTSUSY} also allows the computation of the
  derivatives with respect to the non-CMSSM parameters $m_3$, the soft
  breaking gluino mass, and $h_t$, the top Yukawa coupling.} Points
with larger values for $\Delta$ are more fine-tuned and are in general
regarded as less natural. A fit has been performed to assess the
fine-tuning, using the same observables and LHC limit as for the LHC
fit before, but with {\tt SoftSUSY} as spectrum calculator instead of
{\tt SPheno}.

\begin{figure}[t]
\begin{center}
  \subfigure[]{
    \includegraphics[width=0.47\textwidth,clip=]{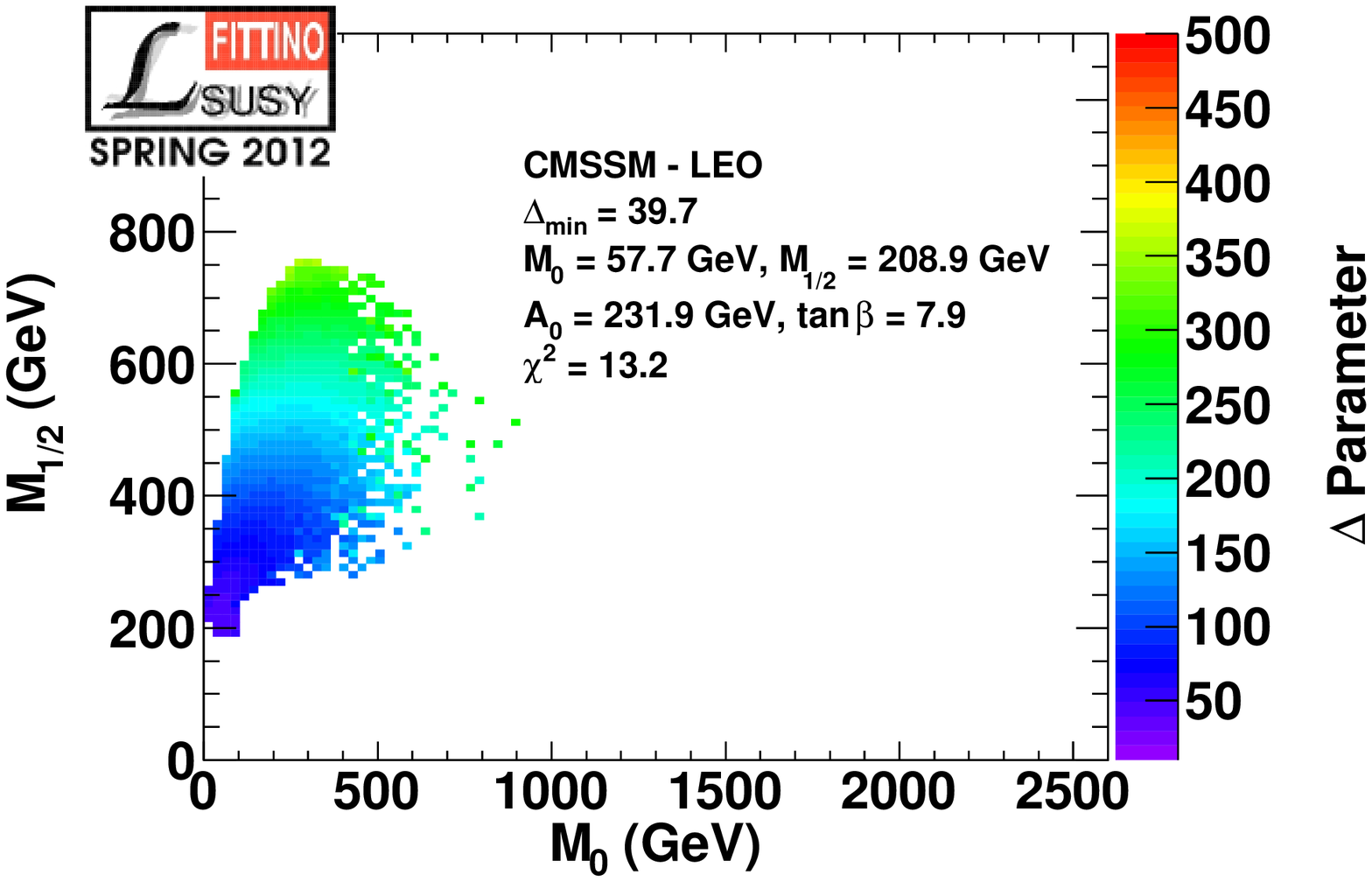}
    \label{fig:TraditionalFTLHC_FT_0fbm0vsm12}
  }
  \subfigure[]{
    \includegraphics[width=0.47\textwidth,clip=]{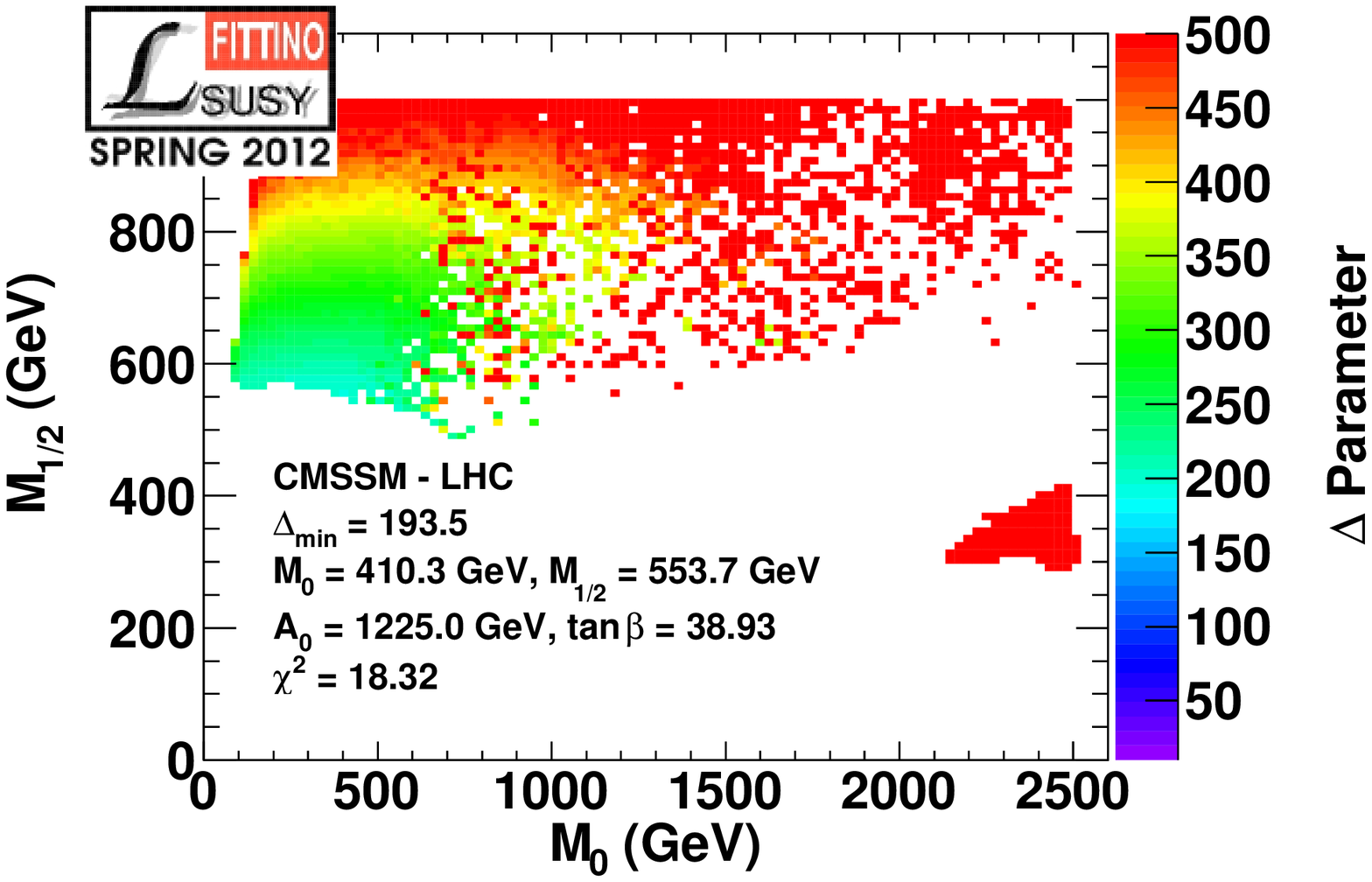}
    \label{fig:TraditionalFTLHC_FT_2fbm0vsm12}
  }
  \subfigure[]{
    \includegraphics[width=0.47\textwidth,clip=]{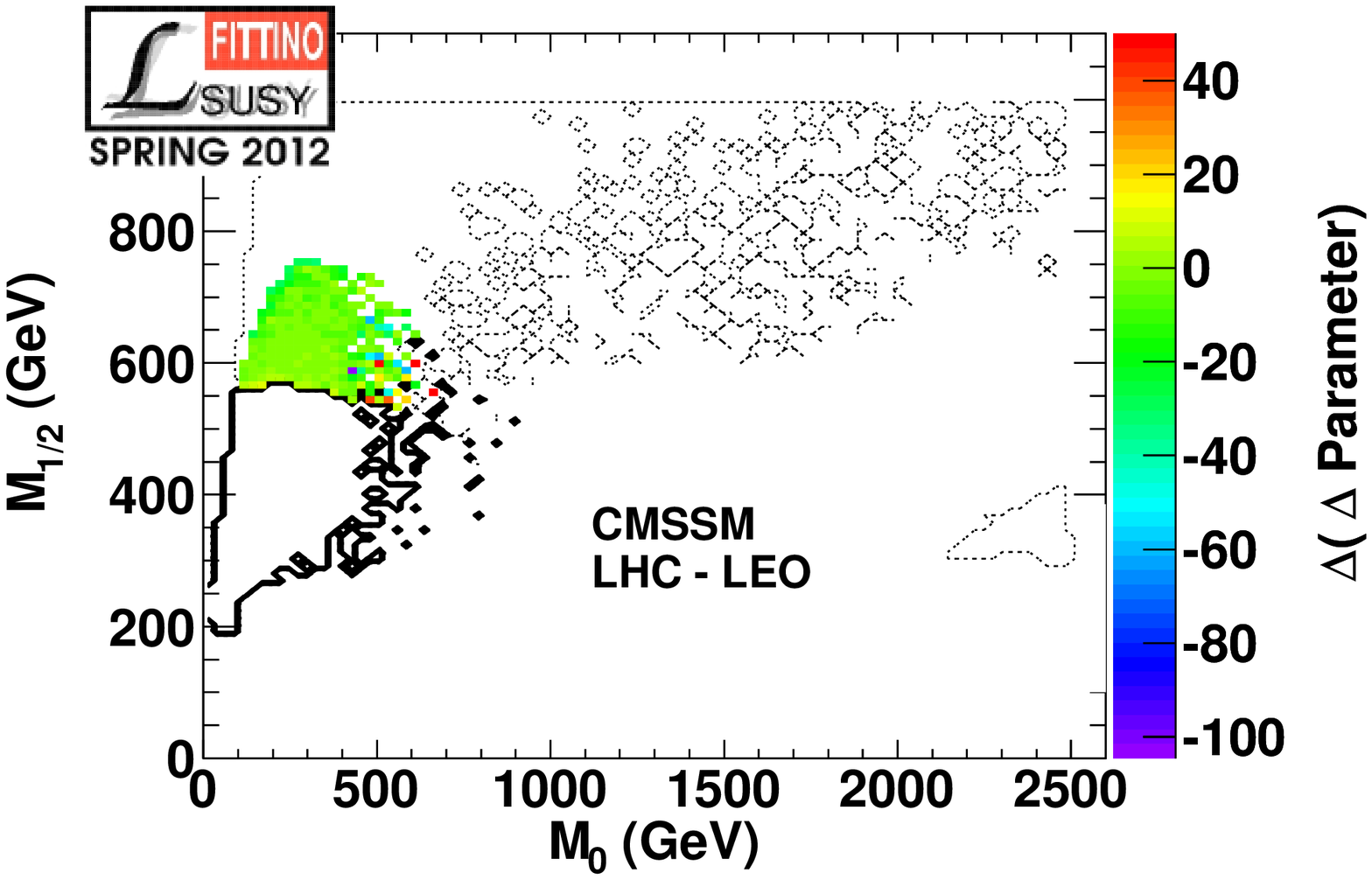}
    \label{fig:TraditionalFTLHC_FT_diffm0vsm12}
  }
%
  \end{center}
  \caption{The minimal amount of fine-tuning as a function of $M_{0}$ 
    and $M_{1/2}$, where $A_{0}$ and $\tan \beta$ are profiled, for 
    (a) the pre-LHC fit and (b) the LHC fit. (c) shows the difference,
    $\mathrm{(b)}\!-\!\mathrm{(a)}$, between both fits, which is mostly
    close to vanishing in the overlap of the $95\%$-CL regions.  The
    last bin on the z-axis in (a), (b), and (c) contains all points
    with $\Delta>500$. The smallest values for the $\Delta$
    parameter are $39.7$ ($193.5$) for the pre-LHC (LHC) fit.
    }\label{fig:TraditionalFTLHC}
\end{figure}

As we have seen, the LHC exclusion leads to an overall decrease of the
fit quality. Despite having a more constrained system, this worse fit
can lead to more points in the 2$\sigma$ allowed range around the
best-fit point. In particular, for a fixed value in the
$M_0$-$M_{1/2}$ plane a larger range of $A_0$ and $\tan\beta$
values can be allowed. In Fig.~\ref{fig:TraditionalFTLHC} we plot the
lowest values of $\Delta$ in the $M_0$-$M_{1/2}$ plane, where the
parameters $A_0$ and $\tan\beta$ have been profiled.
In Fig.~\ref{fig:TraditionalFTLHC_FT_0fbm0vsm12}, $\Delta$ is shown
for the pre-LHC fit.  The lowest amounts of fine-tuning ($\geq39.7$)
are observed for low values of $M_{0}$ and $M_{1/2}$, while in
particular for high values of $M_{1/2}$ the $\Delta$-parameter reaches
larger values, over 300.\footnote{We also considered the fine-tuning
as a function of $M_{0}$ and $M_{1/2}$ for the \textit{best-fitting}
values of $A_{0}$ and $\tan\beta$ in each bin. The results are
indistinguishable from Fig.~\ref{fig:TraditionalFTLHC_FT_0fbm0vsm12}.}

Fig.~\ref{fig:TraditionalFTLHC_FT_2fbm0vsm12} shows the fine-tuning
taking into account the LHC exclusion with a dataset of $5\,\textrm
{fb}^{-1}$. This fit introduces points with a significantly larger
fine-tuning ($\Delta>500$) compared to the pre-LHC fit and the
point of lowest fine-tuning ($\Delta=193.5$), is significantly more
fine-tuned than in the pre-LHC fit.
This is because the fit pushes one to higher values of
$M_0,\,M_{1/2}$. Again overall in the LHC fit, low values of $M_{0}$
and $M_{1/2}$ allow for a lower fine-tuning, while for larger values
for both parameters, in particular in the focus point region, a higher
fine-tuning is observed (see also discussion in
Section~\ref{sec:results_af}).

Fig.~\ref{fig:TraditionalFTLHC_FT_diffm0vsm12} shows the difference,
$\mathrm{(b)}\!-\!\mathrm{(a)}$, between the two fits in the region
that is included in the $2\sigma$-contours of both the pre-LHC as well
as the LHC fit. The solid/dashed lines indicate the
part of the parameter space which is not excluded at the $95\%$-CL in
the pre-LHC/LHC fit, but which is excluded at the
$95\%$-CL in the LHC/pre-LHC fit. In the
$(M_{0},M_{1/2})$ plane, only a few points with a significant change
in the fine-tuning show up, while for the majority of points which are
not excluded in both fits, the fine-tuning is comparable.

We find slightly larger minimal values of $\Delta$ compared to other
studies, for instance \cite{Cassel:2011tg}. However, the values we
find are not dramatically larger and our fits are constrained by a
larger set of observables, such that deviations are expected. While 
we find that with earlier LHC limits (${\cal L}_{\rm int} \sim 1\textrm{fb}^{-1}$), 
using the criteria of \cite{Cassel:2011tg}, points with an acceptable 
amount of fine-tuning ($\Delta < 100$) remain in reasonable agreement 
with all precision measurements, the minimal value of $\Delta$ is 
pushed above this threshold by the up to date LHC results.


\subsection{Correlation of fundamental parameters}\label{sec_results_newft}

The fine tuning measure $\Delta$ described in the previous section has
many merits from a theoretical point of view. However, from an
experimental point of view, there is one major shortcoming: The
absolute change of an arbitrarily selected observable such as $m_Z$ is
experimentally of limited interest. The experimentally more meaningful
quantity is the change of the observable relative to its
uncertainty. Furthermore, experimentally we are interested in all
observables chosen to calculate the $\chi^2$, not only in one
particular quantity of special theoretical interest.

Therefore, we propose a different view on looking at the naturalness
of a model in the light of all observables and all
uncertainties, inspired by the fine tuning. There, if one prediction
for an observable depends strongly on varying one parameter, the
agreement of all measurements with the predictions must be restored by
varying other parameters in a correlated way. Therefore, we examine
the \textit{correlation} of the fundamental parameters in the fit as
an additional information about how highly the parameters have to be
tuned with respect to each other, measured in the form of the quantity
$\varrho_{\rm max}$ defined below, in order to fulfill the same level of
agreement with all data in the fit. Thus, all observables in the fit
contribute with their current experimental precision to $\varrho_{\rm max}$. 

A priori, the fundamental parameters of the tested models are
independent. If however, starting from a point in parameter space with
a quality of fit $\chi^2$, if one parameter is varied, the
experimental bounds may force a correlated change in one of the other
parameters, in order to maintain an equally good fit. A first order
measure for the dependence is Pearson's product-moment correlation
coefficient.  For each pair of parameters, $P_i,\,P_j\in\{
M_0,\,M_{1/2},\,A_0,\,\tan\beta\}$, it is defined by
\begin{eqnarray}
\varrho_{ij} \equiv \left\langle \frac{\left(P_i - \langle P_i\rangle \right)\cdot\left( P_j - \langle P_j\rangle\right)}{\sigma_{P_i}\sigma_{P_{j}}}\right\rangle\,.
\label{eq:newFTCorrCoeff}
\end{eqnarray}
$\langle P_i\rangle$ is the average value for the parameter $P_i$ and
$\sigma_{P_i}$ the standard deviation of the Parameter $P_i$. Therefore
$|\varrho_{ij}|\leq1$.  

The set of values of the parameters $\{P_i\}$ used for the computation
of the average and standard deviation at a fixed point \textbf{P} in the 
parameter space is defined by the value of $\chi^2$ at this point, 
$\chi^2_{\mathbf{P}}$. We consider all points in the MCMC scan with a quality of 
fit in the range $[\chi^2_{\mathbf{P}},\chi^2_{\mathbf{P}}+\Delta\chi^2]$, 
where $\Delta\chi^2=0.001$. These are very 
thin slices. All points included twice or more often in the MCMC scan are 
taken into account only once. If less than 6 points exist within the $\Delta\chi^2 <
0.001$ boundary the tested point is assigned no value and excluded
from the further analysis.

For each point in the $95\%$-CL region, we now compute the quantity
\begin{eqnarray}
\varrho_{\mathrm{max}} = \textrm{max}_{ij} \left( \left |\varrho_{ij} \right| \right ),   
\label{eq:newFTDeltaParameter}
\end{eqnarray}
considering all 6 combinations of $(P_i,P_j)$. For a fixed value of
$M_0$ and $M_{1/2}$, and with $A_0$, $\tan\beta$ profiled, we then
compute the minimum of $\varrho_{\mathrm{max}}$, $[\varrho_{\mathrm
{max}}]_{\mathrm{min}}$, over the $95\%$-CL region. Large values of
$[\varrho_{\mathrm{max}}]_{\mathrm{min}}$ mean the fit is highly 
correlated.

A model highly constrained by the observables with their given
precision will naturally yield $\varrho_{\rm max}\approx1$ near the point
with the exact $\chi^2$ minimum. Depending on the model, observables
and precision, it may however show high (\textit{e.g.} near 1) or low values of
$[\varrho_{\mathrm{max}}]_{\mathrm{min}}$ over the full allowed
$\chi^2$ range, depending on how much the parameters need to be
fine-tuned with respect to each other in order to achieve a given
agreement between the model predictions and all data, taking the
experimental and theoretical precision fully into account.

$[\varrho_{\rm max}]_{\mathrm{min}}$ can describe features of the analyzed
parameter space at very different scales. For low values of $\chi^2$
its value is dominantly built from the small scale structure of the
parameter space. For larger values of $\chi^2$, all parameters 
vary over a larger range and thus the included parameter space grows
significantly and correlations on a more global scale are reflected. 

In order to have a baseline comparison, we have first implemented this
procedure \textit{without} the experimental constraints considered
throughout this paper. Instead we have explored the correlation where
only a minimum set of requirements has been used to constrain the
parameter space: we require (a) the lightest neutralino to be the LSP,
(b) consistent radiative electroweak symmetry breaking and (c) the
absence of tachyons. In order to have a meaningful comparison, the
correlation between two parameters is not calculated over an infinite
parameter range, but instead in a rectangular subset of the parameter
space, the boundaries of which are defined by the maximum and minimum
values of each of the four parameters as found in the $95\%$-CL region
of the LHC fit of the same model. After performing this procedure we
find the maximum correlation between parameters with these minimal
constraints is found to be smaller than $0.2$. This is what the
following results should be compared against.

\begin{figure}[t]
  \begin{center}
  \subfigure[]{  
    \includegraphics[width=0.47\textwidth,clip=]{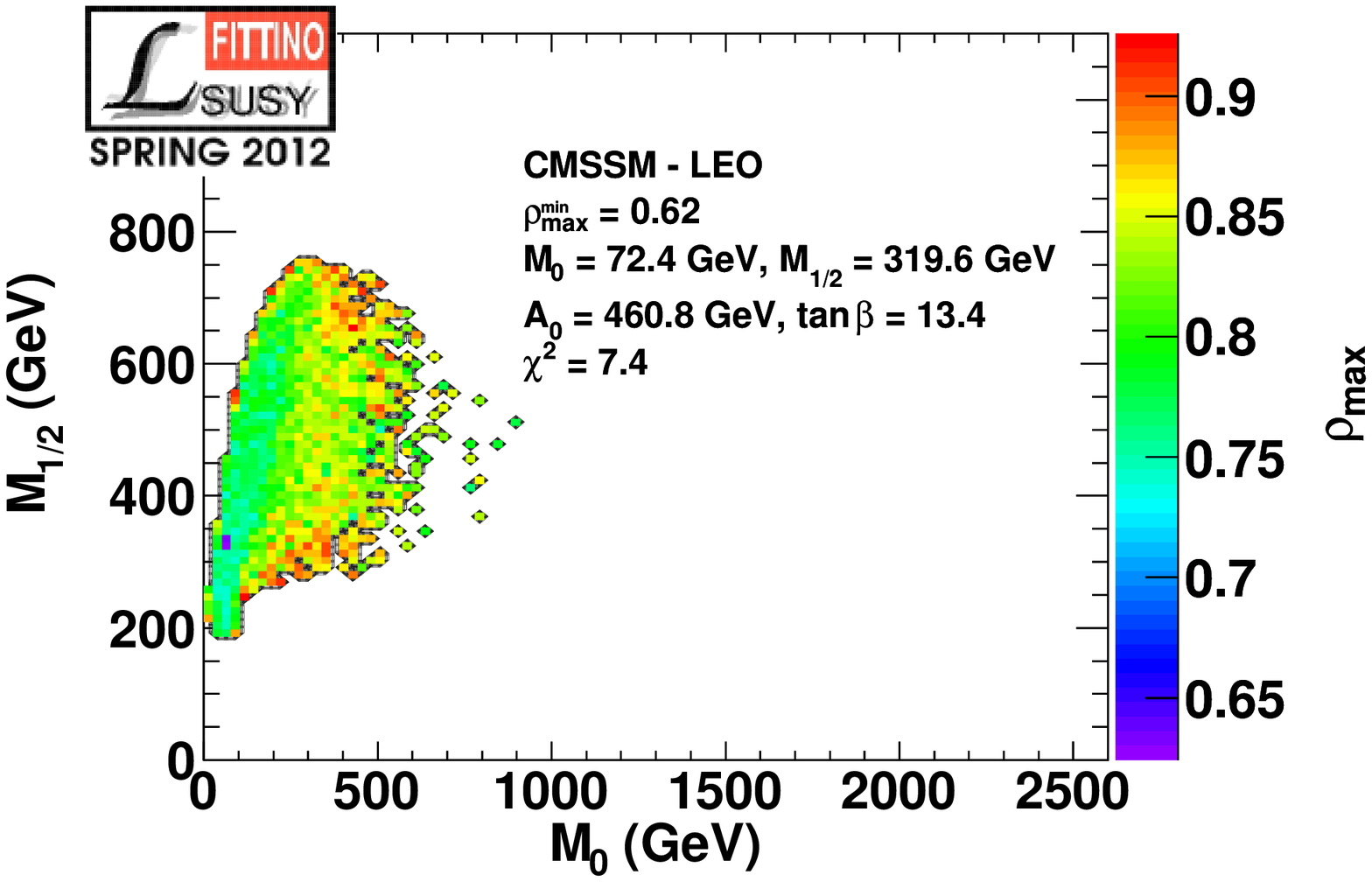}
    \label{fig:NewFTLHC_0fbm0vsm12}
  }
  \subfigure[]{  
    \includegraphics[width=0.47\textwidth,clip=]{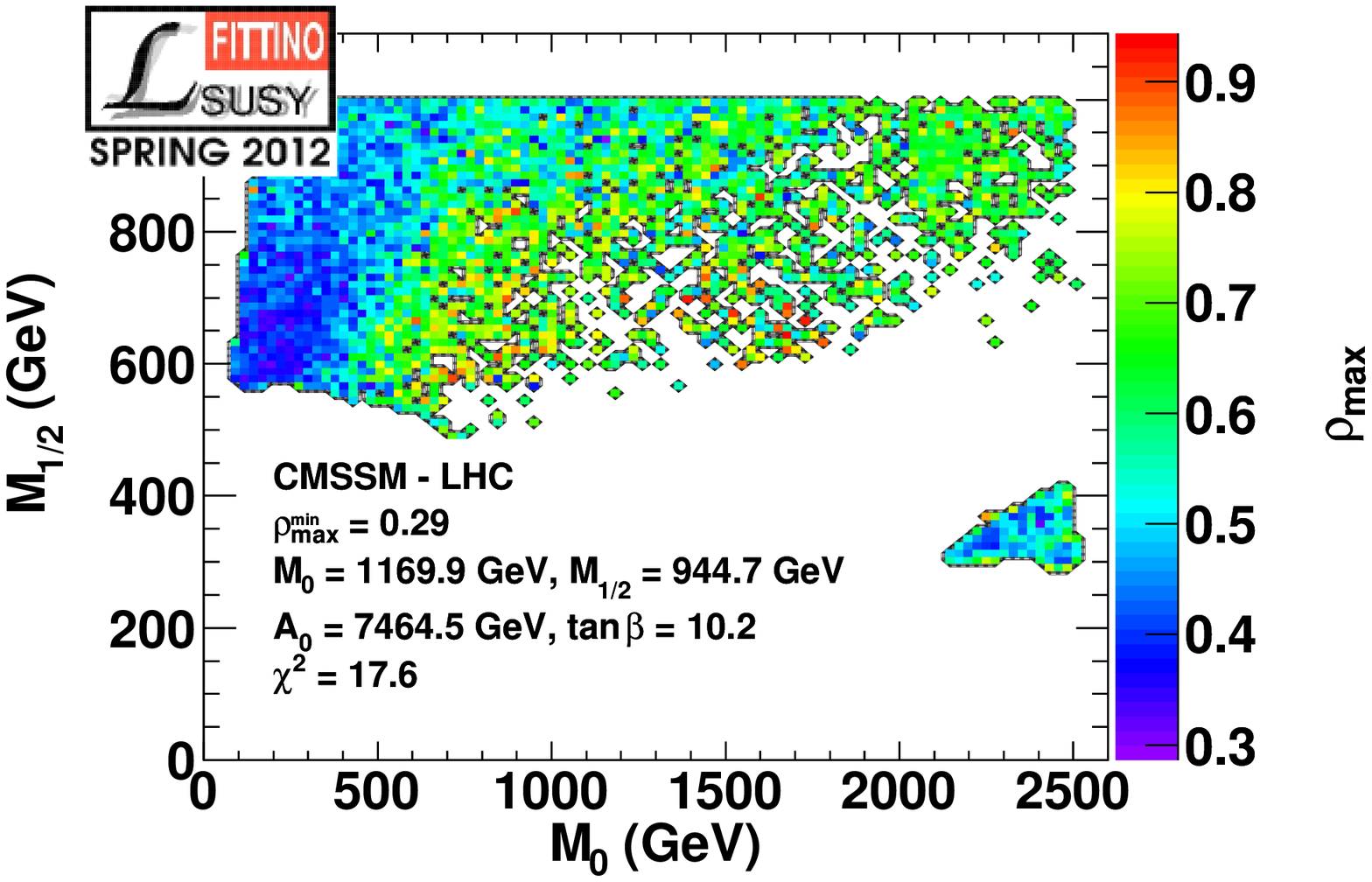}
    \label{fig:NewFTLHC_2fbm0vsm12}
  }
  \subfigure[]{  
    \includegraphics[width=0.47\textwidth,clip=]{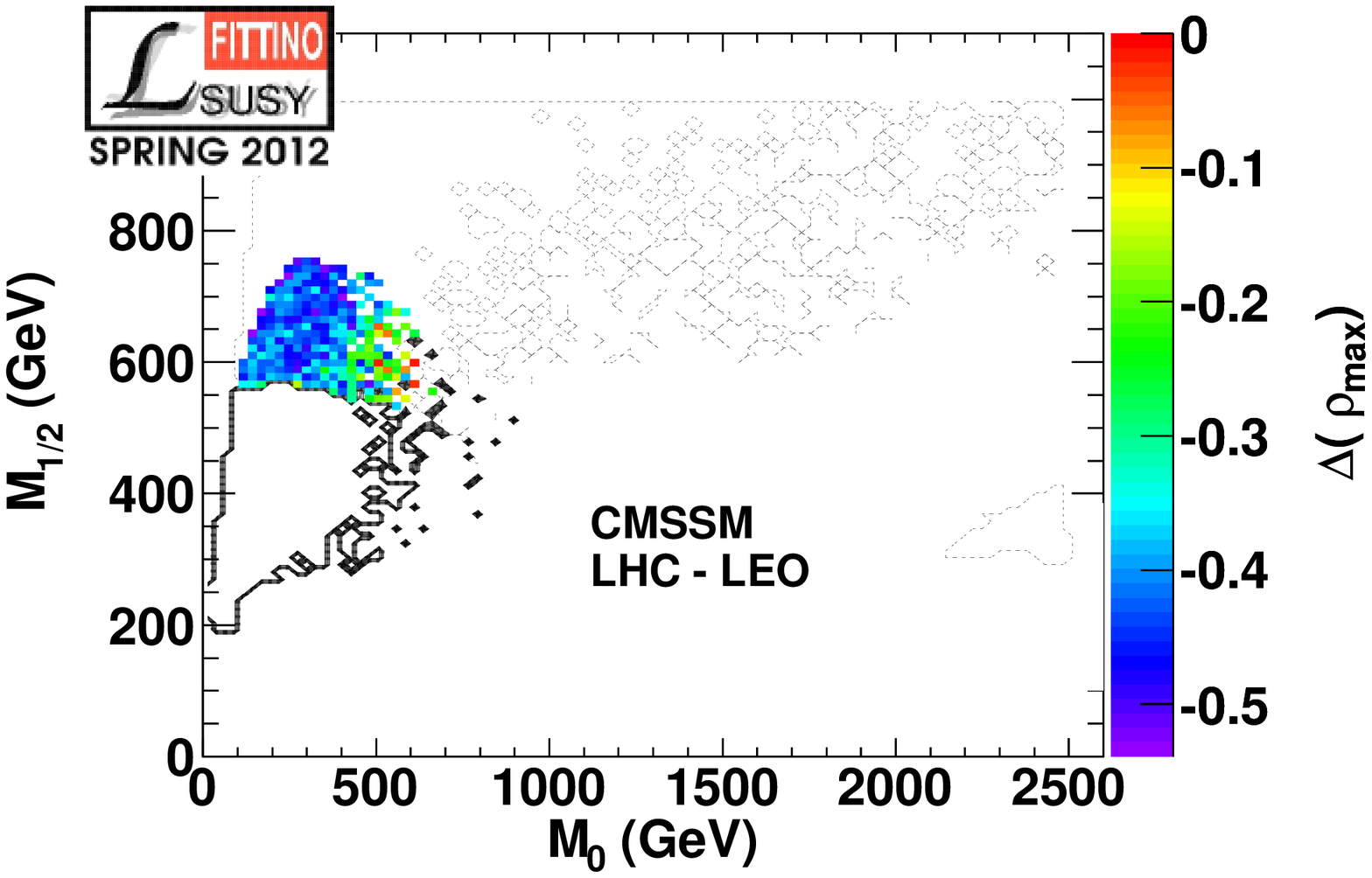}
    \label{fig:NewFTLHC_diffm0vsm12}
  }
  \end{center}
  \caption{Minimum value of $\varrho_{\rm max}$ as a function of $M_{0}$ 
  and $M_{1/2}$, where $A_{0}$ and $\tan\beta$ are profiled. (a) shows 
  the minimum  correlation in the $95\%$-CL region of the pre-LHC fit, 
  (b) shows the minimum correlation in the $95\%$-CL region of the LHC 
  fit, (c) shows the difference of both distributions. For nearly all 
  values of $M_{0}$ and $M_{1/2}$, the pre-LHC fit shows a larger minimum 
  correlation. The smallest values of $\varrho_{\rm max}$ 
  are 0.62 (0.29) for the pre-LHC (LHC) fit.
  }\label{fig:NewFTLHC}
\end{figure}

The results of this study including the pre-LHC and LHC constraints
are summarized in Fig.~\ref{fig:NewFTLHC}. In
Fig.~\ref{fig:NewFTLHC_0fbm0vsm12} we show $[\varrho_{\mathrm
{max}}]_{\mathrm{min}}$ in the $M_0$-$M_{1/2}$ plane, when including
only the pre-LHC constraints. $A_0$ and $\tan\beta$ are profiled. We see
that due to the now restricted parameter space the correlations are
all above 0.75, and well above the 0.2 considered above. Moreover the
correlation increases when increasing $M_{1/2}$ and in particular
when increasing $M_0$

In Fig.~\ref{fig:NewFTLHC_2fbm0vsm12} the LHC limits are now also
taken into account.  The colored region is significantly extended
because the quality of the fit has gone down and the $2\sigma$ region
is larger. At the same time, because $A_0$ and $\tan\beta$ can extend
over wider ranges, the correlation is markedly decreased compared to
Fig.~\ref{fig:NewFTLHC_0fbm0vsm12}. It is however still everywhere
above the baseline value of 0.2.

\begin{figure}[t]
\begin{center}
  \subfigure[]{  
    \includegraphics[width=0.47\textwidth,clip=]{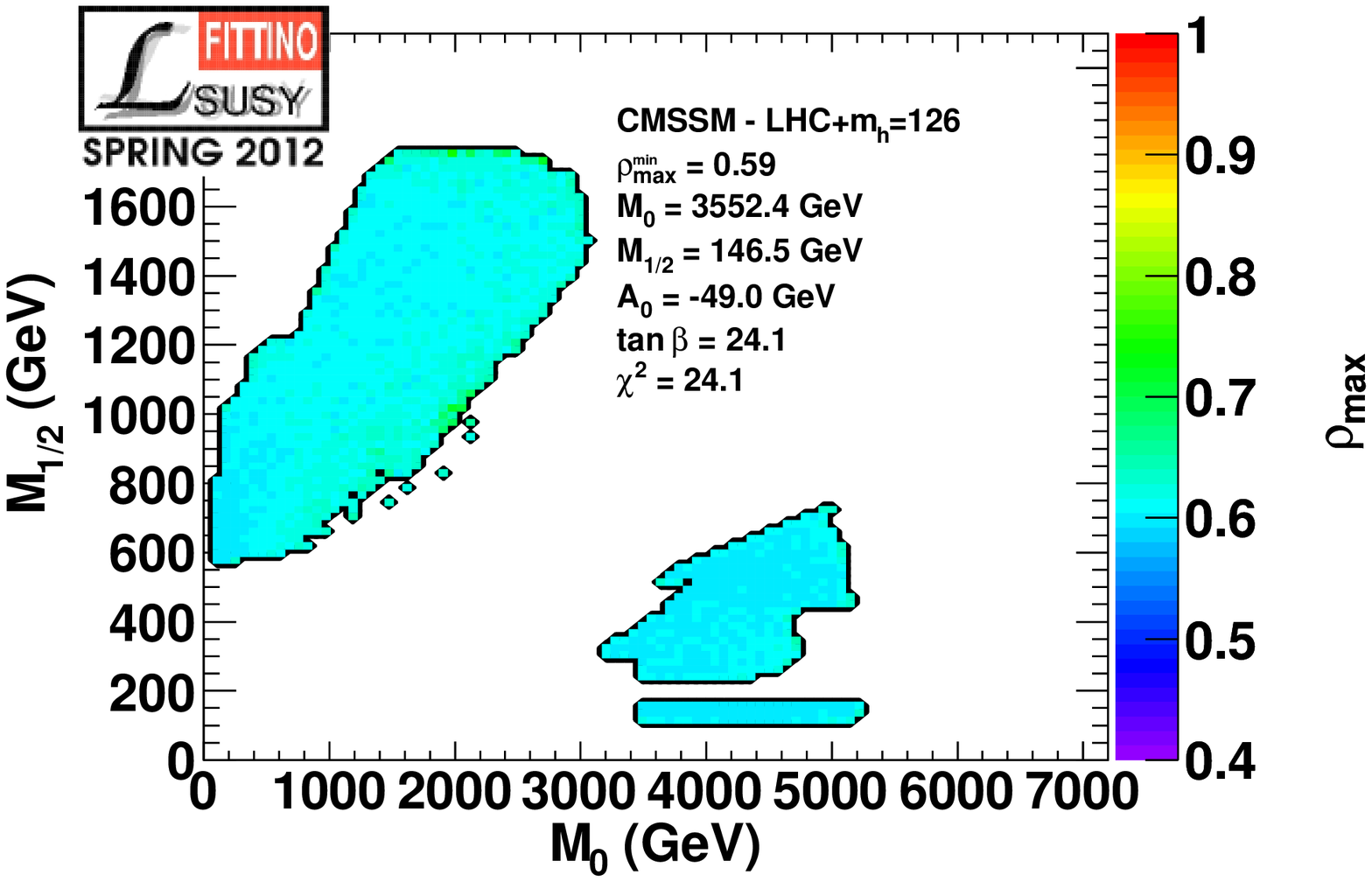}
    \label{fig:NewFTwithHiggs_m0vsm12}
  }
  \subfigure[]{  
    \includegraphics[width=0.47\textwidth,clip=]{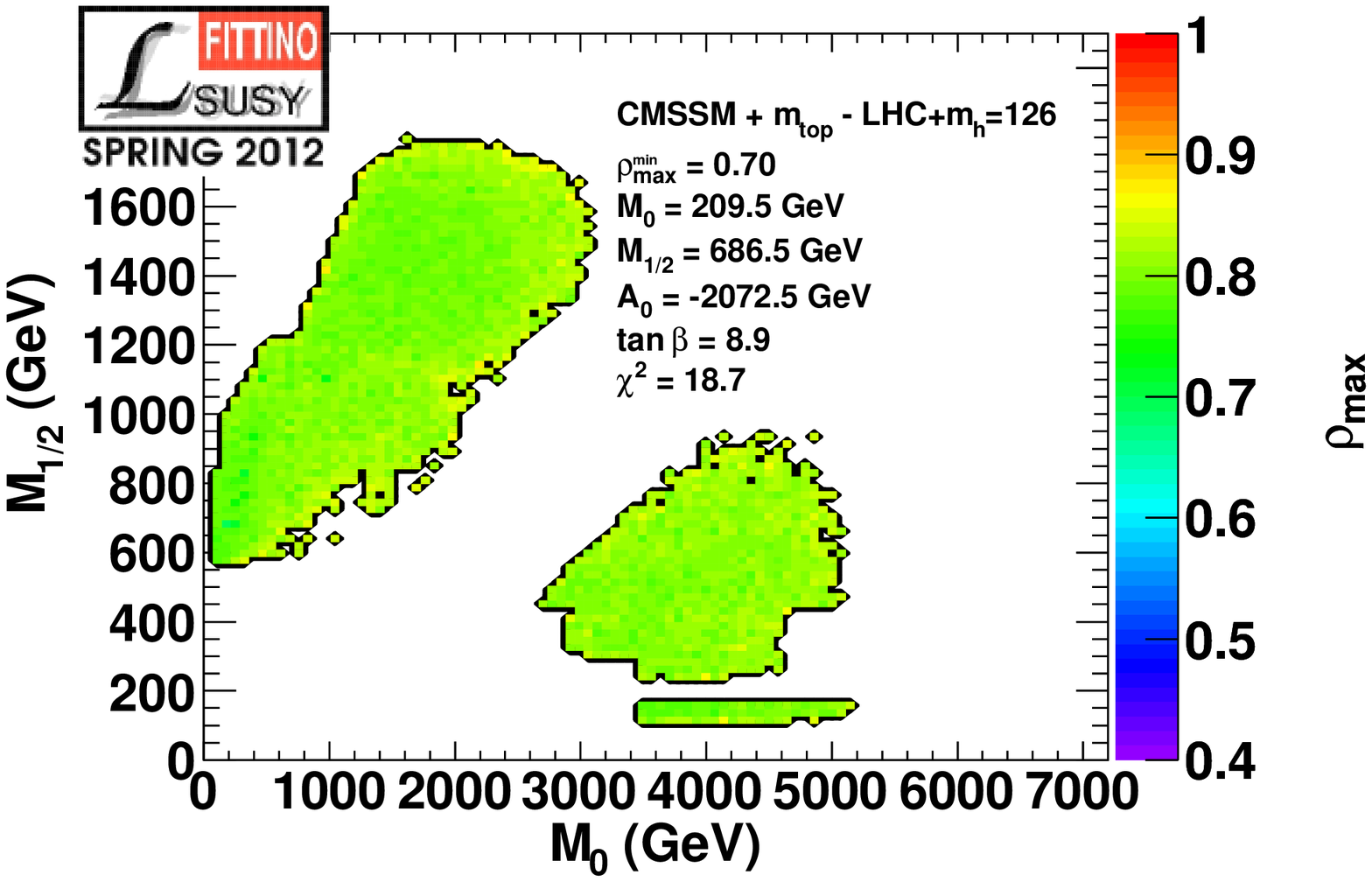}
    \label{fig:NewFTwithHiggs_topm0vsm12}
  }
  \subfigure[]{  
    \includegraphics[width=0.47\textwidth,clip=]{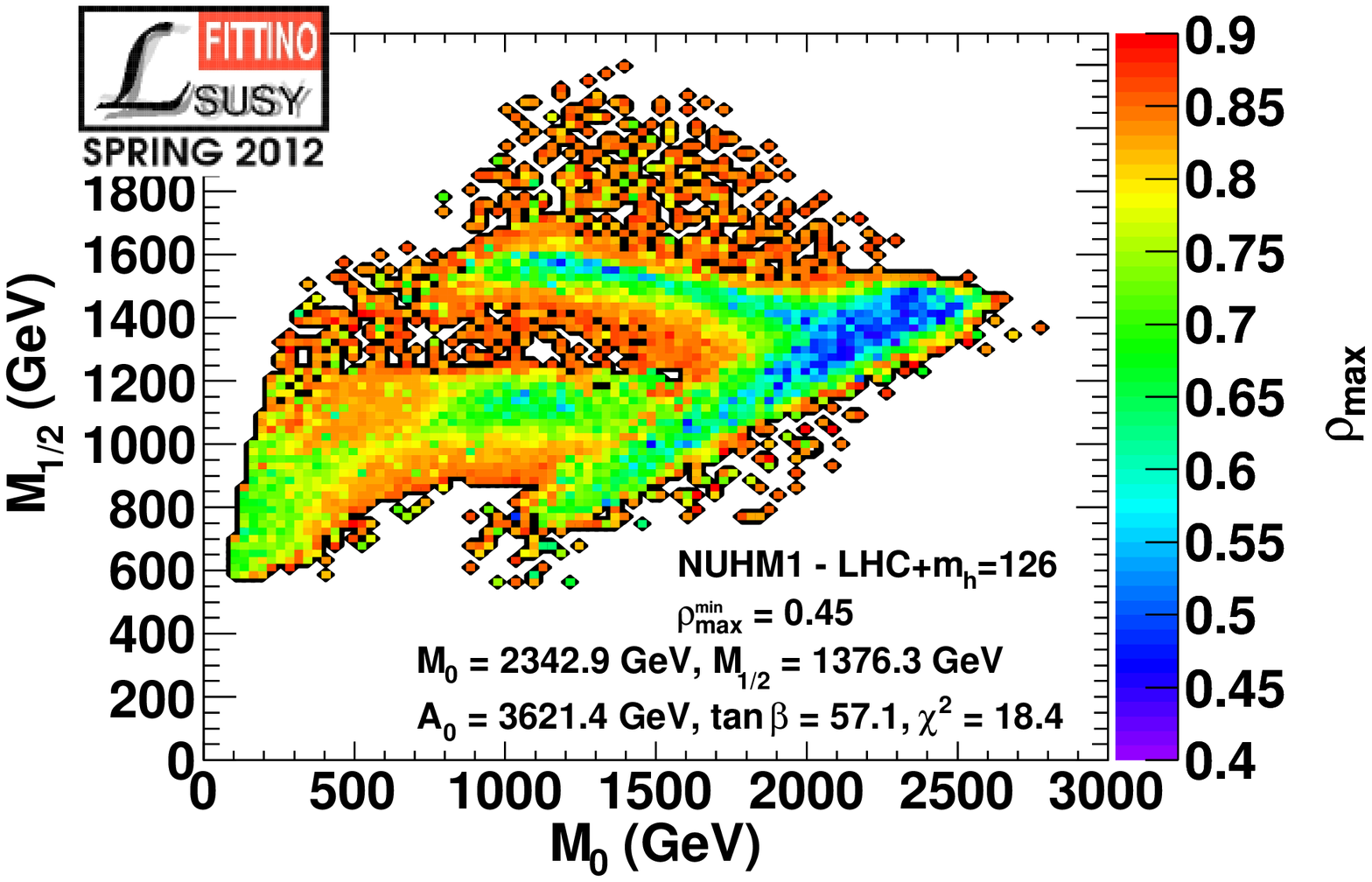}
    \label{fig:NewFTwithHiggs_nuhm1m0vsm12}
  }
\end{center}
  \caption{Minimum value of $\varrho_{\rm max}$ as a function of $M_{0}$
  and $M_{1/2}$, where $A_{0}$ and $\tan\beta$ are profiled. (a) and
  (b) show the minimum correlation for a fit of the CMSSM model with a
  fixed (free) $m_{t}$. For all points, the minimum correlation
  increases for a free $m_{t}$. The minimum value for $[\varrho_
  {\rm max}]_{\mathrm{min}}$ is $0.59$ ($0.70$) for the fit with
  fixed (free) $m_{t}$. In addition, (c) shows the $\varrho_{\rm max}$
  as a function of $M_{0}$ and $M_{1/2}$ for a fit of the NUHM1 model,
  where $A_{0}$, $\tan \beta$ and $m_{H^0}$ are profiled. The minimum
  value of $\varrho_{\rm max}$ for this fit is $0.45$. All fits use $m_{h}
  = 126$ GeV.  }\label{fig:NewFTwithHiggs}
\end{figure}

In Fig.~\ref{fig:NewFTwithHiggs}, we consider slightly modified
scenarios, with one additional fit parameter. In all cases we have
fixed the lightest Higgs mass to $m_h=126\,$GeV. In
Fig.~\ref{fig:NewFTwithHiggs_m0vsm12}
[Fig.~\ref{fig:NewFTwithHiggs_topm0vsm12}] we show the minimal
correlation for a fixed (free) top mass, $m_t$. Due to the fixed Higgs
mass the quality of the fit degrades significantly. The 2$\sigma$
region is thus now extends to very high values of $M_0$. For the free
top mass the correlations are slightly higher than for the fixed case.
In Fig.~\ref{fig:NewFTwithHiggs_nuhm1m0vsm12} we show the correlations
in the case of the previously described model NUHM1. The minimum value
of the correlation for this fit is 0.45. In all cases shown in
Fig.~\ref{fig:NewFTwithHiggs}, $A_{0}$ and $\tan \beta$, as well as
$m_{t}$ and $M_{h}$ where applicable, are profiled.

\begin{figure}[t]
  \begin{center}
    \subfigure[]{  
      \includegraphics[width=0.47\textwidth,clip=]{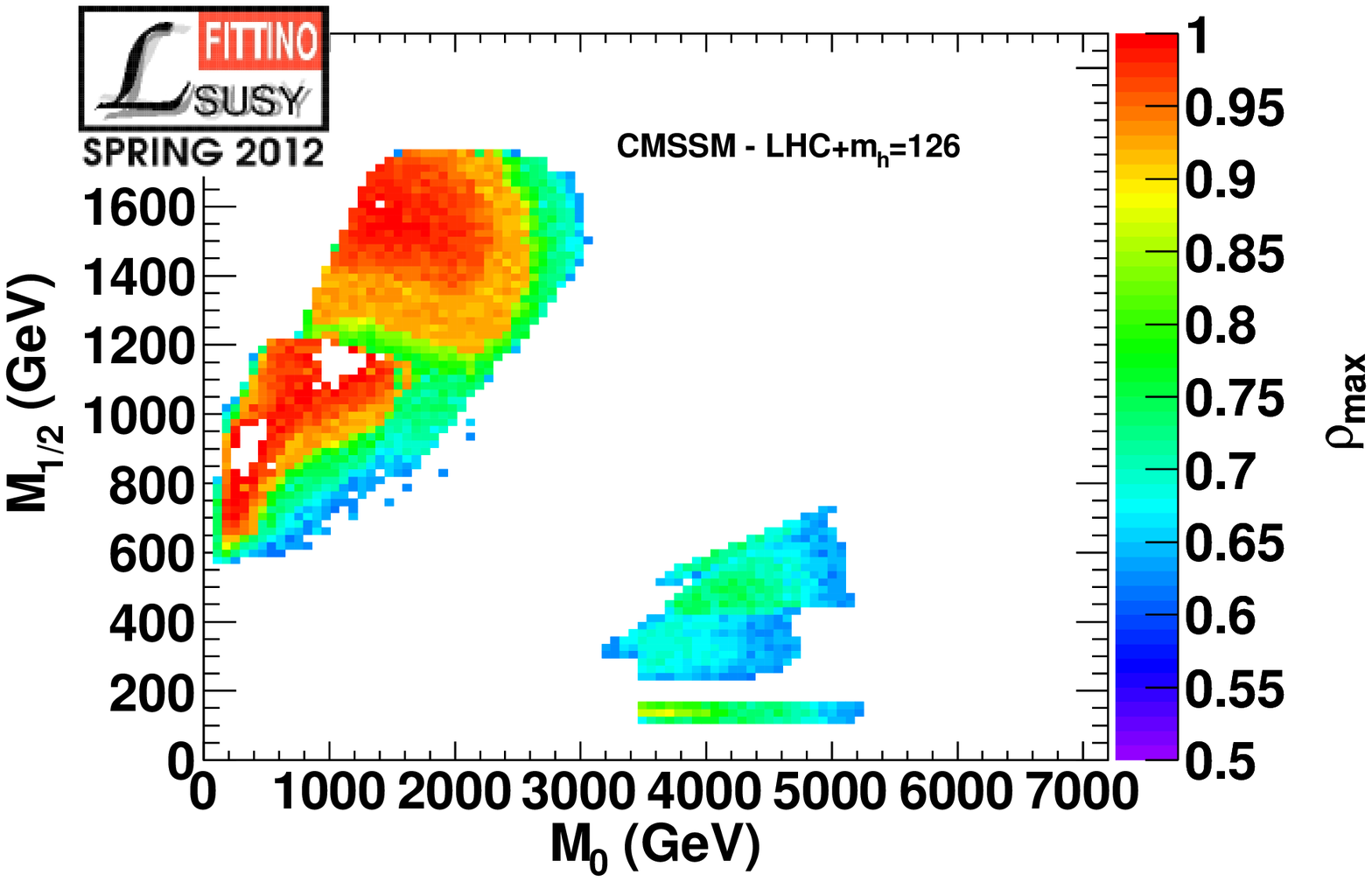}
      \label{fig:NewFTatLowestChi2_lowest}
    }
    \subfigure[]{  
      \includegraphics[width=0.47\textwidth,clip=]{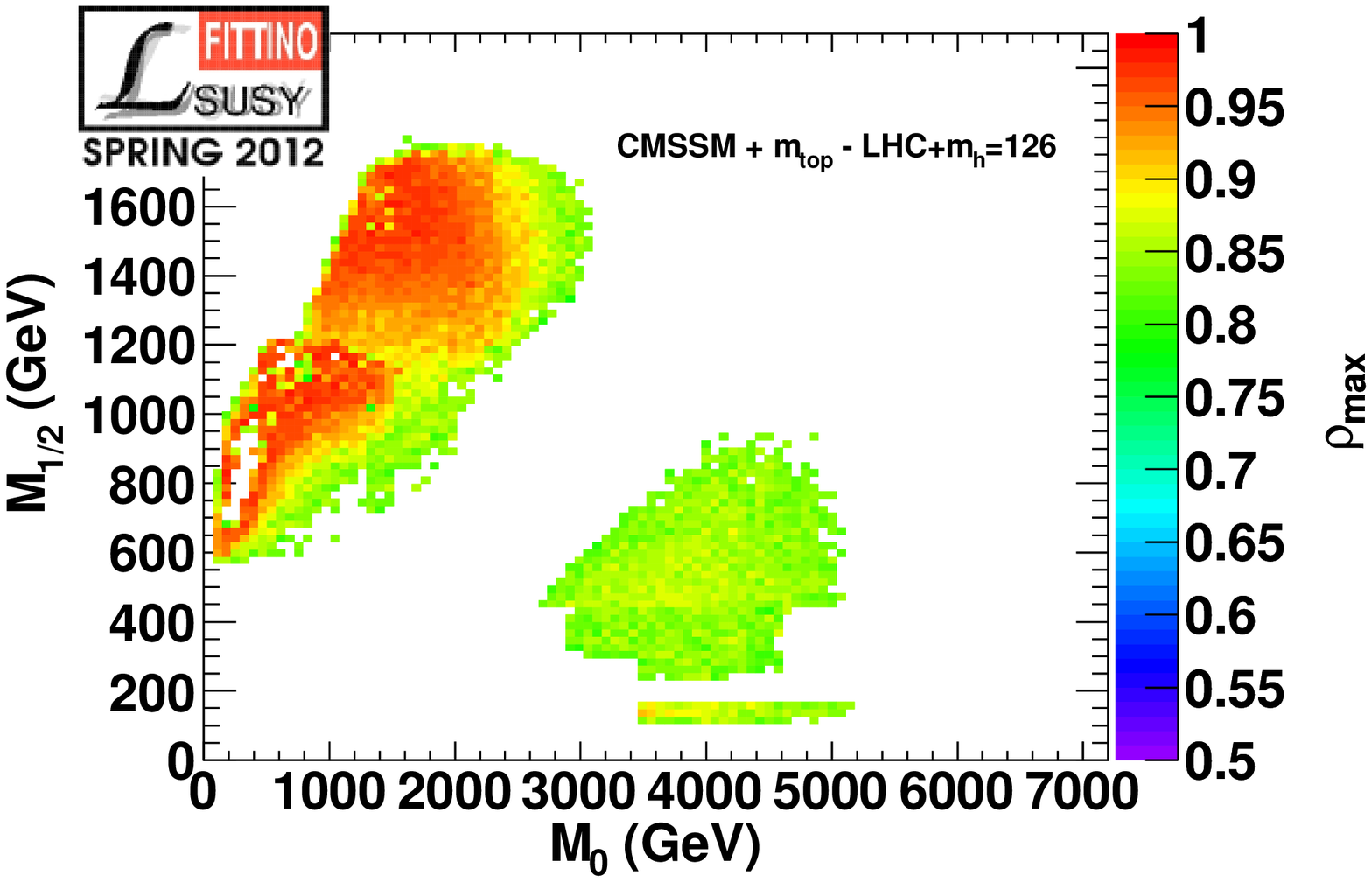}
      \label{fig:NewFTatLowestChi2_mtoplowest}
    }
    \subfigure[]{  
      \includegraphics[width=0.47\textwidth,clip=]{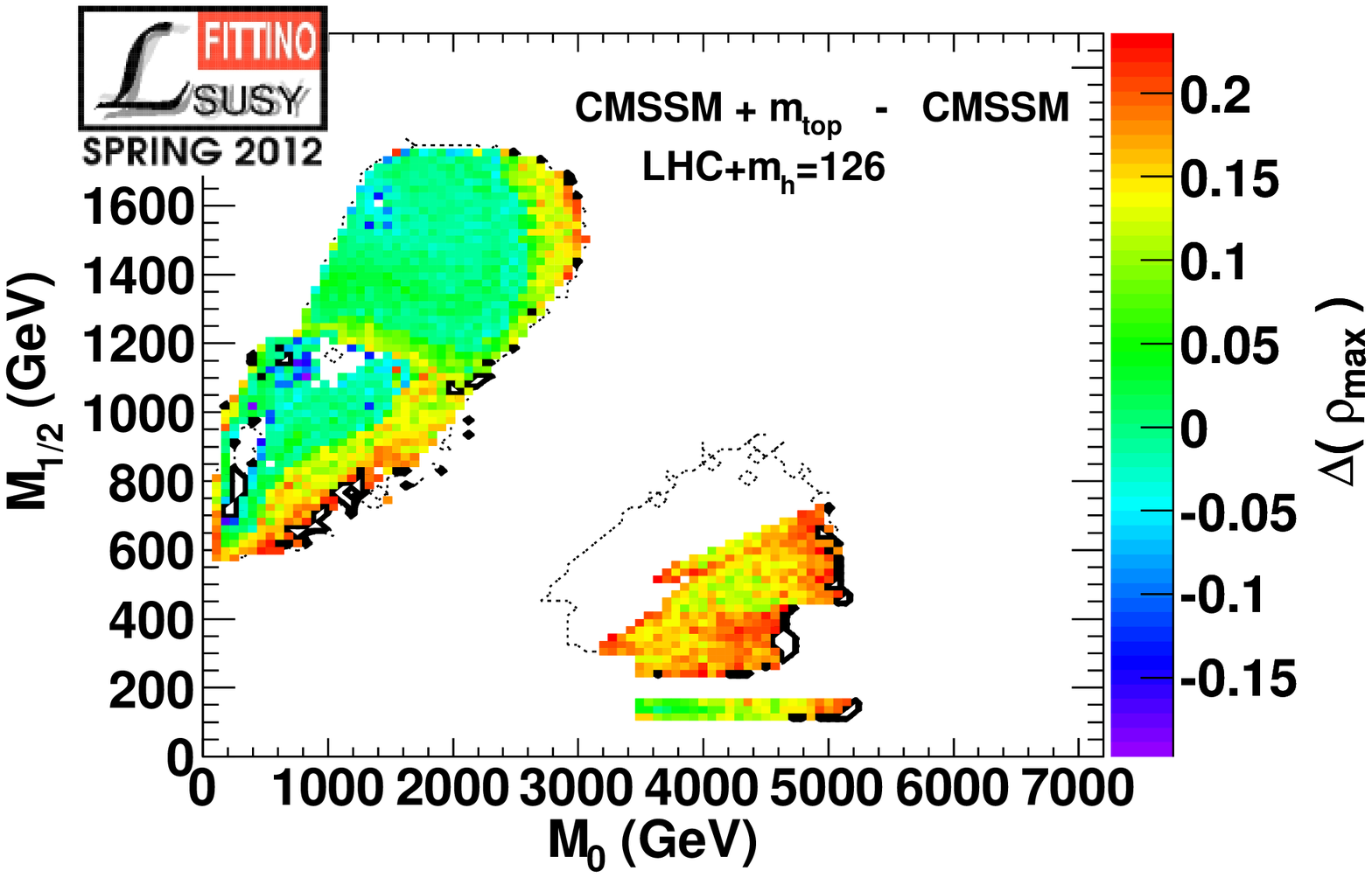}
      \label{fig:NewFTatLowestChi2_mtoplowestdiff}
    }
  \end{center}
  \caption{The value of $\varrho_{\rm max}$ as a function of $M_{0}$ and
  $M_{1/2}$ at the point with the lowest $\chi^2$, where $A_{0}$ and
  $\tan\beta$ are profiled. (a) shows the minimum value for a fit of
  the CMSSM model where $m_{t}$ is fixed, (b) shows the minimum
  value for a fit of the CMSSM model where $m_{t}$ is free. (c)
  shows the difference of the two distributions. Near the best-fit
  point, the minimum of $\varrho_{\rm max}$ is increased if $m_{t}$ is
  fixed, while at the boundaries of the $95\%$-CL region the minimum
  value is decreased. Both fits are constrained by $m_h = 126$
  GeV.}\label{fig:NewFTatLowestChi2}
\end{figure}

As discussed above, $\varrho_{\rm max}$ may vary significantly for similar
values of $M_{0}$ and $M_{1/2}$, if additional points with worse
agreement between model predictions and measurements are reintroduced
by adding another observable. This was so far taken into account by
letting $\chi^2$ vary over the full accessible range, and then
computing $[\varrho_{\rm max}]_{\mathrm{min}}$. It is also interesting to
simply consider the best-fit value of $\chi^2$ for fixed values of
$M_0,\,M_{1/2}$. For the fits of the CMSSM model with $m_h = 126$ GeV,
this is shown in Fig.~\ref{fig:NewFTatLowestChi2}.
Figs.~\ref{fig:NewFTatLowestChi2_lowest}/\ref{fig:NewFTatLowestChi2_mtoplowest}
show $\varrho_{\rm max}$ in the $M_{0}$-$M_{1/2}$ plane at the point with
the lowest $\chi^2$ per bin for the $m_{t}$=fixed/$m_{t}$ = free
fit. Figure \ref{fig:NewFTatLowestChi2_mtoplowestdiff} shows the
difference between the two fits. A negative(positive) entry indicates
that with $m_{t}$=free($m_{t}$=fixed) a lower correlation between
the parameters is found. In particular close to the minimum, $m_{t}$
= free allows for a lower correlation, while in other regions the
maximum correlation between the varied parameters increases if the top
mass is free. White areas indicate that at the corresponding point
with the lowest $\chi^2$ there were less than 6 points within $\Delta
\chi^2 < 0.001$ such that no meaningful correlation could be
calculated.

\subsection{Including $Q$ as a nuisance parameter}
\label{sec:results_Q}

When fitting constrained models such as the CMSSM, the SUSY parameters
are fitted at the GUT scale. The predictions for the observables are
calculated from the MSSM Lagrangian calculated at a scale of
$Q=1$\,TeV where we fixed the scale according to the SPA conventions
\cite{AguilarSaavedra:2005pw}.  Theoretical uncertainties on the
predictions are included in the calculation of the $\chi^2$, as noted
in Section~\ref{sec:input}. However, this does not included the
parametric uncertainties arising from the fact that the masses are
calculated using two-loop RGEs and one-loop (two-loop) threshold
corrections for the sparticle (Higgs) masses.
It is generally assumed in previous work in the
literature (see \textit{e.g.}~\cite{Bechtle:2011dm}), that these uncertainties
are negligible or at least sub-dominant for the observables described
in Section~\ref{sec:input}.

\begin{figure}[t]
  \begin{center}
    \includegraphics[width=0.49\textwidth,clip=]{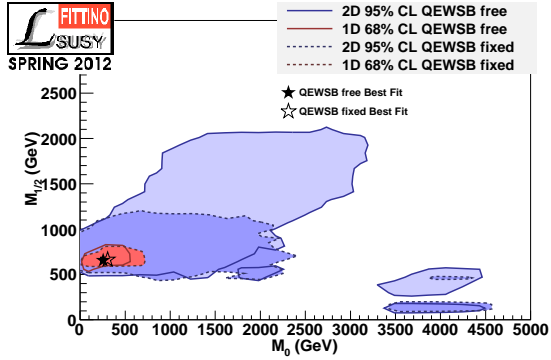}
  \end{center}
  \caption{Fits including LHC exclusions, with and without $Q$ as a
    nuisance parameter. Details about the best fit point are given in
    Tab.~\ref{tab:fitsummary}. Results are shown in the
    ($M_0,M_{1/2}$) plain. The upper limit on the SUSY parameter scale is
    strongly lifted upwards once the RGE uncertainty is explicitly
    included in the fit by floating the scale $Q$.}\label{fig:QEWSB}
\end{figure}

However, there is a straight-forward possibility to include these
previously neglected uncertainties: The scale $Q$, where the
spectrum is calculated, can be included
as a nuisance parameter into the fit. It is allowed to float within
$\frac{1}{2}\sqrt{m_{{\tilde t}_1}m_{{\tilde
      t}_2}}<Q<2\sqrt{m_{{\tilde t}_1}m_{{\tilde t}_2}}$. Then, the
parameter uncertainties on the physical parameters of the model can be
obtained by profiling over the hidden dimensions including $Q$. As
shown in Fig.~\ref{fig:QEWSB} for a fit including LHC exclusion,
\texttt{HiggsBounds} limits and the full {\texttt{AstroFit}} limits,
this has a sizable effect on the upper limit of the available
parameter space, which is dominantly determined by $(g-2)_{\mu}$ and
$\Omega_{DM}$. The reason for this large increase is the flatness
of the $\chi^2$ distribution which implies that already a small change
$\Delta \chi^2$ can lead to the observed changes for the allowed 
regions. However, the best regions hardly changes.
This result suggests that in future fits, the RGE
uncertainty should be included as well.

 We note for completeness that for very large values of
$M_0$ and/or $M_{1/2}$ one obtains 
1 TeV $< \frac{1}{2}\sqrt{m_{{\tilde t}_1}m_{{\tilde t}_2}}$.
However, in these regions the theoretical uncertainties for 
the mass calculations in codes likes \texttt{SPheno} or \texttt{SoftSUSY}
is most likely larger than the scale variations suggests because here
one should use a multi-scale approach as discussed in
\cite{Baer:2005pv,Box:2007ss,Box:2008xu} 
in contrast to the single scale approach of the
current implementations \cite{Allanach:2001kg,SPheno}.

\section{Conclusions}
\label{sec:conclusion}

We have presented a global fit of the constrained MSSM and a
non-universal Higgs mass model (NUHM1) to low-energy precision
measurements, dark matter observables and, in particular, current LHC
exclusion limits from direct SUSY searches in the zero-lepton plus
jets and missing transverse energy channel.  A restricted model with
universal soft SUSY breaking parameters at the unification scale like
the CMSSM does not provide a good description of all observables. The
best fit predicts typical sparticle masses in the TeV-range, albeit
with large errors.  We find that the description of the low-energy
observables, $(g-2)_{\mu}$ in particular, and the non-observation of
SUSY at the LHC become increasingly incompatible within the
CMSSM. Note that the LHC exclusion in the zero-lepton, jets plus
$E_T^{\rm miss}$ channel mainly constrain the squark and gluino
masses, while $(g-2)_{\mu}$ and other low-energy observables are
sensitive to the color-neutral sparticles. Supersymmetric models with
common scalar and gaugino masses like the CMSSM connect these two
sectors and thus lead to an increasing tension between $(g-2)_{\mu}$
and the LHC search results, which would not necessarily be present in
more general SUSY scenarios.

In general, there is an intricate interplay between the different
contributions to the global fit, and an accurate estimate of the LHC
exclusions also in parameter regions away from the published 95\%\,CL
contours is mandatory. We have thus employed an elaborate simulation
of recent LHC search analyses based on state-of-the-art Monte-Carlo
tools and a public detector simulation. We note that it appears
challenging to provide a fast and accurate implementation of LHC
results in a global fit for more general SUSY models with a larger
number of parameters.

A very strong constraint on SUSY can be expected from a measurement of
the light Higgs boson mass. Within the global CMSSM fit the light
Higgs scalar is Standard-Model-like, with a preferred mass around $m_h
\approx 117$\,GeV. We have studied the impact of a potential Higgs
observation on the global fit and find that a light Higgs boson with
mass $m_h\gtrsim125$\,GeV is hardly compatible with the other data
within the CMSSM. The large sparticle mass scales needed to
accommodate Higgs bosons with $m_h\gtrsim125$\,GeV drive the fit into
a region which is incompatible with $(g-2)_{\mu}$ and $B$-physics
observables.  This tension is reduced in models like the NUHM1 where
the Higgs sector is decoupled from the squark and slepton
sector. Still, even in the NUHM1, the interplay of ${\cal
  B}(B_s\to\mu\mu)$ and the lightest Higgs boson mass provides strong
constraints. We have argued that important further constraints on SUSY
models can be expected from improved limits on ${\cal
  B}(B_s\to\mu\mu)$ and, in particular, a measurement of the Higgs
branching fractions or ratios of branching fractions. For the NUHM1,
the search for pseudoscalar Higgs bosons in the channel $A^0 \to
\tau\tau$ will be another interesting input.

We have discussed in detail the implications of the CMSSM fit for
direct and indirect dark matter searches. While direct dark matter
detection claims from CoGeNT or DAMA/LIBRA cannot be accommodated
within the CMSSM, the current upper direct detection limits provided
by XENON do not restrict the CMSSM parameter space beyond the
constraints from low-energy observables, LHC limits and the dark
matter relic density.  Projections of the direct dark matter detection
limits based on a non-observation by XENON1T, on the other
hand, would provide a further important probe of the available SUSY
parameter space. The current indirect detection upper limits as
implemented in {\tt AstroFit} are still too weak to have an impact on
the CMSSM fit. However, a more comprehensive future treatment of
indirect detection limits may have the potential to probe the SUSY
parameter space in regions which are complementary to direct detection
constraints.

In addition, we have studied the fine-tuning within the SUSY models in
two different ways. For the fine-tuning parameter $\Delta$, measuring
the relative dependence of $m_Z$ on all parameters, we find a strong
increase between the fit without and with LHC direct search
limits. For the CMSSM, a minimal value of $\Delta=193.5$ is found for
the LHC fit, which is generally regarded as unattractive. In addition,
we introduce a new concept in the form of the relative precision at
which the parameters need to be fine-tuned with respect to each other,
in order to achieve the same overall agreement with all relevant
observables relative to their uncertainties. Here, we find that within
the $2\sigma$ uncertainty range the maximal correlations amongst
parameters are still in the range of $\rho_{max}\lesssim0.6$, hinting
at an acceptable level of correlations. However, the correlation
naturally increases strongly at the very minimum of the fit.


To further explore the implications of the LHC searches for
supersymmetry it is mandatory to include a larger set of LHC
signatures, including those which directly probe the non-colored
sparticles, and to consider global fits of less restricted SUSY models
beyond the CMSSM. The challenge for such future global analyses will
be to provide a fast and accurate implementation of the LHC results,
if possible for an accurate combination of different search channels,
for more general SUSY models.

\section*{Acknowledgments} We thank Kevin Kr\"oninger and Ben O'Leary
for valuable discussions, and Wolfgang Ehrenfeld for invaluable
computing support. We also thank the Helmholtz Alliance and DESY for
providing Computing Infrastructure at the National Analysis
Facility. This work has been supported in part by the Helmholtz
Alliance ``Physics at the Terascale'', the DFG SFB/TR9 ``Computational
Particle Physics'', the DFG SFB 676 ``Particles, Strings and the Early
Universe'', the European Community's Marie-Curie Research Training
Network under contract MRTN-CT-2006-035505 ``Tools and Precision
Calculations for Physics Discoveries at Colliders'', the DFG
Emmy-Noether grant BR 3954/1-1, the DFG Emmy-Noether grant HE 5560/1-1
and the Helmholtz Young Investigator Grant VH-NG-303. MK thanks the
CERN TH unit for hospitality. HD thanks SCIPP at UC Santa Cruz for
hospitality during this work.

\end{document}